\documentclass[reprint,
preprintnumbers,
 amsmath,amssymb,
 aps,
prb,
]{revtex4-2}
\usepackage{graphicx}
\usepackage{dcolumn}
\usepackage{multirow}
\usepackage{dsfont}
\usepackage{braket}
\usepackage{verbatim}
\usepackage{xcolor}

\usepackage{bm}
\usepackage{hyperref}
\hypersetup{
    colorlinks=true,
    linkcolor=blue,
    citecolor=blue,
    urlcolor=blue
}

\newtheorem{theorem}{Theorem}

\begin{document}

\preprint{MIT-CTP/5651}
\title{Robust and Parallel Control of Many Qubits}

\author{Wenjie Gong}
\author{Soonwon Choi}%
\affiliation{%
Center for Theoretical Physics, Massachusetts Institute of Technology, Cambridge, Massachusetts 02139, USA
}%

\date{\today}

\begin{abstract}
The rapid growth in size of quantum devices demands efficient ways to control them, which is challenging for systems with thousands of qubits or more. Here, we present a simple yet powerful solution: robust, site-dependent control of an arbitrary number of qubits in parallel with only minimal local tunability of the driving field.
Inspired by recent experimental advances, we consider access to only one of three constrained local control capabilities: local control of either the phase or amplitude of the beam at each qubit, or individual $Z$ rotations. In each case, we devise parallelizable composite pulse sequences to realize arbitrary single-qubit unitaries robust against quasistatic amplitude and frequency fluctuations.
Numerical demonstration shows that our approach outperforms existing sequences such as BB1 and CORPSE in almost all regimes considered, achieving average fidelity~$>0.999$ under a decoherence rate of~$\sim 10^{-5}$, even with a few percent amplitude and frequency error.
Our results indicate that even for very large qubit ensembles, accurate, individual manipulation can be achieved despite substantial control inhomogeneity.   
\end{abstract}

\maketitle
Over the past decade, rapid progress in quantum technology has enabled digital quantum processors to push computational limits \cite{Arute2019, Kim2023, Zhu2023, Bluvstein2022} and analog quantum simulators to unveil insights into diverse many-body phenomena \cite{Bernien2017, Lukin2019, Semeghini2021, Choi2023, Farrell2023, Mi2022, Gonzlez-Cuadra2025}.
Such investigations involved devices with hundreds of qubits, and it is soon expected that quantum systems with an order of magnitude more qubits will appear~\cite{IBM}. This poses an immediate and pressing challenge: How can we efficiently control a very large number of qubits in realistic settings?

The fast, arbitrary, and robust manipulation of individual qubits within a larger ensemble is difficult in many quantum platforms. 
In superconducting quantum registers, a considerable number of wires is needed to manage the control and readout of each qubit---a requirement that grows increasingly impractical with larger system sizes \cite{Kjaergaard2020, Siddiqi2021}.
Platforms based on quantum optics, including neutral atom tweezer arrays and trapped ions, can partially circumvent this ``wire problem'' by manipulating qubits with laser beams, thus enabling scalable control through optical technology~\cite{Brown2016, Graham2023, Shih2021, Pogorelov2021, Wang2023, Cesa2023, Zhang2023, Bluvstein2026, Zhou_2025, Ransford2025}. 

Though a variety of technical feats, including fast switching rates, high contrast, and low cross-talk, must be achieved to build large-scale systems of optically addressed qubits, multiple array architectures capable of manipulating thousands of different sites have already been demonstrated \cite{SchlosserPRL2023, PauseOptica24, Norcia2024, Gyger2024, Manetsch2024, Chiu2025}. Nevertheless, implementing high-fidelity, locally-controlled unitaries in parallel under realistic conditions remains an outstanding experimental challenge. In particular, in optical platforms, control is accomplished with only a single or few laser beams, making the full individual manipulation of each qubit a complex task to engineer~\cite{Binai2023, Christen2022}. For example, conventional technologies such as acousto-optic deflectors or spatial light modulators are all limited in their degree of local tunability~\cite{Binai2023}.

In this work, we develop a method to efficiently execute robust, arbitrary, site-dependent single-qubit unitaries in parallel for a large number of qubits. Specifically, inspired by recent experimental developments~\cite{Schindler2013, Lee2016, Polloreno2022, Barnes2022, Shaw2023, Labuhn2014, deLeseleuc2017, Eckner2023, Graham2022}, we consider three different settings of restricted controls, where one can apply pulses to a qubit ensemble with only minimal local tunability (Fig. 1). Under each such setting, we present the shortest pulse sequences that realize arbitrary single-qubit unitaries in parallel.
Furthermore, we devise \emph{parallelizable} composite pulse sequences \cite{Levitt1986, Merrill2012, Brown2004, Hugh2005, Cummins2000, Cummins2003, Torosov2014, Torosov2018, Wimperis1994, Gevorgyan2021, KukitaPRA2022, Kukita2023, Gevorgyan2023, Fromonteil2023} that are robust against both pulse amplitude and off-resonance (detuning) error to account for control imperfections in realistic systems; in large ensembles of qubits, such control errors arise inevitably due to inhomogeneity. 
 \begin{figure}
    \centering
    \includegraphics[width = 0.8\columnwidth]{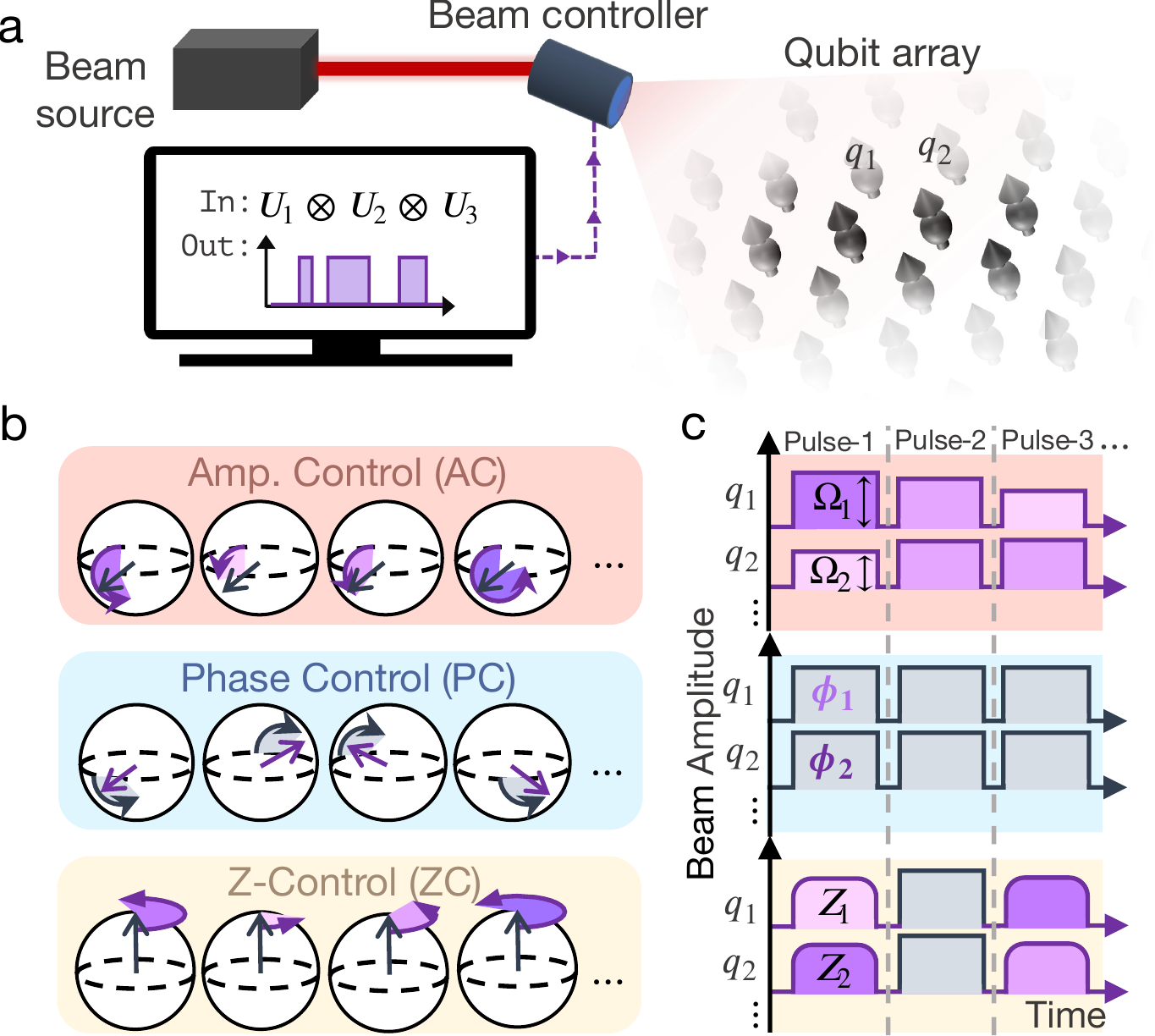}
    \caption{
    {\bf a.} Many qubits are manipulated in parallel, effecting unitaries $U_1\otimes U_2\otimes \cdots$, by applying a sequence of global control beams with restricted tunability.  
    {\bf b.} We consider three kinds of local control (purple): the phase (PC) or amplitude (AC) of the beam can be individually controlled, or individual $Z$-rotations can be applied (ZC). {\bf c.} Under each type of local control, parallelizable operations are realized by simultaneous pulses on all qubits with a certain locally variable parameter.
    }
    \label{fig:intfig}
\end{figure}

To construct our robust pulse sequences, we combine three key ideas: concatenation, pulse primitives, and numerical search. Concatenation \cite{Bando2013} is a technique whereby two pulse sequences which compensate for different errors can be merged to form a doubly-robust sequence. Pulse sequences robust against a single type of error---the building blocks of concatenation---are accordingly deemed pulse primitives. We introduce several classes of pulse primitives and, in the absence of an analytical form, further employ gradient-based optimization to numerically search for robust sequences. With these ideas, we create novel pulse sequences that outperform existing composite pulses, when applicable, at almost any level of amplitude and off-resonance error in the presence of decoherence. Our results thus enable efficient, parallel, and high-fidelity quantum information processing on large, near-term quantum systems.

\textbf{Optimal, Parallel, Arbitrary Unitaries under Global Driving---}\label{sec:basicopt} Our objective is to realize an arbitrary unitary $\otimes_i\,U_i$ in parallel across all qubits $i$ under a global driving field, where $\{U_i\}\,\in$ SU(2) is given as input. In the frame rotating at the frequency of the field, the Hamiltonian of a resonantly driven qubit takes the form $\mathcal{H}\,=\,\frac{\Omega}{2}\,(\cos(\phi)\hat\sigma_{x}\,+\,\sin(\phi)\hat\sigma_y)$, with Rabi frequency $\Omega$, the phase of the field $\phi$, and the Pauli operators $\hat \sigma_{x(y)}$. A single pulse of the field is then given by a SU(2) unitary rotation in the $x-y$ plane. We adopt the notation for a pulse $[\theta]_{\phi}\,\equiv\,\exp(-i \frac{\theta}{2}\hat\sigma_{\phi}),\;\hat \sigma_{\phi}\,\equiv\,\cos(\phi)\hat\sigma_{x}\,+\,\sin(\phi)\,\hat\sigma_y$, which describes a rotation around $\phi$ in the $x\,-\,y$ plane by an angle $\theta$. We further define SU(2) rotations about the axes of the Bloch sphere as $X(\varphi),\,Y(\varphi)$, and $Z(\varphi)$.

When a global beam is applied to the qubit ensemble, we consider one of three local control capabilities: individual tunability of the phase $\{\phi_i \}$ of the beam (phase control, or PC);  individual tunability of the beam intensity $\{\Omega_i\}$ (amplitude control, or AC); or individual rotations about $\hat z$ (Z-control, or ZC) (Fig.~\ref{fig:intfig}{\bf b,c}).
These schemes are motivated by common experimental approaches; for instance, ZC is enabled by local AC-Stark shifts, a standard practice in trapped-ion systems \cite{Schindler2013, Lee2016, Polloreno2022}, while PC \cite{Barnes2022, Shaw2023}, AC \cite{Barnes2022}, and ZC \cite{Labuhn2014, deLeseleuc2017, Eckner2023, Graham2022} have also been demonstrated in tweezer arrays. 

We now present the optimal sequence to implement $\bigotimes_i\,U_i$ in each scenario. For AC, as we have site-dependent control of rotation amplitudes, this can be trivially achieved by three pulses
\begin{align}\label{eq:arbuac}
U^{AC}(\alpha_i, \beta_i, \gamma_i)
&\equiv {[\alpha_i]}_0{[\beta_i]}_{\frac{\pi}{2}}{[\gamma_i]}_0.
\end{align}
The unitary at each qubit $i$ is then parameterized by $XYX$ Euler angles, $U^{AC}(\alpha_i,\beta_i,\gamma_i)\,=\,X(\alpha_i)Y(\beta_i)X(\gamma_i)$, while the phases (rotation axes) are globally fixed. For PC, we can similarly realize
\begin{align}\label{eq:arbupc}
\begin{split}
U^{PC}(\alpha_i, \beta_i, \gamma_i) &\equiv [\pi/2]_{\alpha_i} [\pi]_{\frac{\gamma_i +\alpha_i - \beta_i}{2}}[\pi/2]_{\gamma_i},
\end{split}
\end{align}
where an explicit calculation shows the equivalence $U^{PC}(\alpha_i,\beta_i,\gamma_i)\,=\,Z(\alpha_i)  Y(\beta_i)Z(2\pi - \gamma_i)$, thus capturing all unitaries with ZYZ Euler angles \cite{Chen2021}.  
Finally, the sequence for ZC also follows from ZYZ Euler angles,
\begin{align}\label{eq:arbuzc}
\begin{split}
    U^{ZC}(\alpha_i, \beta_i, \gamma_i) \equiv Z(\alpha_i) [\pi/2]_\pi Z(\beta_i) [\pi/2]_0  Z(\gamma_i) 
\end{split}
\end{align}
with $U^{ZC}(\alpha_i,\beta_i,\gamma_i)\,=\,Z(\alpha_i)Y(\beta_i)Z(\gamma_i)$. The optimality of these sequences can be shown by simple parameter counting. As an arbitrary unitary requires three parameters and we have one individually tunable variable per pulse, the shortest number of pulses is three for AC and PC. For ZC, the three unitary parameters correspond to three individual $Z$ rotations separated by global pulses, leading to an optimal sequence length of five. 

\textbf{Designing Robust Pulse Sequences---}\label{sec:cpreview} Realistic systems inevitably suffer from various imperfections. When such errors are coherent and quasistatic, or approximately constant, across the duration of a pulse sequence, their effect can be compensated via a cleverly designed ``error-correcting" sequence---such a robust sequence is known as a composite pulse. More concretely, suppose that as a result of a small, static error $\zeta$, imperfect rotations $[\theta]^\zeta_{\phi}\,=\,[\theta]_{\phi}\,+\,O(\zeta)$ are implemented. A composite pulse is a sequence of erroneous pulses that together approximate some ideal target unitary $\mathcal{U}$, $[{\theta_1}]^{\zeta}_{\phi_1}\,\dots\, [{\theta_k}]^{\zeta}_{\phi_k}\,=\,\mathcal{U}\,+\,O(\zeta^{n+1}),$
where the pulse areas $\{\theta_j\}$ and phases $\{\phi_j\}$ are chosen to achieve error cancellation up to order $O(\zeta^{n})$. In principle, composite pulses accurate to arbitrarily high order $n$ can be constructed \cite{Brown2004, Low2013}, with their length $k$ scaling optimally as $k\,=\,O(n)$ \cite{supp, Low2013}. However, exceedingly lengthy operations are undesirable due to fast time-dependent noise, which leads to decoherence. An inherent trade-off therefore exists between speed and accuracy.

Two prominent sources of coherent, quasistatic error are slow amplitude and frequency drifts of the resonant driving field. Amplitude error, or $\epsilon$-error, manifests as a multiplicative error in the pulse area, $\theta\,\to\,\theta(1\,+\,\epsilon)$, in which $\epsilon\,=\,\frac{d\Omega}{\Omega}$ with $d\Omega$ a small error in the Rabi frequency.
Off-resonance error, or $\delta$-error, leads to a small detuning $\Delta$, which modifies the rotating frame Hamiltonian as $\mathcal{H}\,=\,\frac{\Omega}{2}(\hat \sigma_{\phi}\,+\,\frac{\Delta}{\Omega}\hat\sigma_z)$. Pulses then become $[\theta]^\delta_\phi\,=\,\exp(-i\frac{\theta}{2}(\hat \sigma_{\phi}\,+\,\delta \hat \sigma_z))$, with $\delta\,=\,\frac{\Delta}{\Omega}$ the off-resonance fraction. We seek to derive composite pulses for each of the basic sequences Eqs. \eqref{eq:arbuac}-\eqref{eq:arbuzc}  which compensate for $\epsilon$ and $\delta$.

A variety of $\epsilon$- and $\delta$-robust composite pulses already exist, with some of the most celebrated including BB1 \cite{Wimperis1994} and CORPSE \cite{Cummins2003}. To construct a robust, parallel, arbitrary unitary, we might first attempt to directly replace each of the individual pulses in our optimal solutions with an existing composite pulse. However, this approach of direct replacement encounters two main issues.
First, these existing composite pulses are generally not compatible with the constraints imposed by the local control methods considered in this work. Second, the length of such sequences, measured in either the number of pulses $k$ or the total pulse area $T\,=\,\sum_j\theta_j$, may be excessive, then increasing vulnerability to other sources of error such as decoherence. In order to address these challenges, we design novel $\epsilon$-, $\delta$-, and 
$\epsilon$,$\delta$-robust pulse sequences for each of Eqs. \eqref{eq:arbuac}-\eqref{eq:arbuzc}.

\textit{Concatenation---}Pulse sequences, or primitives, that compensate for either $\epsilon$ or $\delta$ can be concatenated together to form a sequence simultaneously robust against both errors, as proposed in Ref.~\cite{Bando2013}. 
Broadly, concatenation consists of replacing each pulse in an $\epsilon$-robust sequence with a $\delta$-robust composite pulse, or vice versa. However, such a naive substitution can be problematic if the inner composite pulse disturbs the original error-correcting structure of the outer sequence. Thus, a suitable $\delta$-robust inner composite pulse $\mathcal{C_\delta}$ must \emph{preserve} the effect of $\epsilon$, the error it is not designed to correct, a property called \emph{residual error preserving in} $\epsilon$ ($\epsilon$-REP)~\cite{Bando2013}. An overall sequence which compensates to first order in both $\epsilon$ and $\delta$ can then be obtained by swapping each pulse in an $\epsilon$-robust primitive sequence for a $\delta$-robust \textit{and} $\epsilon$-REP composite pulse. The same holds vice versa in $\delta$ and $\epsilon$.

\textit{Pulse Primitives---} With concatenation in mind, we develop appropriate pulse primitives individually robust against $\epsilon$ or $\delta$. 
For first-order $\delta$ error compensation, we develop a composite pulse deemed SCORE1 (Switchback for Compensating Off-Resonance Error), whose intuition is given in SM~\cite{supp}. SCORE1 realizes target rotations $\mathcal{U}\,=\,[\theta]_{\phi}$, and is compatible with all of PC, ZC, and AC, taking the form
\begin{align}\label{eq:score1}
    \text{SCORE1}(\theta, \phi) \equiv  [\vartheta_1]_{\phi+\pi}[\theta + 2\vartheta_1]_{\phi}[\vartheta_1]_{\phi+\pi}  
\end{align}
with $\text{SCORE1}(\theta,\phi;\delta)\,=\,[\theta]_\phi\,+\,O(\delta^2)$ for 
$\vartheta_1\,=\,\pi-\frac{\theta}{2}-\sin^{-1}(\frac{1}{2}\sin\frac{\theta}{2})$.
Generalizations of SCORE1 for higher orders up to $n\,=\,4$ are also listed in the SM~\cite{supp}. Our higher-order SCORE pulse has sequence length that scales linearly with the order of error compensation; to our knowledge, this is the first such $\delta$-robust sequence in the literature, and may find use in settings beyond the scope of this manuscript.
Our SCORE pulse family is appropriate for concatenation and can directly replace each pulse $[\theta]_\phi$ in the basic sequences Eqs.~\eqref{eq:arbuac}-\eqref{eq:arbuzc} for parallel, $\delta$-robust arbitrary unitaries.

We turn to $\epsilon$-compensating pulse primitives.
Under PC, we insert $[2\pi]_{\phi_j}$ pulses into the sequence we wish to make robust \cite{Brown2004, Low2013, Ichikawa2014}. Applying this method to Eq.~\eqref{eq:arbupc} yields a pulse sequence UP1 (Unitaries under Phase control for first-order $\epsilon$ compensation), 
\begin{align}\label{eq:up1}
    \text{UP1} (\alpha, \beta, \gamma) \equiv [\pi/2]_{\alpha} [2\pi]_{\phi_1}  [\pi]_{\frac{\gamma + \alpha-\beta}{2}} [2\pi]_{\phi_{2}}[\pi/2]_{\gamma},
\end{align}
where $\text{UP1}(\alpha,\beta,\gamma;\epsilon)\,=\,U^{PC}(\alpha, \beta,\gamma)\,+\,O(\epsilon^{2})$ realizes arbitrary SU(2) unitaries for appropriate $\phi_1,\,\phi_2$. 
The solution for $\phi_1$, $\phi_2$, as well as its higher-order extensions UP$n$ are given in the SM~\cite{supp}. 
This construction can be understood by noting that in the absence of error, the additional $[2\pi]_{\phi_j}$ rotations will preserve the original sequence. Moreover, when $\epsilon\,\neq\,0$, erroneous $[2\pi]^\epsilon_{\phi_j}$ pulses can be factored as $[2\pi]_{\phi_j} [2\pi\epsilon]_{\phi_j}$; we then use these pure error terms $[2\pi \epsilon]_{\phi_j}$ as a resource to cancel errors in the original sequence through an appropriate choice of $\phi_j$. 

To develop $\epsilon$-compensating pulse primitives under ZC, we introduce in Theorem \ref{thm:trs} a pulse identity allowing us to generate a class of $\epsilon$-robust sequences from a single sequence:
\begin{theorem}\label{thm:trs}
Under amplitude error in the global drive, a given time-reversal symmetric composite sequence 
\begin{align*}
    [{\theta_1}]^\epsilon_{\phi_1}... [{\theta_k}]^\epsilon_{\phi_k}[{\theta_k}]^\epsilon_{\phi_k}... [{\theta_1}]^\epsilon_{\phi_1} = [{\theta}]_{\phi} + O(\epsilon^{n+1})
\end{align*}
implies the class of composite sequences 
\begin{align*}
Z(\alpha) \; [{\theta_1}]^\epsilon_{\phi_1}... [{\theta_k}]^\epsilon_{\phi_k} \; Z(\beta) \; [{\theta_k}]^\epsilon_{\phi_k}...  [{\theta_1}]^\epsilon_{\phi_1} \; Z(\gamma)\\
= Z(\alpha)\; [\theta/2]_{\phi} \; Z(\beta) \; [\theta/2]_{\phi} \; Z(\gamma) + O(\epsilon^{n+1}).
\end{align*}
\end{theorem}
The proof is given in the SM~\cite{supp}. As a consequence of Theorem \ref{thm:trs}, $\epsilon$-robust sequences for Eq. \eqref{eq:arbuzc} immediately follow from any time-reversal symmetric composite $\pi$ pulse.

For AC, we design RA1 (Rotations under Amplitude control for first order $\epsilon$-compensation),
\begin{align}\label{eq:ra1}
    \text{RA1}(\theta, \phi)\equiv [{\vartheta_1}]_{\phi + \frac{\pi}{2}} [\pi]_{\phi} [{2\vartheta_1}]_{\phi + \frac{\pi}{2}} [\pi]_{\phi } [{\vartheta_1}]_{\phi + \frac{\pi}{2}} [\theta]_{\phi } .
\end{align}
Here, $\text{RA1}(\theta, \phi; \epsilon)\,=\,[\theta]_{\phi}\,+\,O(\epsilon^2)$ for $\vartheta_{1}\,=\,\cos^{-1}(-\frac{\theta}{2\pi})$, implementing arbitrary rotations.
To our best knowledge, RA1 is the first $\epsilon$-robust composite pulse which employs global phases and variable pulse areas. The derivation procedure of RA1, as well as a second-order extension of this sequence, is presented in the SM~\cite{supp}.
 
\textit{Numerical Search---} In the absence of any appropriate pulse primitives, we develop a numerical method based on gradient-based optimization to find robust pulse sequences. This numerical search is primarily adapted for AC, in which our individually controllable parameters--- the pulse areas---are also subject to error, and the lack of a perfect tuning knob makes robust sequence design more challenging. Specifically, we use gradient ascent to find the pulse areas of a sequence ansatz which maximize fidelity to the target unitary. To ensure robustness, we conduct this maximization over a range of $(\epsilon,\delta,\cdots)$ errors. A detailed explanation of our numerical method is provided in the SM \cite{supp}.



\textbf{Results---}
We can now comprehensively state our developed pulse sequences for parallelizable, arbitrary unitaries robust against $\epsilon,\,\delta$, or both under each type of individual control. A concise list of these sequences is provided in Table \ref{tab:comppul}; for a graphical schematic of these sequences, see the SM~\cite{supp}.

\begin{table}
    \centering
    \begin{tabular}{c|c|c|c}
     Control & $\epsilon$-robust &  $\delta$-robust &  $\epsilon$, $\delta$-robust \\\hline \hline
    PC  & UP1  & SCORE1*  & SCORE1inUP1 \\\hline
    \multirow{2}{*}{$\text{ZC}^\dagger$} & UZ1 & \multirow{2}{*}{SCORE1*} & \multirow{2}{*}{SCORE1inUZ1} \\
    & sUZ1 & & \\\hline
    \multirow{2}{*}{AC} & RA1*   & \multirow{2}{*}{SCORE1*}& SCORE1inRA1* \\
    & numerical & &numerical
\end{tabular}
    \caption{A summary of pulse sequences introduced in this work. In the $\epsilon$-robust column, the first letter of each sequence indicates either target unitaries or rotations, while the second specifies either PC, ZC, or AC. The number stipulates the order of error compensation. As a general-purpose $\delta$-robust composite pulse, SCORE does not follow this convention. ${}^\dagger$Sequences are presented for $\epsilon_s\,\ll\,\epsilon,\delta$. Robust sequences for large $\epsilon_s$ are found numerically. *Robust unitaries are achieved by directly replacing each pulse in the corresponding basic sequence (Eqs. ~\eqref{eq:arbuac}-\eqref{eq:arbuzc}) with the robust pulse. }
    \label{tab:comppul}
\end{table}

\textit{Phase Control and Amplitude Control---} \label{sec:pcacseq} Under each of PC and AC,  $\epsilon$- or $\delta$-robust unitaries are achieved via our previous pulse primitives. Doubly-robust sequences are then obtained by concatenating the two primitives (Table \ref{tab:comppul}): for instance, concatenating SCORE1 in UP1 leads to the sequence SCORE1inUP1. For AC, we additionally employ our numerical method to find shorter $\epsilon$- and doubly-robust sequences~\cite{supp}.

\textit{Z Control---}\label{sec:dcseq}
Unlike PC and AC, ZC assumes the ability to apply local $Z$ rotations, which are generally subject to additional error beyond $\epsilon$ and $\delta$. We thus discuss ZC in further detail. Specifically, the form of the local $Z$ error naturally depends on the specific experimental implementation of ZC. In multiple platforms, local $Z(\varphi_i)$ rotations have been realized through AC Stark shifts induced by a far-detuned field with varying intensity \cite{Schindler2013, Lee2016, Eckner2023}. Under off-resonant driving at Rabi frequency $\Omega$, qubit evolution in the rotating frame becomes $\exp(-i \mathcal{H}t)\,=\,\exp(-i\frac{1}{2}\frac{\Omega^{2}t}{2\Delta}\sigma_z)\,\equiv\,Z(\varphi)$, with $\varphi\,=\,\frac{\Omega^2}{2\Delta}t$. Thus, small drifts in the Rabi frequency $\Omega\,+\,d\Omega$ during local gates will lead to a local Stark error ($\epsilon_s$ error) $\varphi\,\to\,\varphi(1\,+\,\epsilon_s),\,\epsilon_{s}\,=\,2\frac{d\Omega}{\Omega}\,+\,(\frac{d\Omega}{\Omega})^2$, in addition to $\epsilon$ and $\delta$ error in the global pulses. We therefore address three different regimes: $\epsilon_{s}\,\gg\,\epsilon,\,\delta$, when local errors dominate; $\epsilon_{s}\,\ll\,\epsilon,\,\delta$, when global errors dominate; and $\epsilon_{s}\,\approx\,\epsilon,\,\delta$, when all errors contribute.

When $\epsilon_{s}\,\gg\,\epsilon,\,\delta$, our global pulses are almost perfect, and we can treat $[\frac{\pi}{2}]_{\pi}Z(\varphi_i)[\frac{\pi}{2}]_{0}\,=\,Y(\varphi_i)$, where $\varphi_i$ may contain $\epsilon_s$ error. In this case, $\epsilon_s$-compensation then directly maps to $\epsilon$-compensation under AC via a rotation of Eq.~\eqref{eq:arbuzc} by $Y(-\frac{\pi}{2})$, and we can obtain $\epsilon_s$-robust pulse sequences using a version of our numerical method. When $\epsilon_{s}\,\approx\,\epsilon,\delta$, solutions robust against  $\epsilon_s$,$\,$$\epsilon$, and $\delta$ can similarly be found numerically \cite{supp}. 

When $\epsilon_{s}\,\ll\,\epsilon,\delta$, our local $Z$ rotations are approximately perfect, and Theorem \ref{thm:trs} allows us to immediately construct $\epsilon$-robust pulse primitives for $U^{ZC}$ from any time-reversal symmetric composite $\pi$ pulse. Based on the composite $\pi$ pulses $[2\pi]_{\phi_1}^\epsilon{[\pi]}_0^\epsilon{[2\pi]}_{\phi_1}^{\epsilon}\,=\,[\pi]_{0}\,+\,O(\epsilon^2)$ and $[\pi]_{\phi_1}^{\epsilon}[\pi]_{\phi_2}^{\epsilon}[\pi]_{\phi_1}^{\epsilon}\,=\,[\pi]_{0}\,+\,O(\epsilon^{2})$, we respectively devise two different sequences UZ1 and sUZ1 (Unitaries under Z control for first-order $\epsilon$ compensation), 
\begin{align}
    \text{UZ1} &\equiv Z(\alpha) [2\pi]_{\bar{\phi}_1} [\pi/2]_\pi Z(\beta)[\pi/2]_0 [2\pi]_{\phi_{1}}Z(\gamma), \label{eq:uz1}\\
    \text{sUZ1} &\equiv Z(\alpha) [\pi]_{\bar{\phi}_1} [\pi/2]_{\bar{\phi}_2} Z(\beta) [\pi/2]_{\phi_2} [\pi]_{\phi_{1}} Z(\gamma),\label{eq:suz1}
\end{align}
with $\bar{\phi}_{j}\,=\,\phi_j+\pi$. The second sequence sUZ1 is shorter in total pulse area. $\text{UZ1}(\alpha,\beta,\gamma;\epsilon)\,=\,U^{ZC}(\alpha,\beta,\gamma)\,+\,O(\epsilon^{2})$ for $\phi_{1}\,=\,\cos^{-1}(-\frac{1}{4})$, and $\text{sUZ1}(\alpha,\beta,\gamma;\epsilon)\,=\,U^{ZC}(\alpha,\beta,\gamma)\,+\,O(\epsilon^2)$ for $(\phi_1,\phi_2)\,=\,(\frac{\pi}{3},\frac{5\pi}{3})$.
While sUZ1 is shorter, UZ1 better serves as a sequence for concatenation~\cite{supp}.
Order $n$ extensions of both sequences follow straightforwardly \cite{Gevorgyan2021, Jones2013, supp}, and solutions for the higher-order versions of UZ1 and sUZ1 up to order $n = 5$ are given in the SM~\cite{supp}. 

As before, $\delta$-robust unitaries are achieved by directly replacing each $\frac{\pi}{2}$ pulse in Eq.~\eqref{eq:arbuzc} with a SCORE primitive, while double compensating sequences are given by concatenation.

\textbf{Numerical Demonstration---}\label{sec:numsim}
We now numerically test our pulse sequences, benchmarking them against widely-adopted existing composite pulses such as BB1~\cite{Wimperis1994} and CORPSE~\cite{Cummins2000}. In practice, various sources of incoherent errors---which cannot be mitigated by composite pulses---also arise over the duration of a pulse sequence and reduce gate fidelity. To capture this effect, we evaluate the average gate fidelity in the presence of decoherence modeled as a depolarizing channel~\cite{supp}: 
\begin{eqnarray}\label{eq:fidgate}
     \mathcal{F} = e^{-\gamma t} \mathbb{E}_\psi  \big[ |\braket{\psi|\mathcal{U}^\dagger \mathcal{U}(\epsilon, \delta)|\psi}|^2\big] + (1- e^{-\gamma t})/2. \nonumber
\end{eqnarray}
Here, $\mathbb{E}_\psi$ denotes averaging over uniformly random initial states, $\mathcal{U}$ is a target unitary, $\mathcal{U}(\epsilon,\delta)$ is a pulse sequence under imperfections, and $\gamma$ is the decoherence rate. 

\begin{figure}
    \centering
    \includegraphics[width=1\columnwidth]{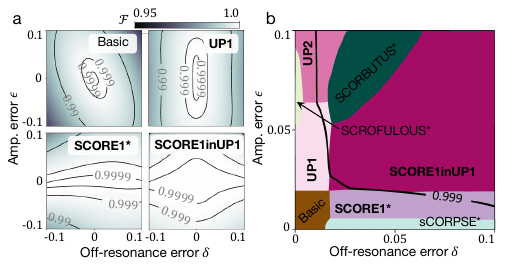}
    \caption{{\bf a.} Fidelity contours at $\gamma = 0$ for our basic sequence in Eq.~\eqref{eq:arbupc}, as well as various parallelizable PC composite pulse sequences averaged over target $\mathcal{U}\,\in\,[H, Z(\frac{\pi}{4}),\,X(\frac{\pi}{2}),\,Y(\frac{\pi}{2})]$. 
    {\bf b.} Phase diagram of the pulse sequence that maximizes $\mathcal{F}$ at each $(\epsilon, \delta)$ under PC, drawn for $\gamma\,=\,5 \times 10^{-5} \Omega $. Our newly developed pulse sequences, indicated in bold, occupy most of the parameter range considered here. As $\gamma$ increases, the basic and first-order sequences will grow to dominate the phase space. See the SM~\cite{supp} for phase diagrams for ZC and AC, as well as for more parameter regimes.
    }
    \label{fig:fidex}
\end{figure}

We compute $\mathcal{F}$ over a wide range of $(\epsilon,\delta,\gamma)$ for both our sequences and other composite pulses, where $\mathcal{F}$ is averaged over the choice $\mathcal{U}\,\in\,{[H, Z(\frac{\pi}{4}),X(\frac{\pi}{2}),Y(\frac{\pi}{2})]}$. Here, $H$ is the Hadamard gate, and $Z(\frac{\pi}{4})$ is the $T$ gate up to a phase.
$H$ and $T$, along with any entangling gate, form an universal gate set.
The result is plotted as fidelity contours in $\epsilon-\delta$ plane, as shown in Fig~\ref{fig:fidex}\textbf{a} for our various PC sequences at $\gamma =0$.

We further construct ``phase diagrams" indicating which sequence, among those considered, achieves the highest $\mathcal{F}$ at each point in $(\epsilon,\,\delta,\,\gamma)$ phase space. To make these diagrams, $\mathcal{F}$ is averaged four-fold to lie in the positive $(\epsilon,\delta)$ quadrant. Though each phase diagram depends on the ensemble of target unitaries $\mathcal{U}$, we generically find that our sequences for arbitrary unitaries outperform existing composite pulses almost everywhere in phase space. As an example, we show the phase diagram for PC sequences at $\gamma\,=\,5\times 10^{-5}\,\Omega$ (Fig.~\ref{fig:fidex}\textbf{b}), and more extensive numerical results are presented in the SM~\cite{supp}. Fig.~\ref{fig:fidex}\textbf{b} indicates that our PC sequences enable performance at fidelities $>$ 0.999, even in the presence of quasistatic laser amplitude or detuning error up to $\sim\,10\%$.

\textbf{Discussion---}
We have developed a method of efficient, parallel, and highly robust control of an arbitrarily large qubit ensemble. Quasistatic errors can be efficiently suppressed by our parallelizable composite pulse sequences, even without knowledge of the precise amount of error present. Consequently, the Rabi frequencies and detunings of individual qubits do not need to be calibrated beyond a percent level in practical situations. Indeed, the simultaneous calibration and removal of spatial inhomogeneity becomes increasingly demanding, or practically infeasible, for larger systems of tens or hundreds of thousands of qubits. Our approach lifts this challenge, having immediate impact for current experiments. Moreover, our method also effectively mitigates other coherent sources of imperfection, such as quasistatic correlated errors. 

Although our method may increase vulnerability to incoherent errors such as phase noise and scattering, recent developments indicate that suitably designed qubit encoding and error correction schemes can efficiently contain some decay via conversion into erasure-biased errors~\cite{Ma2023, Kang2023}. Since such error correction techniques may be less effective for coherent control errors \cite{Jandura2023}, the robust control methods presented here complement erasure conversion. When implemented in conjunction, both coherent and incoherent errors can be suppressed in large qubit arrays.

Experimental implementations of our pulses may also benefit from additional closed-loop optimization, perhaps using machine learning \cite{Shi2023}, to suppress other platform-specific sources of error. 
Finally, we note that while our work primarily concerns optical controls, a similar approach might also be useful for superconducting qubit systems owing to the reduced number of calibrated operations~\cite{McKay2016, Chen2021}.

\begin{acknowledgments}
We would like thank Adam Shaw, Ran Finkelstein, Joonhee Choi, and Manuel Endres for sharing their early experimental results, which inspired this work.
We also thank Adam Shaw, Ran Finkelstein, Harry Zhou, Dolev Bluvstein, Marcin Kalinowski, Samuel Li, Sophie Li, Sasha Geim, Manuel Endres, Joonhee Choi, Tracy Northup, and Jeff Thompson for helpful discussion, insights, and providing feedback on earlier version of this manuscript.
This work is supported in parts by
NSF CUA (PHY-1734011), QLCI program (2016245), and CAREER Award (DMR-2237244).
W.G is supported by the Paul and Daisy Soros Fellowship for New Americans, and the Hertz Foundation Fellowship. 
\end{acknowledgments}
\bibliography{ddrefs}

\begin{thebibliography}{69}%
\makeatletter
\providecommand \@ifxundefined [1]{%
 \@ifx{#1\undefined}
}%
\providecommand \@ifnum [1]{%
 \ifnum #1\expandafter \@firstoftwo
 \else \expandafter \@secondoftwo
 \fi
}%
\providecommand \@ifx [1]{%
 \ifx #1\expandafter \@firstoftwo
 \else \expandafter \@secondoftwo
 \fi
}%
\providecommand \natexlab [1]{#1}%
\providecommand \enquote  [1]{``#1''}%
\providecommand \bibnamefont  [1]{#1}%
\providecommand \bibfnamefont [1]{#1}%
\providecommand \citenamefont [1]{#1}%
\providecommand \href@noop [0]{\@secondoftwo}%
\providecommand \href [0]{\begingroup \@sanitize@url \@href}%
\providecommand \@href[1]{\@@startlink{#1}\@@href}%
\providecommand \@@href[1]{\endgroup#1\@@endlink}%
\providecommand \@sanitize@url [0]{\catcode `\\12\catcode `\$12\catcode `\&12\catcode `\#12\catcode `\^12\catcode `\_12\catcode `\%12\relax}%
\providecommand \@@startlink[1]{}%
\providecommand \@@endlink[0]{}%
\providecommand \url  [0]{\begingroup\@sanitize@url \@url }%
\providecommand \@url [1]{\endgroup\@href {#1}{\urlprefix }}%
\providecommand \urlprefix  [0]{URL }%
\providecommand \Eprint [0]{\href }%
\providecommand \doibase [0]{https://doi.org/}%
\providecommand \selectlanguage [0]{\@gobble}%
\providecommand \bibinfo  [0]{\@secondoftwo}%
\providecommand \bibfield  [0]{\@secondoftwo}%
\providecommand \translation [1]{[#1]}%
\providecommand \BibitemOpen [0]{}%
\providecommand \bibitemStop [0]{}%
\providecommand \bibitemNoStop [0]{.\EOS\space}%
\providecommand \EOS [0]{\spacefactor3000\relax}%
\providecommand \BibitemShut  [1]{\csname bibitem#1\endcsname}%
\let\auto@bib@innerbib\@empty
\bibitem [{\citenamefont {Arute}\ \emph {et~al.}(2019)\citenamefont {Arute}, \citenamefont {Arya}, \citenamefont {Babbush}, \citenamefont {Bacon}, \citenamefont {Bardin}, \citenamefont {Barends}, \citenamefont {Biswas}, \citenamefont {Boixo}, \citenamefont {Brandao}, \citenamefont {Buell}, \citenamefont {Burkett}, \citenamefont {Chen}, \citenamefont {Chen}, \citenamefont {Chiaro}, \citenamefont {Collins}, \citenamefont {Courtney}, \citenamefont {Dunsworth}, \citenamefont {Farhi}, \citenamefont {Foxen}, \citenamefont {Fowler}, \citenamefont {Gidney}, \citenamefont {Giustina}, \citenamefont {Graff}, \citenamefont {Guerin}, \citenamefont {Habegger}, \citenamefont {Harrigan}, \citenamefont {Hartmann}, \citenamefont {Ho}, \citenamefont {Hoffmann}, \citenamefont {Huang}, \citenamefont {Humble}, \citenamefont {Isakov}, \citenamefont {Jeffrey}, \citenamefont {Jiang}, \citenamefont {Kafri}, \citenamefont {Kechedzhi}, \citenamefont {Kelly}, \citenamefont {Klimov}, \citenamefont {Knysh}, \citenamefont {Korotkov},
  \citenamefont {Kostritsa}, \citenamefont {Landhuis}, \citenamefont {Lindmark}, \citenamefont {Lucero}, \citenamefont {Lyakh}, \citenamefont {Mandrà}, \citenamefont {McClean}, \citenamefont {McEwen}, \citenamefont {Megrant}, \citenamefont {Mi}, \citenamefont {Michielsen}, \citenamefont {Mohseni}, \citenamefont {Mutus}, \citenamefont {Naaman}, \citenamefont {Neeley}, \citenamefont {Neill}, \citenamefont {Niu}, \citenamefont {Ostby}, \citenamefont {Petukhov}, \citenamefont {Platt}, \citenamefont {Quintana}, \citenamefont {Rieffel}, \citenamefont {Roushan}, \citenamefont {Rubin}, \citenamefont {Sank}, \citenamefont {Satzinger}, \citenamefont {Smelyanskiy}, \citenamefont {Sung}, \citenamefont {Trevithick}, \citenamefont {Vainsencher}, \citenamefont {Villalonga}, \citenamefont {White}, \citenamefont {Yao}, \citenamefont {Yeh}, \citenamefont {Zalcman}, \citenamefont {Neven},\ and\ \citenamefont {Martinis}}]{Arute2019}%
  \BibitemOpen
  \bibfield  {author} {\bibinfo {author} {\bibfnamefont {F.}~\bibnamefont {Arute}}, \bibinfo {author} {\bibfnamefont {K.}~\bibnamefont {Arya}}, \bibinfo {author} {\bibfnamefont {R.}~\bibnamefont {Babbush}}, \bibinfo {author} {\bibfnamefont {D.}~\bibnamefont {Bacon}}, \bibinfo {author} {\bibfnamefont {J.~C.}\ \bibnamefont {Bardin}}, \bibinfo {author} {\bibfnamefont {R.}~\bibnamefont {Barends}}, \bibinfo {author} {\bibfnamefont {R.}~\bibnamefont {Biswas}}, \bibinfo {author} {\bibfnamefont {S.}~\bibnamefont {Boixo}}, \bibinfo {author} {\bibfnamefont {F.~G.}\ \bibnamefont {Brandao}}, \bibinfo {author} {\bibfnamefont {D.~A.}\ \bibnamefont {Buell}}, \emph {et~al.},\ }\bibinfo {title} {Quantum supremacy using a programmable superconducting processor},\ \href {https://doi.org/10.1038/s41586-019-1666-5} {\bibfield  {journal} {\bibinfo  {journal} {Nature}\ }\textbf {\bibinfo {volume} {574}},\ \bibinfo {pages} {505} (\bibinfo {year} {2019})}\BibitemShut {NoStop}%
\bibitem [{\citenamefont {Kim}\ \emph {et~al.}(2023)\citenamefont {Kim}, \citenamefont {Eddins}, \citenamefont {Anand}, \citenamefont {Wei}, \citenamefont {van~den Berg}, \citenamefont {Rosenblatt}, \citenamefont {Nayfeh}, \citenamefont {Wu}, \citenamefont {Zaletel}, \citenamefont {Temme},\ and\ \citenamefont {Kandala}}]{Kim2023}%
  \BibitemOpen
  \bibfield  {author} {\bibinfo {author} {\bibfnamefont {Y.}~\bibnamefont {Kim}}, \bibinfo {author} {\bibfnamefont {A.}~\bibnamefont {Eddins}}, \bibinfo {author} {\bibfnamefont {S.}~\bibnamefont {Anand}}, \bibinfo {author} {\bibfnamefont {K.~X.}\ \bibnamefont {Wei}}, \bibinfo {author} {\bibfnamefont {E.}~\bibnamefont {van~den Berg}}, \bibinfo {author} {\bibfnamefont {S.}~\bibnamefont {Rosenblatt}}, \bibinfo {author} {\bibfnamefont {H.}~\bibnamefont {Nayfeh}}, \bibinfo {author} {\bibfnamefont {Y.}~\bibnamefont {Wu}}, \bibinfo {author} {\bibfnamefont {M.}~\bibnamefont {Zaletel}}, \bibinfo {author} {\bibfnamefont {K.}~\bibnamefont {Temme}}, \emph {et~al.},\ }\bibinfo {title} {Evidence for the utility of quantum computing before fault tolerance},\ \href {https://doi.org/10.1038/s41586-023-06096-3} {\bibfield  {journal} {\bibinfo  {journal} {Nature}\ }\textbf {\bibinfo {volume} {618}},\ \bibinfo {pages} {500} (\bibinfo {year} {2023})}\BibitemShut {NoStop}%
\bibitem [{\citenamefont {Zhu}\ \emph {et~al.}(2023)\citenamefont {Zhu}, \citenamefont {Kahanamoku-Meyer}, \citenamefont {Lewis}, \citenamefont {Noel}, \citenamefont {Katz}, \citenamefont {Harraz}, \citenamefont {Wang}, \citenamefont {Risinger}, \citenamefont {Feng}, \citenamefont {Biswas}, \citenamefont {Egan}, \citenamefont {Gheorghiu}, \citenamefont {Nam}, \citenamefont {Vidick}, \citenamefont {Vazirani}, \citenamefont {Yao}, \citenamefont {Cetina},\ and\ \citenamefont {Monroe}}]{Zhu2023}%
  \BibitemOpen
  \bibfield  {author} {\bibinfo {author} {\bibfnamefont {D.}~\bibnamefont {Zhu}}, \bibinfo {author} {\bibfnamefont {G.~D.}\ \bibnamefont {Kahanamoku-Meyer}}, \bibinfo {author} {\bibfnamefont {L.}~\bibnamefont {Lewis}}, \bibinfo {author} {\bibfnamefont {C.}~\bibnamefont {Noel}}, \bibinfo {author} {\bibfnamefont {O.}~\bibnamefont {Katz}}, \bibinfo {author} {\bibfnamefont {B.}~\bibnamefont {Harraz}}, \bibinfo {author} {\bibfnamefont {Q.}~\bibnamefont {Wang}}, \bibinfo {author} {\bibfnamefont {A.}~\bibnamefont {Risinger}}, \bibinfo {author} {\bibfnamefont {L.}~\bibnamefont {Feng}}, \bibinfo {author} {\bibfnamefont {D.}~\bibnamefont {Biswas}}, \emph {et~al.},\ }\bibinfo {title} {Interactive cryptographic proofs of quantumness using mid-circuit measurements},\ \href {https://doi.org/10.1038/s41567-023-02162-9} {\bibfield  {journal} {\bibinfo  {journal} {Nature Physics}\ }\textbf {\bibinfo {volume} {19}},\ \bibinfo {pages} {1725–1731} (\bibinfo {year} {2023})}\BibitemShut {NoStop}%
\bibitem [{\citenamefont {Bluvstein}\ \emph {et~al.}(2022)\citenamefont {Bluvstein}, \citenamefont {Levine}, \citenamefont {Semeghini}, \citenamefont {Wang}, \citenamefont {Ebadi}, \citenamefont {Kalinowski}, \citenamefont {Keesling}, \citenamefont {Maskara}, \citenamefont {Pichler}, \citenamefont {Greiner}, \citenamefont {Vuletić},\ and\ \citenamefont {Lukin}}]{Bluvstein2022}%
  \BibitemOpen
  \bibfield  {author} {\bibinfo {author} {\bibfnamefont {D.}~\bibnamefont {Bluvstein}}, \bibinfo {author} {\bibfnamefont {H.}~\bibnamefont {Levine}}, \bibinfo {author} {\bibfnamefont {G.}~\bibnamefont {Semeghini}}, \bibinfo {author} {\bibfnamefont {T.~T.}\ \bibnamefont {Wang}}, \bibinfo {author} {\bibfnamefont {S.}~\bibnamefont {Ebadi}}, \bibinfo {author} {\bibfnamefont {M.}~\bibnamefont {Kalinowski}}, \bibinfo {author} {\bibfnamefont {A.}~\bibnamefont {Keesling}}, \bibinfo {author} {\bibfnamefont {N.}~\bibnamefont {Maskara}}, \bibinfo {author} {\bibfnamefont {H.}~\bibnamefont {Pichler}}, \bibinfo {author} {\bibfnamefont {M.}~\bibnamefont {Greiner}}, \emph {et~al.},\ }\bibinfo {title} {A quantum processor based on coherent transport of entangled atom arrays},\ \href {https://doi.org/10.1038/s41586-022-04592-6} {\bibfield  {journal} {\bibinfo  {journal} {Nature}\ }\textbf {\bibinfo {volume} {604}},\ \bibinfo {pages} {451} (\bibinfo {year} {2022})}\BibitemShut {NoStop}%
\bibitem [{\citenamefont {Bernien}\ \emph {et~al.}(2017)\citenamefont {Bernien}, \citenamefont {Schwartz}, \citenamefont {Keesling}, \citenamefont {Levine}, \citenamefont {Omran}, \citenamefont {Pichler}, \citenamefont {Choi}, \citenamefont {Zibrov}, \citenamefont {Endres}, \citenamefont {Greiner}, \citenamefont {Vuletić},\ and\ \citenamefont {Lukin}}]{Bernien2017}%
  \BibitemOpen
  \bibfield  {author} {\bibinfo {author} {\bibfnamefont {H.}~\bibnamefont {Bernien}}, \bibinfo {author} {\bibfnamefont {S.}~\bibnamefont {Schwartz}}, \bibinfo {author} {\bibfnamefont {A.}~\bibnamefont {Keesling}}, \bibinfo {author} {\bibfnamefont {H.}~\bibnamefont {Levine}}, \bibinfo {author} {\bibfnamefont {A.}~\bibnamefont {Omran}}, \bibinfo {author} {\bibfnamefont {H.}~\bibnamefont {Pichler}}, \bibinfo {author} {\bibfnamefont {S.}~\bibnamefont {Choi}}, \bibinfo {author} {\bibfnamefont {A.~S.}\ \bibnamefont {Zibrov}}, \bibinfo {author} {\bibfnamefont {M.}~\bibnamefont {Endres}}, \bibinfo {author} {\bibfnamefont {M.}~\bibnamefont {Greiner}}, \emph {et~al.},\ }\bibinfo {title} {Probing many-body dynamics on a 51-atom quantum simulator},\ \href {https://doi.org/10.1038/nature24622} {\bibfield  {journal} {\bibinfo  {journal} {Nature}\ }\textbf {\bibinfo {volume} {551}},\ \bibinfo {pages} {579–584} (\bibinfo {year} {2017})}\BibitemShut {NoStop}%
\bibitem [{\citenamefont {Lukin}\ \emph {et~al.}(2019)\citenamefont {Lukin}, \citenamefont {Rispoli}, \citenamefont {Schittko}, \citenamefont {Tai}, \citenamefont {Kaufman}, \citenamefont {Choi}, \citenamefont {Khemani}, \citenamefont {Léonard},\ and\ \citenamefont {Greiner}}]{Lukin2019}%
  \BibitemOpen
  \bibfield  {author} {\bibinfo {author} {\bibfnamefont {A.}~\bibnamefont {Lukin}}, \bibinfo {author} {\bibfnamefont {M.}~\bibnamefont {Rispoli}}, \bibinfo {author} {\bibfnamefont {R.}~\bibnamefont {Schittko}}, \bibinfo {author} {\bibfnamefont {M.~E.}\ \bibnamefont {Tai}}, \bibinfo {author} {\bibfnamefont {A.~M.}\ \bibnamefont {Kaufman}}, \bibinfo {author} {\bibfnamefont {S.}~\bibnamefont {Choi}}, \bibinfo {author} {\bibfnamefont {V.}~\bibnamefont {Khemani}}, \bibinfo {author} {\bibfnamefont {J.}~\bibnamefont {Léonard}},\ and\ \bibinfo {author} {\bibfnamefont {M.}~\bibnamefont {Greiner}},\ }\bibinfo {title} {Probing entanglement in a many-body–localized system},\ \href {https://doi.org/10.1126/science.aau0818} {\bibfield  {journal} {\bibinfo  {journal} {Science}\ }\textbf {\bibinfo {volume} {364}},\ \bibinfo {pages} {256} (\bibinfo {year} {2019})}\BibitemShut {NoStop}%
\bibitem [{\citenamefont {Semeghini}\ \emph {et~al.}(2021)\citenamefont {Semeghini}, \citenamefont {Levine}, \citenamefont {Keesling}, \citenamefont {Ebadi}, \citenamefont {Wang}, \citenamefont {Bluvstein}, \citenamefont {Verresen}, \citenamefont {Pichler}, \citenamefont {Kalinowski}, \citenamefont {Samajdar}, \citenamefont {Omran}, \citenamefont {Sachdev}, \citenamefont {Vishwanath}, \citenamefont {Greiner}, \citenamefont {Vuletić},\ and\ \citenamefont {Lukin}}]{Semeghini2021}%
  \BibitemOpen
  \bibfield  {author} {\bibinfo {author} {\bibfnamefont {G.}~\bibnamefont {Semeghini}}, \bibinfo {author} {\bibfnamefont {H.}~\bibnamefont {Levine}}, \bibinfo {author} {\bibfnamefont {A.}~\bibnamefont {Keesling}}, \bibinfo {author} {\bibfnamefont {S.}~\bibnamefont {Ebadi}}, \bibinfo {author} {\bibfnamefont {T.~T.}\ \bibnamefont {Wang}}, \bibinfo {author} {\bibfnamefont {D.}~\bibnamefont {Bluvstein}}, \bibinfo {author} {\bibfnamefont {R.}~\bibnamefont {Verresen}}, \bibinfo {author} {\bibfnamefont {H.}~\bibnamefont {Pichler}}, \bibinfo {author} {\bibfnamefont {M.}~\bibnamefont {Kalinowski}}, \bibinfo {author} {\bibfnamefont {R.}~\bibnamefont {Samajdar}}, \emph {et~al.},\ }\bibinfo {title} {Probing topological spin liquids on a programmable quantum simulator},\ \href {https://doi.org/10.1126/science.abi8794} {\bibfield  {journal} {\bibinfo  {journal} {Science}\ }\textbf {\bibinfo {volume} {374}},\ \bibinfo {pages} {1242} (\bibinfo {year} {2021})}\BibitemShut {NoStop}%
\bibitem [{\citenamefont {Choi}\ \emph {et~al.}(2023)\citenamefont {Choi}, \citenamefont {Shaw}, \citenamefont {Madjarov}, \citenamefont {Xie}, \citenamefont {Finkelstein}, \citenamefont {Covey}, \citenamefont {Cotler}, \citenamefont {Mark}, \citenamefont {Huang}, \citenamefont {Kale}, \citenamefont {Pichler}, \citenamefont {Brandão}, \citenamefont {Choi},\ and\ \citenamefont {Endres}}]{Choi2023}%
  \BibitemOpen
  \bibfield  {author} {\bibinfo {author} {\bibfnamefont {J.}~\bibnamefont {Choi}}, \bibinfo {author} {\bibfnamefont {A.~L.}\ \bibnamefont {Shaw}}, \bibinfo {author} {\bibfnamefont {I.~S.}\ \bibnamefont {Madjarov}}, \bibinfo {author} {\bibfnamefont {X.}~\bibnamefont {Xie}}, \bibinfo {author} {\bibfnamefont {R.}~\bibnamefont {Finkelstein}}, \bibinfo {author} {\bibfnamefont {J.~P.}\ \bibnamefont {Covey}}, \bibinfo {author} {\bibfnamefont {J.~S.}\ \bibnamefont {Cotler}}, \bibinfo {author} {\bibfnamefont {D.~K.}\ \bibnamefont {Mark}}, \bibinfo {author} {\bibfnamefont {H.-Y.}\ \bibnamefont {Huang}}, \bibinfo {author} {\bibfnamefont {A.}~\bibnamefont {Kale}}, \emph {et~al.},\ }\bibinfo {title} {Preparing random states and benchmarking with many-body quantum chaos},\ \href {https://doi.org/10.1038/s41586-022-05442-1} {\bibfield  {journal} {\bibinfo  {journal} {Nature}\ }\textbf {\bibinfo {volume} {613}},\ \bibinfo {pages} {468} (\bibinfo {year} {2023})}\BibitemShut {NoStop}%
\bibitem [{\citenamefont {Farrell}\ \emph {et~al.}(2024)\citenamefont {Farrell}, \citenamefont {Illa}, \citenamefont {Ciavarella},\ and\ \citenamefont {Savage}}]{Farrell2023}%
  \BibitemOpen
  \bibfield  {author} {\bibinfo {author} {\bibfnamefont {R.~C.}\ \bibnamefont {Farrell}}, \bibinfo {author} {\bibfnamefont {M.}~\bibnamefont {Illa}}, \bibinfo {author} {\bibfnamefont {A.~N.}\ \bibnamefont {Ciavarella}},\ and\ \bibinfo {author} {\bibfnamefont {M.~J.}\ \bibnamefont {Savage}},\ }\bibinfo {title} {Scalable circuits for preparing ground states on digital quantum computers: The {S}chwinger model vacuum on 100 qubits},\ \href {https://doi.org/10.1103/PRXQuantum.5.020315} {\bibfield  {journal} {\bibinfo  {journal} {PRX Quantum}\ }\textbf {\bibinfo {volume} {5}},\ \bibinfo {pages} {020315} (\bibinfo {year} {2024})}\BibitemShut {NoStop}%
\bibitem [{\citenamefont {Mi}\ \emph {et~al.}(2022)\citenamefont {Mi}, \citenamefont {Sonner}, \citenamefont {Niu}, \citenamefont {Lee}, \citenamefont {Foxen}, \citenamefont {Acharya}, \citenamefont {Aleiner}, \citenamefont {Andersen}, \citenamefont {Arute}, \citenamefont {Arya}, \citenamefont {Asfaw}, \citenamefont {Atalaya}, \citenamefont {Bardin}, \citenamefont {Basso}, \citenamefont {Bengtsson}, \citenamefont {Bortoli}, \citenamefont {Bourassa}, \citenamefont {Brill}, \citenamefont {Broughton}, \citenamefont {Buckley}, \citenamefont {Buell}, \citenamefont {Burkett}, \citenamefont {Bushnell}, \citenamefont {Chen}, \citenamefont {Chiaro}, \citenamefont {Collins}, \citenamefont {Conner}, \citenamefont {Courtney}, \citenamefont {Crook}, \citenamefont {Debroy}, \citenamefont {Demura}, \citenamefont {Dunsworth}, \citenamefont {Eppens}, \citenamefont {Erickson}, \citenamefont {Faoro}, \citenamefont {Farhi}, \citenamefont {Fatemi}, \citenamefont {Flores}, \citenamefont {Forati}, \citenamefont {Fowler},
  \citenamefont {Giang}, \citenamefont {Gidney}, \citenamefont {Gilboa}, \citenamefont {Giustina}, \citenamefont {Dau}, \citenamefont {Gross}, \citenamefont {Habegger}, \citenamefont {Harrigan}, \citenamefont {Hoffmann}, \citenamefont {Hong}, \citenamefont {Huang}, \citenamefont {Huff}, \citenamefont {Huggins}, \citenamefont {Ioffe}, \citenamefont {Isakov}, \citenamefont {Iveland}, \citenamefont {Jeffrey}, \citenamefont {Jiang}, \citenamefont {Jones}, \citenamefont {Kafri}, \citenamefont {Kechedzhi}, \citenamefont {Khattar}, \citenamefont {Kim}, \citenamefont {Kitaev}, \citenamefont {Klimov}, \citenamefont {Klots}, \citenamefont {Korotkov}, \citenamefont {Kostritsa}, \citenamefont {Kreikebaum}, \citenamefont {Landhuis}, \citenamefont {Laptev}, \citenamefont {Lau}, \citenamefont {Lee}, \citenamefont {Laws}, \citenamefont {Liu}, \citenamefont {Locharla}, \citenamefont {Martin}, \citenamefont {McClean}, \citenamefont {McEwen}, \citenamefont {Costa}, \citenamefont {Miao}, \citenamefont {Mohseni}, \citenamefont
  {Montazeri}, \citenamefont {Morvan}, \citenamefont {Mount}, \citenamefont {Mruczkiewicz}, \citenamefont {Naaman}, \citenamefont {Neeley}, \citenamefont {Neill}, \citenamefont {Newman}, \citenamefont {O’Brien}, \citenamefont {Opremcak}, \citenamefont {Petukhov}, \citenamefont {Potter}, \citenamefont {Quintana}, \citenamefont {Rubin}, \citenamefont {Saei}, \citenamefont {Sank}, \citenamefont {Sankaragomathi}, \citenamefont {Satzinger}, \citenamefont {Schuster}, \citenamefont {Shearn}, \citenamefont {Shvarts}, \citenamefont {Strain}, \citenamefont {Su}, \citenamefont {Szalay}, \citenamefont {Vidal}, \citenamefont {Villalonga}, \citenamefont {Vollgraff-Heidweiller}, \citenamefont {White}, \citenamefont {Yao}, \citenamefont {Yeh}, \citenamefont {Yoo}, \citenamefont {Zalcman}, \citenamefont {Zhang}, \citenamefont {Zhu}, \citenamefont {Neven}, \citenamefont {Bacon}, \citenamefont {Hilton}, \citenamefont {Lucero}, \citenamefont {Babbush}, \citenamefont {Boixo}, \citenamefont {Megrant}, \citenamefont {Chen},
  \citenamefont {Kelly}, \citenamefont {Smelyanskiy}, \citenamefont {Abanin},\ and\ \citenamefont {Roushan}}]{Mi2022}%
  \BibitemOpen
  \bibfield  {author} {\bibinfo {author} {\bibfnamefont {X.}~\bibnamefont {Mi}}, \bibinfo {author} {\bibfnamefont {M.}~\bibnamefont {Sonner}}, \bibinfo {author} {\bibfnamefont {M.~Y.}\ \bibnamefont {Niu}}, \bibinfo {author} {\bibfnamefont {K.~W.}\ \bibnamefont {Lee}}, \bibinfo {author} {\bibfnamefont {B.}~\bibnamefont {Foxen}}, \bibinfo {author} {\bibfnamefont {R.}~\bibnamefont {Acharya}}, \bibinfo {author} {\bibfnamefont {I.}~\bibnamefont {Aleiner}}, \bibinfo {author} {\bibfnamefont {T.~I.}\ \bibnamefont {Andersen}}, \bibinfo {author} {\bibfnamefont {F.}~\bibnamefont {Arute}}, \bibinfo {author} {\bibfnamefont {K.}~\bibnamefont {Arya}}, \emph {et~al.},\ }\bibinfo {title} {Noise-resilient edge modes on a chain of superconducting qubits},\ \href {https://doi.org/10.1126/science.abq5769} {\bibfield  {journal} {\bibinfo  {journal} {Science}\ }\textbf {\bibinfo {volume} {378}},\ \bibinfo {pages} {785} (\bibinfo {year} {2022})}\BibitemShut {NoStop}%
\bibitem [{\citenamefont {González-Cuadra}\ \emph {et~al.}(2025)\citenamefont {González-Cuadra}, \citenamefont {Hamdan}, \citenamefont {Zache}, \citenamefont {Braverman}, \citenamefont {Kornjača}, \citenamefont {Lukin}, \citenamefont {Cantú}, \citenamefont {Liu}, \citenamefont {Wang}, \citenamefont {Keesling}, \citenamefont {Lukin}, \citenamefont {Zoller},\ and\ \citenamefont {Bylinskii}}]{Gonzlez-Cuadra2025}%
  \BibitemOpen
  \bibfield  {author} {\bibinfo {author} {\bibfnamefont {D.}~\bibnamefont {González-Cuadra}}, \bibinfo {author} {\bibfnamefont {M.}~\bibnamefont {Hamdan}}, \bibinfo {author} {\bibfnamefont {T.~V.}\ \bibnamefont {Zache}}, \bibinfo {author} {\bibfnamefont {B.}~\bibnamefont {Braverman}}, \bibinfo {author} {\bibfnamefont {M.}~\bibnamefont {Kornjača}}, \bibinfo {author} {\bibfnamefont {A.}~\bibnamefont {Lukin}}, \bibinfo {author} {\bibfnamefont {S.~H.}\ \bibnamefont {Cantú}}, \bibinfo {author} {\bibfnamefont {F.}~\bibnamefont {Liu}}, \bibinfo {author} {\bibfnamefont {S.-T.}\ \bibnamefont {Wang}}, \bibinfo {author} {\bibfnamefont {A.}~\bibnamefont {Keesling}}, \emph {et~al.},\ }\bibinfo {title} {Observation of string breaking on a (2 + 1){D} {R}ydberg quantum simulator},\ \href {https://doi.org/10.1038/s41586-025-09051-6} {\bibfield  {journal} {\bibinfo  {journal} {Nature}\ }\textbf {\bibinfo {volume} {642}},\ \bibinfo {pages} {321} (\bibinfo {year} {2025})}\BibitemShut {NoStop}%
\bibitem [{IBM()}]{IBM}%
  \BibitemOpen
  \href@noop {} {}\bibinfo {note} {\url{https://www.ibm.com/quantum/roadmap\#content-block-simple-2}}\BibitemShut {NoStop}%
\bibitem [{\citenamefont {Kjaergaard}\ \emph {et~al.}(2020)\citenamefont {Kjaergaard}, \citenamefont {Schwartz}, \citenamefont {Braumüller}, \citenamefont {Krantz}, \citenamefont {Wang}, \citenamefont {Gustavsson},\ and\ \citenamefont {Oliver}}]{Kjaergaard2020}%
  \BibitemOpen
  \bibfield  {author} {\bibinfo {author} {\bibfnamefont {M.}~\bibnamefont {Kjaergaard}}, \bibinfo {author} {\bibfnamefont {M.~E.}\ \bibnamefont {Schwartz}}, \bibinfo {author} {\bibfnamefont {J.}~\bibnamefont {Braumüller}}, \bibinfo {author} {\bibfnamefont {P.}~\bibnamefont {Krantz}}, \bibinfo {author} {\bibfnamefont {J.~I.-J.}\ \bibnamefont {Wang}}, \bibinfo {author} {\bibfnamefont {S.}~\bibnamefont {Gustavsson}},\ and\ \bibinfo {author} {\bibfnamefont {W.~D.}\ \bibnamefont {Oliver}},\ }\bibinfo {title} {Superconducting qubits: Current state of play},\ \href {https://doi.org/10.1146/annurev-conmatphys-031119-050605} {\bibfield  {journal} {\bibinfo  {journal} {Annual Review of Condensed Matter Physics}\ }\textbf {\bibinfo {volume} {11}},\ \bibinfo {pages} {369} (\bibinfo {year} {2020})}\BibitemShut {NoStop}%
\bibitem [{\citenamefont {Siddiqi}(2021)}]{Siddiqi2021}%
  \BibitemOpen
  \bibfield  {author} {\bibinfo {author} {\bibfnamefont {I.}~\bibnamefont {Siddiqi}},\ }\bibinfo {title} {Engineering high-coherence superconducting qubits},\ \href {https://doi.org/10.1038/s41578-021-00370-4} {\bibfield  {journal} {\bibinfo  {journal} {Nature Reviews Materials}\ }\textbf {\bibinfo {volume} {6}},\ \bibinfo {pages} {875} (\bibinfo {year} {2021})}\BibitemShut {NoStop}%
\bibitem [{\citenamefont {Brown}\ \emph {et~al.}(2016)\citenamefont {Brown}, \citenamefont {Kim},\ and\ \citenamefont {Monroe}}]{Brown2016}%
  \BibitemOpen
  \bibfield  {author} {\bibinfo {author} {\bibfnamefont {K.~R.}\ \bibnamefont {Brown}}, \bibinfo {author} {\bibfnamefont {J.}~\bibnamefont {Kim}},\ and\ \bibinfo {author} {\bibfnamefont {C.}~\bibnamefont {Monroe}},\ }\bibinfo {title} {Co-designing a scalable quantum computer with trapped atomic ions},\ \href {https://doi.org/10.1038/npjqi.2016.34} {\bibfield  {journal} {\bibinfo  {journal} {npj Quantum Information}\ }\textbf {\bibinfo {volume} {2}},\ \bibinfo {pages} {16034} (\bibinfo {year} {2016})}\BibitemShut {NoStop}%
\bibitem [{\citenamefont {Graham}\ \emph {et~al.}(2023)\citenamefont {Graham}, \citenamefont {Oh},\ and\ \citenamefont {Saffman}}]{Graham2023}%
  \BibitemOpen
  \bibfield  {author} {\bibinfo {author} {\bibfnamefont {T.~M.}\ \bibnamefont {Graham}}, \bibinfo {author} {\bibfnamefont {E.}~\bibnamefont {Oh}},\ and\ \bibinfo {author} {\bibfnamefont {M.}~\bibnamefont {Saffman}},\ }\bibinfo {title} {Multiscale architecture for fast optical addressing and control of large-scale qubit arrays},\ \href {https://doi.org/10.1364/AO.484367} {\bibfield  {journal} {\bibinfo  {journal} {Applied Optics}\ }\textbf {\bibinfo {volume} {62}},\ \bibinfo {pages} {3242} (\bibinfo {year} {2023})}\BibitemShut {NoStop}%
\bibitem [{\citenamefont {Shih}\ \emph {et~al.}(2021)\citenamefont {Shih}, \citenamefont {Motlakunta}, \citenamefont {Kotibhaskar}, \citenamefont {Sajjan}, \citenamefont {Hablützel},\ and\ \citenamefont {Islam}}]{Shih2021}%
  \BibitemOpen
  \bibfield  {author} {\bibinfo {author} {\bibfnamefont {C.-Y.}\ \bibnamefont {Shih}}, \bibinfo {author} {\bibfnamefont {S.}~\bibnamefont {Motlakunta}}, \bibinfo {author} {\bibfnamefont {N.}~\bibnamefont {Kotibhaskar}}, \bibinfo {author} {\bibfnamefont {M.}~\bibnamefont {Sajjan}}, \bibinfo {author} {\bibfnamefont {R.}~\bibnamefont {Hablützel}},\ and\ \bibinfo {author} {\bibfnamefont {R.}~\bibnamefont {Islam}},\ }\bibinfo {title} {Reprogrammable and high-precision holographic optical addressing of trapped ions for scalable quantum control},\ \href {https://doi.org/10.1038/s41534-021-00396-0} {\bibfield  {journal} {\bibinfo  {journal} {npj Quantum Information}\ }\textbf {\bibinfo {volume} {7}},\ \bibinfo {pages} {57} (\bibinfo {year} {2021})}\BibitemShut {NoStop}%
\bibitem [{\citenamefont {Pogorelov}\ \emph {et~al.}(2021)\citenamefont {Pogorelov}, \citenamefont {Feldker}, \citenamefont {Marciniak}, \citenamefont {Postler}, \citenamefont {Jacob}, \citenamefont {Krieglsteiner}, \citenamefont {Podlesnic}, \citenamefont {Meth}, \citenamefont {Negnevitsky}, \citenamefont {Stadler}, \citenamefont {Höfer}, \citenamefont {Wächter}, \citenamefont {Lakhmanskiy}, \citenamefont {Blatt}, \citenamefont {Schindler},\ and\ \citenamefont {Monz}}]{Pogorelov2021}%
  \BibitemOpen
  \bibfield  {author} {\bibinfo {author} {\bibfnamefont {I.}~\bibnamefont {Pogorelov}}, \bibinfo {author} {\bibfnamefont {T.}~\bibnamefont {Feldker}}, \bibinfo {author} {\bibfnamefont {C.~D.}\ \bibnamefont {Marciniak}}, \bibinfo {author} {\bibfnamefont {L.}~\bibnamefont {Postler}}, \bibinfo {author} {\bibfnamefont {G.}~\bibnamefont {Jacob}}, \bibinfo {author} {\bibfnamefont {O.}~\bibnamefont {Krieglsteiner}}, \bibinfo {author} {\bibfnamefont {V.}~\bibnamefont {Podlesnic}}, \bibinfo {author} {\bibfnamefont {M.}~\bibnamefont {Meth}}, \bibinfo {author} {\bibfnamefont {V.}~\bibnamefont {Negnevitsky}}, \bibinfo {author} {\bibfnamefont {M.}~\bibnamefont {Stadler}}, \emph {et~al.},\ }\bibinfo {title} {Compact ion-trap quantum computing demonstrator},\ \href {https://doi.org/10.1103/PRXQuantum.2.020343} {\bibfield  {journal} {\bibinfo  {journal} {PRX Quantum}\ }\textbf {\bibinfo {volume} {2}},\ \bibinfo {pages} {020343} (\bibinfo {year} {2021})}\BibitemShut {NoStop}%
\bibitem [{\citenamefont {Wang}\ \emph {et~al.}(2025)\citenamefont {Wang}, \citenamefont {Xu}, \citenamefont {Li}, \citenamefont {Vuletić},\ and\ \citenamefont {Cappellaro}}]{Wang2023}%
  \BibitemOpen
  \bibfield  {author} {\bibinfo {author} {\bibfnamefont {G.}~\bibnamefont {Wang}}, \bibinfo {author} {\bibfnamefont {W.}~\bibnamefont {Xu}}, \bibinfo {author} {\bibfnamefont {C.}~\bibnamefont {Li}}, \bibinfo {author} {\bibfnamefont {V.}~\bibnamefont {Vuletić}},\ and\ \bibinfo {author} {\bibfnamefont {P.}~\bibnamefont {Cappellaro}},\ }\bibinfo {title} {Individual-atom control in an array through phase modulation},\ \href {https://doi.org/10.1103/PhysRevApplied.23.024072} {\bibfield  {journal} {\bibinfo  {journal} {Physical Review Applied}\ }\textbf {\bibinfo {volume} {23}},\ \bibinfo {pages} {024072} (\bibinfo {year} {2025})}\BibitemShut {NoStop}%
\bibitem [{\citenamefont {Cesa}\ and\ \citenamefont {Pichler}(2023)}]{Cesa2023}%
  \BibitemOpen
  \bibfield  {author} {\bibinfo {author} {\bibfnamefont {F.}~\bibnamefont {Cesa}}\ and\ \bibinfo {author} {\bibfnamefont {H.}~\bibnamefont {Pichler}},\ }\bibinfo {title} {Universal quantum computation in globally driven {R}ydberg atom arrays},\ \bibfield  {journal} {\bibinfo  {journal} {Physical Review Letters}\ }\textbf {\bibinfo {volume} {131}},\ \href {https://doi.org/10.1103/physrevlett.131.170601} {10.1103/physrevlett.131.170601} (\bibinfo {year} {2023})\BibitemShut {NoStop}%
\bibitem [{\citenamefont {Zhang}\ \emph {et~al.}(2024)\citenamefont {Zhang}, \citenamefont {Peng}, \citenamefont {Paul},\ and\ \citenamefont {Thompson}}]{Zhang2023}%
  \BibitemOpen
  \bibfield  {author} {\bibinfo {author} {\bibfnamefont {B.}~\bibnamefont {Zhang}}, \bibinfo {author} {\bibfnamefont {P.}~\bibnamefont {Peng}}, \bibinfo {author} {\bibfnamefont {A.}~\bibnamefont {Paul}},\ and\ \bibinfo {author} {\bibfnamefont {J.~D.}\ \bibnamefont {Thompson}},\ }\bibinfo {title} {Scaled local gate controller for optically addressed qubits},\ \href {https://doi.org/10.1364/OPTICA.512155} {\bibfield  {journal} {\bibinfo  {journal} {Optica}\ }\textbf {\bibinfo {volume} {11}},\ \bibinfo {pages} {227} (\bibinfo {year} {2024})}\BibitemShut {NoStop}%
\bibitem [{\citenamefont {Bluvstein}\ \emph {et~al.}(2026)\citenamefont {Bluvstein}, \citenamefont {Geim}, \citenamefont {Li}, \citenamefont {Evered}, \citenamefont {Ataides}, \citenamefont {Baranes}, \citenamefont {Gu}, \citenamefont {Manovitz}, \citenamefont {Xu}, \citenamefont {Kalinowski}, \citenamefont {Majidy}, \citenamefont {Kokail}, \citenamefont {Maskara}, \citenamefont {Trapp}, \citenamefont {Stewart}, \citenamefont {Hollerith}, \citenamefont {Zhou}, \citenamefont {Gullans}, \citenamefont {Yelin}, \citenamefont {Greiner}, \citenamefont {Vuletić}, \citenamefont {Cain},\ and\ \citenamefont {Lukin}}]{Bluvstein2026}%
  \BibitemOpen
  \bibfield  {author} {\bibinfo {author} {\bibfnamefont {D.}~\bibnamefont {Bluvstein}}, \bibinfo {author} {\bibfnamefont {A.~A.}\ \bibnamefont {Geim}}, \bibinfo {author} {\bibfnamefont {S.~H.}\ \bibnamefont {Li}}, \bibinfo {author} {\bibfnamefont {S.~J.}\ \bibnamefont {Evered}}, \bibinfo {author} {\bibfnamefont {J.~P.~B.}\ \bibnamefont {Ataides}}, \bibinfo {author} {\bibfnamefont {G.}~\bibnamefont {Baranes}}, \bibinfo {author} {\bibfnamefont {A.}~\bibnamefont {Gu}}, \bibinfo {author} {\bibfnamefont {T.}~\bibnamefont {Manovitz}}, \bibinfo {author} {\bibfnamefont {M.}~\bibnamefont {Xu}}, \bibinfo {author} {\bibfnamefont {M.}~\bibnamefont {Kalinowski}}, \emph {et~al.},\ }\bibinfo {title} {A fault-tolerant neutral-atom architecture for universal quantum computation},\ \href {https://doi.org/10.1038/s41586-025-09848-5} {\bibfield  {journal} {\bibinfo  {journal} {Nature}\ }\textbf {\bibinfo {volume} {649}},\ \bibinfo {pages} {39} (\bibinfo {year} {2026})}\BibitemShut {NoStop}%
\bibitem [{\citenamefont {Zhou}\ \emph {et~al.}(2025)\citenamefont {Zhou}, \citenamefont {Duckering}, \citenamefont {Zhao}, \citenamefont {Bluvstein}, \citenamefont {Cain}, \citenamefont {Kubica}, \citenamefont {Wang},\ and\ \citenamefont {Lukin}}]{Zhou_2025}%
  \BibitemOpen
  \bibfield  {author} {\bibinfo {author} {\bibfnamefont {H.}~\bibnamefont {Zhou}}, \bibinfo {author} {\bibfnamefont {C.}~\bibnamefont {Duckering}}, \bibinfo {author} {\bibfnamefont {C.}~\bibnamefont {Zhao}}, \bibinfo {author} {\bibfnamefont {D.}~\bibnamefont {Bluvstein}}, \bibinfo {author} {\bibfnamefont {M.}~\bibnamefont {Cain}}, \bibinfo {author} {\bibfnamefont {A.}~\bibnamefont {Kubica}}, \bibinfo {author} {\bibfnamefont {S.-T.}\ \bibnamefont {Wang}},\ and\ \bibinfo {author} {\bibfnamefont {M.~D.}\ \bibnamefont {Lukin}},\ }in\ \href {https://doi.org/10.1145/3695053.3731039} {\emph {\bibinfo {booktitle} {Proceedings of the 52nd Annual International Symposium on Computer Architecture}}},\ \bibinfo {series and number} {SIGARCH ’25}\ (\bibinfo  {publisher} {ACM},\ \bibinfo {year} {2025})\ p.\ \bibinfo {pages} {1432–1448}\BibitemShut {NoStop}%
\bibitem [{\citenamefont {Ransford}\ \emph {et~al.}(2025)\citenamefont {Ransford}, \citenamefont {Allman}, \citenamefont {Arkinstall}, \citenamefont {III}, \citenamefont {Cooper}, \citenamefont {Delaney}, \citenamefont {Dreiling}, \citenamefont {Estey}, \citenamefont {Figgatt}, \citenamefont {Hall}, \citenamefont {Husain}, \citenamefont {Isanaka}, \citenamefont {Kennedy}, \citenamefont {Kotibhaskar}, \citenamefont {Madjarov}, \citenamefont {Mayer}, \citenamefont {Milne}, \citenamefont {Park}, \citenamefont {Reed}, \citenamefont {Ancona}, \citenamefont {Andersen}, \citenamefont {Andres-Martinez}, \citenamefont {Angenent}, \citenamefont {Argueta}, \citenamefont {Arkin}, \citenamefont {Ascarrunz}, \citenamefont {Baker}, \citenamefont {Barnes}, \citenamefont {Bartolotta}, \citenamefont {Berg}, \citenamefont {Besand}, \citenamefont {Bjork}, \citenamefont {Blain}, \citenamefont {Blanchard}, \citenamefont {Blume-Kohout}, \citenamefont {Bohn}, \citenamefont {Borgna}, \citenamefont {Botamanenko}, \citenamefont
  {Boutelle}, \citenamefont {Brown}, \citenamefont {Buckingham}, \citenamefont {Burdick}, \citenamefont {Burton}, \citenamefont {Carey}, \citenamefont {Carron}, \citenamefont {Chambers}, \citenamefont {Children}, \citenamefont {Colussi}, \citenamefont {Crepinsek}, \citenamefont {Cureton}, \citenamefont {Davies}, \citenamefont {Davis}, \citenamefont {DeCross}, \citenamefont {Deen}, \citenamefont {Delaney}, \citenamefont {DelVento}, \citenamefont {DeSalvo}, \citenamefont {Dominy}, \citenamefont {Duncan}, \citenamefont {Eccles}, \citenamefont {Edgington}, \citenamefont {Erickson}, \citenamefont {Erickson}, \citenamefont {Ertsgaard}, \citenamefont {Evans}, \citenamefont {Evans}, \citenamefont {Fabrikant}, \citenamefont {Fischer}, \citenamefont {Foltz}, \citenamefont {Foss-Feig}, \citenamefont {Francois}, \citenamefont {Freyberg}, \citenamefont {Gao}, \citenamefont {Garay}, \citenamefont {Garvin}, \citenamefont {Gaudiosi}, \citenamefont {Gilbreth}, \citenamefont {Giles}, \citenamefont {Glynn}, \citenamefont
  {Graves}, \citenamefont {Hansen}, \citenamefont {Hayes}, \citenamefont {Heidemann}, \citenamefont {Higashi}, \citenamefont {Hilbun}, \citenamefont {Hines}, \citenamefont {Hlavaty}, \citenamefont {Hoffman}, \citenamefont {Hoffman}, \citenamefont {Holliman}, \citenamefont {Hooper}, \citenamefont {Horning}, \citenamefont {Hostetter}, \citenamefont {Hothem}, \citenamefont {Houlton}, \citenamefont {Hout}, \citenamefont {Hutson}, \citenamefont {Jacobs}, \citenamefont {Jacobs}, \citenamefont {Johannsen}, \citenamefont {Johansen}, \citenamefont {Jones}, \citenamefont {Julian}, \citenamefont {Jung}, \citenamefont {Keay}, \citenamefont {Klein}, \citenamefont {Koch}, \citenamefont {Kondo}, \citenamefont {Kong}, \citenamefont {Kosto}, \citenamefont {Lawrence}, \citenamefont {Liefer}, \citenamefont {Lollie}, \citenamefont {Lucchetti}, \citenamefont {Lysne}, \citenamefont {Lytle}, \citenamefont {MacPherson}, \citenamefont {Malm}, \citenamefont {Mather}, \citenamefont {Mathewson}, \citenamefont {Maxwell}, \citenamefont
  {McCaffrey}, \citenamefont {McDougall}, \citenamefont {Mendoza}, \citenamefont {Mills}, \citenamefont {Morrison}, \citenamefont {Narmour}, \citenamefont {Nguyen}, \citenamefont {Nugent}, \citenamefont {Olson}, \citenamefont {Ouellette}, \citenamefont {Parks}, \citenamefont {Peters}, \citenamefont {Petricka}, \citenamefont {Pino}, \citenamefont {Polito}, \citenamefont {Preidl}, \citenamefont {Price}, \citenamefont {Proctor}, \citenamefont {Pugh}, \citenamefont {Ratcliff}, \citenamefont {Raymondson}, \citenamefont {Rhodes}, \citenamefont {Roman}, \citenamefont {Roy}, \citenamefont {Ryan-Anderson}, \citenamefont {Sanchez}, \citenamefont {Sangiolo}, \citenamefont {Sawadski}, \citenamefont {Schaffer}, \citenamefont {Schow}, \citenamefont {Sedlacek}, \citenamefont {Semenenko}, \citenamefont {Shevchuk}, \citenamefont {Shore}, \citenamefont {Siegfried}, \citenamefont {Singhal}, \citenamefont {Sivarajah}, \citenamefont {Skripka}, \citenamefont {Sletten}, \citenamefont {Spaun}, \citenamefont {Sprenkle}, \citenamefont
  {Stoufer}, \citenamefont {Tader}, \citenamefont {Taylor}, \citenamefont {Thompson}, \citenamefont {Tobey}, \citenamefont {Tran}, \citenamefont {Tran}, \citenamefont {Vittorini}, \citenamefont {Volin}, \citenamefont {Walker}, \citenamefont {White}, \citenamefont {Wilson}, \citenamefont {Wolf}, \citenamefont {Wringe}, \citenamefont {Young}, \citenamefont {Zheng}, \citenamefont {Zuraski}, \citenamefont {Baldwin}, \citenamefont {Chernoguzov}, \citenamefont {Gaebler}, \citenamefont {Sanders}, \citenamefont {Neyenhuis}, \citenamefont {Stutz},\ and\ \citenamefont {Bohnet}}]{Ransford2025}%
  \BibitemOpen
  \bibfield  {author} {\bibinfo {author} {\bibfnamefont {A.}~\bibnamefont {Ransford}}, \bibinfo {author} {\bibfnamefont {M.~S.}\ \bibnamefont {Allman}}, \bibinfo {author} {\bibfnamefont {J.}~\bibnamefont {Arkinstall}}, \bibinfo {author} {\bibfnamefont {J.~P.~C.}\ \bibnamefont {III}}, \bibinfo {author} {\bibfnamefont {S.~F.}\ \bibnamefont {Cooper}}, \bibinfo {author} {\bibfnamefont {R.~D.}\ \bibnamefont {Delaney}}, \bibinfo {author} {\bibfnamefont {J.~M.}\ \bibnamefont {Dreiling}}, \bibinfo {author} {\bibfnamefont {B.}~\bibnamefont {Estey}}, \bibinfo {author} {\bibfnamefont {C.}~\bibnamefont {Figgatt}}, \bibinfo {author} {\bibfnamefont {A.}~\bibnamefont {Hall}}, \emph {et~al.},\ }\href {https://arxiv.org/abs/2511.05465} {\bibinfo {title} {Helios: A 98-qubit trapped-ion quantum computer}} (\bibinfo {year} {2025}),\ \Eprint {https://arxiv.org/abs/2511.05465} {arXiv:2511.05465 [quant-ph]} \BibitemShut {NoStop}%
\bibitem [{\citenamefont {Schlosser}\ \emph {et~al.}(2023)\citenamefont {Schlosser}, \citenamefont {Tichelmann}, \citenamefont {Sch\"affner}, \citenamefont {de~Mello}, \citenamefont {Hambach}, \citenamefont {Sch\"utz},\ and\ \citenamefont {Birkl}}]{SchlosserPRL2023}%
  \BibitemOpen
  \bibfield  {author} {\bibinfo {author} {\bibfnamefont {M.}~\bibnamefont {Schlosser}}, \bibinfo {author} {\bibfnamefont {S.}~\bibnamefont {Tichelmann}}, \bibinfo {author} {\bibfnamefont {D.}~\bibnamefont {Sch\"affner}}, \bibinfo {author} {\bibfnamefont {D.~O.}\ \bibnamefont {de~Mello}}, \bibinfo {author} {\bibfnamefont {M.}~\bibnamefont {Hambach}}, \bibinfo {author} {\bibfnamefont {J.}~\bibnamefont {Sch\"utz}},\ and\ \bibinfo {author} {\bibfnamefont {G.}~\bibnamefont {Birkl}},\ }\bibinfo {title} {Scalable multilayer architecture of assembled single-atom qubit arrays in a three-dimensional {T}albot tweezer lattice},\ \href {https://doi.org/10.1103/PhysRevLett.130.180601} {\bibfield  {journal} {\bibinfo  {journal} {Phys. Rev. Lett.}\ }\textbf {\bibinfo {volume} {130}},\ \bibinfo {pages} {180601} (\bibinfo {year} {2023})}\BibitemShut {NoStop}%
\bibitem [{\citenamefont {Pause}\ \emph {et~al.}(2024)\citenamefont {Pause}, \citenamefont {Sturm}, \citenamefont {Mittenb\"{u}hler}, \citenamefont {Amann}, \citenamefont {Preuschoff}, \citenamefont {Sch\"{a}ffner}, \citenamefont {Schlosser},\ and\ \citenamefont {Birkl}}]{PauseOptica24}%
  \BibitemOpen
  \bibfield  {author} {\bibinfo {author} {\bibfnamefont {L.}~\bibnamefont {Pause}}, \bibinfo {author} {\bibfnamefont {L.}~\bibnamefont {Sturm}}, \bibinfo {author} {\bibfnamefont {M.}~\bibnamefont {Mittenb\"{u}hler}}, \bibinfo {author} {\bibfnamefont {S.}~\bibnamefont {Amann}}, \bibinfo {author} {\bibfnamefont {T.}~\bibnamefont {Preuschoff}}, \bibinfo {author} {\bibfnamefont {D.}~\bibnamefont {Sch\"{a}ffner}}, \bibinfo {author} {\bibfnamefont {M.}~\bibnamefont {Schlosser}},\ and\ \bibinfo {author} {\bibfnamefont {G.}~\bibnamefont {Birkl}},\ }\bibinfo {title} {Supercharged two-dimensional tweezer array with more than 1000 atomic qubits},\ \href {https://doi.org/10.1364/OPTICA.513551} {\bibfield  {journal} {\bibinfo  {journal} {Optica}\ }\textbf {\bibinfo {volume} {11}},\ \bibinfo {pages} {222} (\bibinfo {year} {2024})}\BibitemShut {NoStop}%
\bibitem [{\citenamefont {Norcia}\ \emph {et~al.}(2024)\citenamefont {Norcia}, \citenamefont {Kim}, \citenamefont {Cairncross}, \citenamefont {Stone}, \citenamefont {Ryou}, \citenamefont {Jaffe}, \citenamefont {Brown}, \citenamefont {Barnes}, \citenamefont {Battaglino}, \citenamefont {Bohdanowicz}, \citenamefont {Brown}, \citenamefont {Cassella}, \citenamefont {Chen}, \citenamefont {Coxe}, \citenamefont {Crow}, \citenamefont {Epstein}, \citenamefont {Griger}, \citenamefont {Halperin}, \citenamefont {Hummel}, \citenamefont {Jones}, \citenamefont {Kindem}, \citenamefont {King}, \citenamefont {Kotru}, \citenamefont {Lauigan}, \citenamefont {Li}, \citenamefont {Lu}, \citenamefont {Megidish}, \citenamefont {Marjanovic}, \citenamefont {McDonald}, \citenamefont {Mittiga}, \citenamefont {Muniz}, \citenamefont {Narayanaswami}, \citenamefont {Nishiguchi}, \citenamefont {Paule}, \citenamefont {Pawlak}, \citenamefont {Peng}, \citenamefont {Pudenz}, \citenamefont {Pérez}, \citenamefont {Smull}, \citenamefont {Stack},
  \citenamefont {Urbanek}, \citenamefont {van~de Veerdonk}, \citenamefont {Vendeiro}, \citenamefont {Wadleigh}, \citenamefont {Wilkason}, \citenamefont {Wu}, \citenamefont {Xie}, \citenamefont {Zalys-Geller}, \citenamefont {Zhang},\ and\ \citenamefont {Bloom}}]{Norcia2024}%
  \BibitemOpen
  \bibfield  {author} {\bibinfo {author} {\bibfnamefont {M.~A.}\ \bibnamefont {Norcia}}, \bibinfo {author} {\bibfnamefont {H.}~\bibnamefont {Kim}}, \bibinfo {author} {\bibfnamefont {W.~B.}\ \bibnamefont {Cairncross}}, \bibinfo {author} {\bibfnamefont {M.}~\bibnamefont {Stone}}, \bibinfo {author} {\bibfnamefont {A.}~\bibnamefont {Ryou}}, \bibinfo {author} {\bibfnamefont {M.}~\bibnamefont {Jaffe}}, \bibinfo {author} {\bibfnamefont {M.~O.}\ \bibnamefont {Brown}}, \bibinfo {author} {\bibfnamefont {K.}~\bibnamefont {Barnes}}, \bibinfo {author} {\bibfnamefont {P.}~\bibnamefont {Battaglino}}, \bibinfo {author} {\bibfnamefont {T.~C.}\ \bibnamefont {Bohdanowicz}}, \emph {et~al.},\ }\bibinfo {title} {Iterative assembly of 171 {Y}b atom arrays with cavity-enhanced optical lattices},\ \href {https://doi.org/10.1103/PRXQuantum.5.030316} {\bibfield  {journal} {\bibinfo  {journal} {PRX Quantum}\ }\textbf {\bibinfo {volume} {5}},\ \bibinfo {pages} {030316} (\bibinfo {year} {2024})}\BibitemShut {NoStop}%
\bibitem [{\citenamefont {Gyger}\ \emph {et~al.}(2024)\citenamefont {Gyger}, \citenamefont {Ammenwerth}, \citenamefont {Tao}, \citenamefont {Timme}, \citenamefont {Snigirev}, \citenamefont {Bloch},\ and\ \citenamefont {Zeiher}}]{Gyger2024}%
  \BibitemOpen
  \bibfield  {author} {\bibinfo {author} {\bibfnamefont {F.}~\bibnamefont {Gyger}}, \bibinfo {author} {\bibfnamefont {M.}~\bibnamefont {Ammenwerth}}, \bibinfo {author} {\bibfnamefont {R.}~\bibnamefont {Tao}}, \bibinfo {author} {\bibfnamefont {H.}~\bibnamefont {Timme}}, \bibinfo {author} {\bibfnamefont {S.}~\bibnamefont {Snigirev}}, \bibinfo {author} {\bibfnamefont {I.}~\bibnamefont {Bloch}},\ and\ \bibinfo {author} {\bibfnamefont {J.}~\bibnamefont {Zeiher}},\ }\bibinfo {title} {Continuous operation of large-scale atom arrays in optical lattices},\ \href {https://doi.org/10.1103/PhysRevResearch.6.033104} {\bibfield  {journal} {\bibinfo  {journal} {Physical Review Research}\ }\textbf {\bibinfo {volume} {6}},\ \bibinfo {pages} {033104} (\bibinfo {year} {2024})}\BibitemShut {NoStop}%
\bibitem [{\citenamefont {Manetsch}\ \emph {et~al.}(2025)\citenamefont {Manetsch}, \citenamefont {Nomura}, \citenamefont {Bataille}, \citenamefont {Lv}, \citenamefont {Leung},\ and\ \citenamefont {Endres}}]{Manetsch2024}%
  \BibitemOpen
  \bibfield  {author} {\bibinfo {author} {\bibfnamefont {H.~J.}\ \bibnamefont {Manetsch}}, \bibinfo {author} {\bibfnamefont {G.}~\bibnamefont {Nomura}}, \bibinfo {author} {\bibfnamefont {E.}~\bibnamefont {Bataille}}, \bibinfo {author} {\bibfnamefont {X.}~\bibnamefont {Lv}}, \bibinfo {author} {\bibfnamefont {K.~H.}\ \bibnamefont {Leung}},\ and\ \bibinfo {author} {\bibfnamefont {M.}~\bibnamefont {Endres}},\ }\bibinfo {title} {A tweezer array with 6,100 highly coherent atomic qubits},\ \href {https://doi.org/10.1038/s41586-025-09641-4} {\bibfield  {journal} {\bibinfo  {journal} {Nature}\ }\textbf {\bibinfo {volume} {647}},\ \bibinfo {pages} {60} (\bibinfo {year} {2025})}\BibitemShut {NoStop}%
\bibitem [{\citenamefont {Chiu}\ \emph {et~al.}(2025)\citenamefont {Chiu}, \citenamefont {Trapp}, \citenamefont {Guo}, \citenamefont {Abobeih}, \citenamefont {Stewart}, \citenamefont {Hollerith}, \citenamefont {Stroganov}, \citenamefont {Kalinowski}, \citenamefont {Geim}, \citenamefont {Evered}, \citenamefont {Li}, \citenamefont {Lyu}, \citenamefont {Peters}, \citenamefont {Bluvstein}, \citenamefont {Wang}, \citenamefont {Greiner}, \citenamefont {Vuletić},\ and\ \citenamefont {Lukin}}]{Chiu2025}%
  \BibitemOpen
  \bibfield  {author} {\bibinfo {author} {\bibfnamefont {N.-C.}\ \bibnamefont {Chiu}}, \bibinfo {author} {\bibfnamefont {E.~C.}\ \bibnamefont {Trapp}}, \bibinfo {author} {\bibfnamefont {J.}~\bibnamefont {Guo}}, \bibinfo {author} {\bibfnamefont {M.~H.}\ \bibnamefont {Abobeih}}, \bibinfo {author} {\bibfnamefont {L.~M.}\ \bibnamefont {Stewart}}, \bibinfo {author} {\bibfnamefont {S.}~\bibnamefont {Hollerith}}, \bibinfo {author} {\bibfnamefont {P.~L.}\ \bibnamefont {Stroganov}}, \bibinfo {author} {\bibfnamefont {M.}~\bibnamefont {Kalinowski}}, \bibinfo {author} {\bibfnamefont {A.~A.}\ \bibnamefont {Geim}}, \bibinfo {author} {\bibfnamefont {S.~J.}\ \bibnamefont {Evered}}, \emph {et~al.},\ }\bibinfo {title} {Continuous operation of a coherent 3,000-qubit system},\ \href {https://doi.org/10.1038/s41586-025-09596-6} {\bibfield  {journal} {\bibinfo  {journal} {Nature}\ }\textbf {\bibinfo {volume} {646}},\ \bibinfo {pages} {1075} (\bibinfo {year} {2025})}\BibitemShut {NoStop}%
\bibitem [{\citenamefont {Binai-Motlagh}\ \emph {et~al.}(2023)\citenamefont {Binai-Motlagh}, \citenamefont {Day}, \citenamefont {Videnov}, \citenamefont {Greenberg}, \citenamefont {Senko},\ and\ \citenamefont {Islam}}]{Binai2023}%
  \BibitemOpen
  \bibfield  {author} {\bibinfo {author} {\bibfnamefont {A.}~\bibnamefont {Binai-Motlagh}}, \bibinfo {author} {\bibfnamefont {M.~L.}\ \bibnamefont {Day}}, \bibinfo {author} {\bibfnamefont {N.}~\bibnamefont {Videnov}}, \bibinfo {author} {\bibfnamefont {N.}~\bibnamefont {Greenberg}}, \bibinfo {author} {\bibfnamefont {C.}~\bibnamefont {Senko}},\ and\ \bibinfo {author} {\bibfnamefont {R.}~\bibnamefont {Islam}},\ }\bibinfo {title} {A guided light system for agile individual addressing of {B}a{+} qubits with $10^{-4}$ level intensity crosstalk},\ \href {https://doi.org/10.1088/2058-9565/ace6cb} {\bibfield  {journal} {\bibinfo  {journal} {Quantum Science and Technology}\ }\textbf {\bibinfo {volume} {8}},\ \bibinfo {pages} {045012} (\bibinfo {year} {2023})}\BibitemShut {NoStop}%
\bibitem [{\citenamefont {Christen}\ \emph {et~al.}(2025)\citenamefont {Christen}, \citenamefont {Propson}, \citenamefont {Sutula}, \citenamefont {Sattari}, \citenamefont {Choong}, \citenamefont {Panuski}, \citenamefont {Melville}, \citenamefont {Mallek}, \citenamefont {Brabec}, \citenamefont {Hamilton}, \citenamefont {Dixon}, \citenamefont {Menssen}, \citenamefont {Braje}, \citenamefont {Ghadimi},\ and\ \citenamefont {Englund}}]{Christen2022}%
  \BibitemOpen
  \bibfield  {author} {\bibinfo {author} {\bibfnamefont {I.}~\bibnamefont {Christen}}, \bibinfo {author} {\bibfnamefont {T.}~\bibnamefont {Propson}}, \bibinfo {author} {\bibfnamefont {M.}~\bibnamefont {Sutula}}, \bibinfo {author} {\bibfnamefont {H.}~\bibnamefont {Sattari}}, \bibinfo {author} {\bibfnamefont {G.}~\bibnamefont {Choong}}, \bibinfo {author} {\bibfnamefont {C.}~\bibnamefont {Panuski}}, \bibinfo {author} {\bibfnamefont {A.}~\bibnamefont {Melville}}, \bibinfo {author} {\bibfnamefont {J.}~\bibnamefont {Mallek}}, \bibinfo {author} {\bibfnamefont {C.}~\bibnamefont {Brabec}}, \bibinfo {author} {\bibfnamefont {S.}~\bibnamefont {Hamilton}}, \emph {et~al.},\ }\bibinfo {title} {An integrated photonic engine for programmable atomic control},\ \href {https://doi.org/10.1038/s41467-024-55423-3} {\bibfield  {journal} {\bibinfo  {journal} {Nature Communications}\ }\textbf {\bibinfo {volume} {16}},\ \bibinfo {pages} {82} (\bibinfo {year} {2025})}\BibitemShut {NoStop}%
\bibitem [{\citenamefont {Schindler}\ \emph {et~al.}(2013)\citenamefont {Schindler}, \citenamefont {Nigg}, \citenamefont {Monz}, \citenamefont {Barreiro}, \citenamefont {Martinez}, \citenamefont {Wang}, \citenamefont {Quint}, \citenamefont {Brandl}, \citenamefont {Nebendahl}, \citenamefont {Roos}, \citenamefont {Chwalla}, \citenamefont {Hennrich},\ and\ \citenamefont {Blatt}}]{Schindler2013}%
  \BibitemOpen
  \bibfield  {author} {\bibinfo {author} {\bibfnamefont {P.}~\bibnamefont {Schindler}}, \bibinfo {author} {\bibfnamefont {D.}~\bibnamefont {Nigg}}, \bibinfo {author} {\bibfnamefont {T.}~\bibnamefont {Monz}}, \bibinfo {author} {\bibfnamefont {J.~T.}\ \bibnamefont {Barreiro}}, \bibinfo {author} {\bibfnamefont {E.}~\bibnamefont {Martinez}}, \bibinfo {author} {\bibfnamefont {S.~X.}\ \bibnamefont {Wang}}, \bibinfo {author} {\bibfnamefont {S.}~\bibnamefont {Quint}}, \bibinfo {author} {\bibfnamefont {M.~F.}\ \bibnamefont {Brandl}}, \bibinfo {author} {\bibfnamefont {V.}~\bibnamefont {Nebendahl}}, \bibinfo {author} {\bibfnamefont {C.~F.}\ \bibnamefont {Roos}}, \emph {et~al.},\ }\bibinfo {title} {A quantum information processor with trapped ions},\ \href {https://doi.org/10.1088/1367-2630/15/12/123012} {\bibfield  {journal} {\bibinfo  {journal} {New Journal of Physics}\ }\textbf {\bibinfo {volume} {15}},\ \bibinfo {pages} {123012} (\bibinfo {year} {2013})}\BibitemShut {NoStop}%
\bibitem [{\citenamefont {Lee}\ \emph {et~al.}(2016)\citenamefont {Lee}, \citenamefont {Smith}, \citenamefont {Richerme}, \citenamefont {Neyenhuis}, \citenamefont {Hess}, \citenamefont {Zhang},\ and\ \citenamefont {Monroe}}]{Lee2016}%
  \BibitemOpen
  \bibfield  {author} {\bibinfo {author} {\bibfnamefont {A.~C.}\ \bibnamefont {Lee}}, \bibinfo {author} {\bibfnamefont {J.}~\bibnamefont {Smith}}, \bibinfo {author} {\bibfnamefont {P.}~\bibnamefont {Richerme}}, \bibinfo {author} {\bibfnamefont {B.}~\bibnamefont {Neyenhuis}}, \bibinfo {author} {\bibfnamefont {P.~W.}\ \bibnamefont {Hess}}, \bibinfo {author} {\bibfnamefont {J.}~\bibnamefont {Zhang}},\ and\ \bibinfo {author} {\bibfnamefont {C.}~\bibnamefont {Monroe}},\ }\bibinfo {title} {Engineering large stark shifts for control of individual clock state qubits},\ \href {https://doi.org/10.1103/PhysRevA.94.042308} {\bibfield  {journal} {\bibinfo  {journal} {Phys. Rev. A}\ }\textbf {\bibinfo {volume} {94}},\ \bibinfo {pages} {042308} (\bibinfo {year} {2016})}\BibitemShut {NoStop}%
\bibitem [{\citenamefont {Polloreno}\ \emph {et~al.}(2022)\citenamefont {Polloreno}, \citenamefont {Rey},\ and\ \citenamefont {Bollinger}}]{Polloreno2022}%
  \BibitemOpen
  \bibfield  {author} {\bibinfo {author} {\bibfnamefont {A.~M.}\ \bibnamefont {Polloreno}}, \bibinfo {author} {\bibfnamefont {A.~M.}\ \bibnamefont {Rey}},\ and\ \bibinfo {author} {\bibfnamefont {J.~J.}\ \bibnamefont {Bollinger}},\ }\bibinfo {title} {Individual qubit addressing of rotating ion crystals in a penning trap},\ \href {https://doi.org/10.1103/PhysRevResearch.4.033076} {\bibfield  {journal} {\bibinfo  {journal} {Physical Review Research}\ }\textbf {\bibinfo {volume} {4}},\ \bibinfo {pages} {033076} (\bibinfo {year} {2022})}\BibitemShut {NoStop}%
\bibitem [{\citenamefont {Barnes}\ \emph {et~al.}(2022)\citenamefont {Barnes}, \citenamefont {Battaglino}, \citenamefont {Bloom}, \citenamefont {Cassella}, \citenamefont {Coxe}, \citenamefont {Crisosto}, \citenamefont {King}, \citenamefont {Kondov}, \citenamefont {Kotru}, \citenamefont {Larsen}, \citenamefont {Lauigan}, \citenamefont {Lester}, \citenamefont {McDonald}, \citenamefont {Megidish}, \citenamefont {Narayanaswami}, \citenamefont {Nishiguchi}, \citenamefont {Notermans}, \citenamefont {Peng}, \citenamefont {Ryou}, \citenamefont {Wu},\ and\ \citenamefont {Yarwood}}]{Barnes2022}%
  \BibitemOpen
  \bibfield  {author} {\bibinfo {author} {\bibfnamefont {K.}~\bibnamefont {Barnes}}, \bibinfo {author} {\bibfnamefont {P.}~\bibnamefont {Battaglino}}, \bibinfo {author} {\bibfnamefont {B.~J.}\ \bibnamefont {Bloom}}, \bibinfo {author} {\bibfnamefont {K.}~\bibnamefont {Cassella}}, \bibinfo {author} {\bibfnamefont {R.}~\bibnamefont {Coxe}}, \bibinfo {author} {\bibfnamefont {N.}~\bibnamefont {Crisosto}}, \bibinfo {author} {\bibfnamefont {J.~P.}\ \bibnamefont {King}}, \bibinfo {author} {\bibfnamefont {S.~S.}\ \bibnamefont {Kondov}}, \bibinfo {author} {\bibfnamefont {K.}~\bibnamefont {Kotru}}, \bibinfo {author} {\bibfnamefont {S.~C.}\ \bibnamefont {Larsen}}, \emph {et~al.},\ }\bibinfo {title} {Assembly and coherent control of a register of nuclear spin qubits},\ \bibfield  {journal} {\bibinfo  {journal} {Nature Communications}\ }\textbf {\bibinfo {volume} {13}},\ \href {https://doi.org/10.1038/s41467-022-29977-z} {10.1038/s41467-022-29977-z} (\bibinfo {year} {2022})\BibitemShut {NoStop}%
\bibitem [{\citenamefont {Shaw}\ \emph {et~al.}(2024)\citenamefont {Shaw}, \citenamefont {Finkelstein}, \citenamefont {Tsai}, \citenamefont {Scholl}, \citenamefont {Yoon}, \citenamefont {Choi},\ and\ \citenamefont {Endres}}]{Shaw2023}%
  \BibitemOpen
  \bibfield  {author} {\bibinfo {author} {\bibfnamefont {A.~L.}\ \bibnamefont {Shaw}}, \bibinfo {author} {\bibfnamefont {R.}~\bibnamefont {Finkelstein}}, \bibinfo {author} {\bibfnamefont {R.~B.-S.}\ \bibnamefont {Tsai}}, \bibinfo {author} {\bibfnamefont {P.}~\bibnamefont {Scholl}}, \bibinfo {author} {\bibfnamefont {T.~H.}\ \bibnamefont {Yoon}}, \bibinfo {author} {\bibfnamefont {J.}~\bibnamefont {Choi}},\ and\ \bibinfo {author} {\bibfnamefont {M.}~\bibnamefont {Endres}},\ }\bibinfo {title} {Multi-ensemble metrology by programming local rotations with atom movements},\ \href {https://doi.org/10.1038/s41567-023-02323-w} {\bibfield  {journal} {\bibinfo  {journal} {Nature Physics}\ }\textbf {\bibinfo {volume} {20}},\ \bibinfo {pages} {195} (\bibinfo {year} {2024})}\BibitemShut {NoStop}%
\bibitem [{\citenamefont {Labuhn}\ \emph {et~al.}(2014)\citenamefont {Labuhn}, \citenamefont {Ravets}, \citenamefont {Barredo}, \citenamefont {Béguin}, \citenamefont {Nogrette}, \citenamefont {Lahaye},\ and\ \citenamefont {Browaeys}}]{Labuhn2014}%
  \BibitemOpen
  \bibfield  {author} {\bibinfo {author} {\bibfnamefont {H.}~\bibnamefont {Labuhn}}, \bibinfo {author} {\bibfnamefont {S.}~\bibnamefont {Ravets}}, \bibinfo {author} {\bibfnamefont {D.}~\bibnamefont {Barredo}}, \bibinfo {author} {\bibfnamefont {L.}~\bibnamefont {Béguin}}, \bibinfo {author} {\bibfnamefont {F.}~\bibnamefont {Nogrette}}, \bibinfo {author} {\bibfnamefont {T.}~\bibnamefont {Lahaye}},\ and\ \bibinfo {author} {\bibfnamefont {A.}~\bibnamefont {Browaeys}},\ }\bibinfo {title} {Single-atom addressing in microtraps for quantum-state engineering using {R}ydberg atoms},\ \href {https://doi.org/10.1103/PhysRevA.90.023415} {\bibfield  {journal} {\bibinfo  {journal} {Phys. Rev. A}\ }\textbf {\bibinfo {volume} {90}},\ \bibinfo {pages} {023415} (\bibinfo {year} {2014})}\BibitemShut {NoStop}%
\bibitem [{\citenamefont {de~Léséleuc}\ \emph {et~al.}(2017)\citenamefont {de~Léséleuc}, \citenamefont {Barredo}, \citenamefont {Lienhard}, \citenamefont {Browaeys},\ and\ \citenamefont {Lahaye}}]{deLeseleuc2017}%
  \BibitemOpen
  \bibfield  {author} {\bibinfo {author} {\bibfnamefont {S.}~\bibnamefont {de~Léséleuc}}, \bibinfo {author} {\bibfnamefont {D.}~\bibnamefont {Barredo}}, \bibinfo {author} {\bibfnamefont {V.}~\bibnamefont {Lienhard}}, \bibinfo {author} {\bibfnamefont {A.}~\bibnamefont {Browaeys}},\ and\ \bibinfo {author} {\bibfnamefont {T.}~\bibnamefont {Lahaye}},\ }\bibinfo {title} {Optical control of the resonant dipole-dipole interaction between {R}ydberg atoms},\ \href {https://doi.org/10.1103/PhysRevLett.119.053202} {\bibfield  {journal} {\bibinfo  {journal} {Physical Review Letters}\ }\textbf {\bibinfo {volume} {119}},\ \bibinfo {pages} {053202} (\bibinfo {year} {2017})}\BibitemShut {NoStop}%
\bibitem [{\citenamefont {Eckner}\ \emph {et~al.}(2023)\citenamefont {Eckner}, \citenamefont {Darkwah~Oppong}, \citenamefont {Cao}, \citenamefont {Young}, \citenamefont {Milner}, \citenamefont {Robinson}, \citenamefont {Ye},\ and\ \citenamefont {Kaufman}}]{Eckner2023}%
  \BibitemOpen
  \bibfield  {author} {\bibinfo {author} {\bibfnamefont {W.~J.}\ \bibnamefont {Eckner}}, \bibinfo {author} {\bibfnamefont {N.}~\bibnamefont {Darkwah~Oppong}}, \bibinfo {author} {\bibfnamefont {A.}~\bibnamefont {Cao}}, \bibinfo {author} {\bibfnamefont {A.~W.}\ \bibnamefont {Young}}, \bibinfo {author} {\bibfnamefont {W.~R.}\ \bibnamefont {Milner}}, \bibinfo {author} {\bibfnamefont {J.~M.}\ \bibnamefont {Robinson}}, \bibinfo {author} {\bibfnamefont {J.}~\bibnamefont {Ye}},\ and\ \bibinfo {author} {\bibfnamefont {A.~M.}\ \bibnamefont {Kaufman}},\ }\bibinfo {title} {Realizing spin squeezing with {R}ydberg interactions in an optical clock},\ \href {https://doi.org/10.1038/s41586-023-06360-6} {\bibfield  {journal} {\bibinfo  {journal} {Nature}\ }\textbf {\bibinfo {volume} {621}},\ \bibinfo {pages} {734–739} (\bibinfo {year} {2023})}\BibitemShut {NoStop}%
\bibitem [{\citenamefont {Graham}\ \emph {et~al.}(2022)\citenamefont {Graham}, \citenamefont {Song}, \citenamefont {Scott}, \citenamefont {Poole}, \citenamefont {Phuttitarn}, \citenamefont {Jooya}, \citenamefont {Eichler}, \citenamefont {Jiang}, \citenamefont {Marra}, \citenamefont {Grinkemeyer}, \citenamefont {Kwon}, \citenamefont {Ebert}, \citenamefont {Cherek}, \citenamefont {Lichtman}, \citenamefont {Gillette}, \citenamefont {Gilbert}, \citenamefont {Bowman}, \citenamefont {Ballance}, \citenamefont {Campbell}, \citenamefont {Dahl}, \citenamefont {Crawford}, \citenamefont {Blunt}, \citenamefont {Rogers}, \citenamefont {Noel},\ and\ \citenamefont {Saffman}}]{Graham2022}%
  \BibitemOpen
  \bibfield  {author} {\bibinfo {author} {\bibfnamefont {T.~M.}\ \bibnamefont {Graham}}, \bibinfo {author} {\bibfnamefont {Y.}~\bibnamefont {Song}}, \bibinfo {author} {\bibfnamefont {J.}~\bibnamefont {Scott}}, \bibinfo {author} {\bibfnamefont {C.}~\bibnamefont {Poole}}, \bibinfo {author} {\bibfnamefont {L.}~\bibnamefont {Phuttitarn}}, \bibinfo {author} {\bibfnamefont {K.}~\bibnamefont {Jooya}}, \bibinfo {author} {\bibfnamefont {P.}~\bibnamefont {Eichler}}, \bibinfo {author} {\bibfnamefont {X.}~\bibnamefont {Jiang}}, \bibinfo {author} {\bibfnamefont {A.}~\bibnamefont {Marra}}, \bibinfo {author} {\bibfnamefont {B.}~\bibnamefont {Grinkemeyer}}, \emph {et~al.},\ }\bibinfo {title} {Multi-qubit entanglement and algorithms on a neutral-atom quantum computer},\ \href {https://doi.org/10.1038/s41586-022-04603-6} {\bibfield  {journal} {\bibinfo  {journal} {Nature}\ }\textbf {\bibinfo {volume} {604}},\ \bibinfo {pages} {457–462} (\bibinfo {year} {2022})}\BibitemShut {NoStop}%
\bibitem [{\citenamefont {Levitt}(1986)}]{Levitt1986}%
  \BibitemOpen
  \bibfield  {author} {\bibinfo {author} {\bibfnamefont {M.~H.}\ \bibnamefont {Levitt}},\ }\bibinfo {title} {Composite pulses},\ \href {https://doi.org/10.1016/0079-6565(86)80005-X} {\bibfield  {journal} {\bibinfo  {journal} {Progress in Nuclear Magnetic Resonance Spectroscopy}\ }\textbf {\bibinfo {volume} {18}},\ \bibinfo {pages} {61} (\bibinfo {year} {1986})}\BibitemShut {NoStop}%
\bibitem [{\citenamefont {Merrill}\ and\ \citenamefont {Brown}(2012)}]{Merrill2012}%
  \BibitemOpen
  \bibfield  {author} {\bibinfo {author} {\bibfnamefont {J.~T.}\ \bibnamefont {Merrill}}\ and\ \bibinfo {author} {\bibfnamefont {K.~R.}\ \bibnamefont {Brown}},\ }\href@noop {} {} (\bibinfo {year} {2012}),\ \Eprint {https://arxiv.org/abs/1203.6392} {arXiv:1203.6392 [quant-ph]} \BibitemShut {NoStop}%
\bibitem [{\citenamefont {Brown}\ \emph {et~al.}(2004)\citenamefont {Brown}, \citenamefont {Harrow},\ and\ \citenamefont {Chuang}}]{Brown2004}%
  \BibitemOpen
  \bibfield  {author} {\bibinfo {author} {\bibfnamefont {K.~R.}\ \bibnamefont {Brown}}, \bibinfo {author} {\bibfnamefont {A.~W.}\ \bibnamefont {Harrow}},\ and\ \bibinfo {author} {\bibfnamefont {I.~L.}\ \bibnamefont {Chuang}},\ }\bibinfo {title} {Arbitrarily accurate composite pulse sequences},\ \href {https://doi.org/10.1103/PhysRevA.70.052318} {\bibfield  {journal} {\bibinfo  {journal} {Phys. Rev. A}\ }\textbf {\bibinfo {volume} {70}},\ \bibinfo {pages} {052318} (\bibinfo {year} {2004})}\BibitemShut {NoStop}%
\bibitem [{\citenamefont {Mc~Hugh}\ and\ \citenamefont {Twamley}(2005)}]{Hugh2005}%
  \BibitemOpen
  \bibfield  {author} {\bibinfo {author} {\bibfnamefont {D.}~\bibnamefont {Mc~Hugh}}\ and\ \bibinfo {author} {\bibfnamefont {J.}~\bibnamefont {Twamley}},\ }\bibinfo {title} {Sixth-order robust gates for quantum control},\ \href {https://doi.org/10.1103/PhysRevA.71.012327} {\bibfield  {journal} {\bibinfo  {journal} {Phys. Rev. A}\ }\textbf {\bibinfo {volume} {71}},\ \bibinfo {pages} {012327} (\bibinfo {year} {2005})}\BibitemShut {NoStop}%
\bibitem [{\citenamefont {Cummins}\ and\ \citenamefont {Jones}(2000)}]{Cummins2000}%
  \BibitemOpen
  \bibfield  {author} {\bibinfo {author} {\bibfnamefont {H.~K.}\ \bibnamefont {Cummins}}\ and\ \bibinfo {author} {\bibfnamefont {J.~A.}\ \bibnamefont {Jones}},\ }\bibinfo {title} {Use of composite rotations to correct systematic errors in {NMR} quantum computation},\ \href {https://doi.org/10.1088/1367-2630/2/1/006} {\bibfield  {journal} {\bibinfo  {journal} {New Journal of Physics}\ }\textbf {\bibinfo {volume} {2}},\ \bibinfo {pages} {006} (\bibinfo {year} {2000})}\BibitemShut {NoStop}%
\bibitem [{\citenamefont {Cummins}\ \emph {et~al.}(2003)\citenamefont {Cummins}, \citenamefont {Llewellyn},\ and\ \citenamefont {Jones}}]{Cummins2003}%
  \BibitemOpen
  \bibfield  {author} {\bibinfo {author} {\bibfnamefont {H.~K.}\ \bibnamefont {Cummins}}, \bibinfo {author} {\bibfnamefont {G.}~\bibnamefont {Llewellyn}},\ and\ \bibinfo {author} {\bibfnamefont {J.~A.}\ \bibnamefont {Jones}},\ }\bibinfo {title} {Tackling systematic errors in quantum logic gates with composite rotations},\ \href {https://doi.org/10.1103/PhysRevA.67.042308} {\bibfield  {journal} {\bibinfo  {journal} {Phys. Rev. A}\ }\textbf {\bibinfo {volume} {67}},\ \bibinfo {pages} {042308} (\bibinfo {year} {2003})}\BibitemShut {NoStop}%
\bibitem [{\citenamefont {Torosov}\ and\ \citenamefont {Vitanov}(2014)}]{Torosov2014}%
  \BibitemOpen
  \bibfield  {author} {\bibinfo {author} {\bibfnamefont {B.~T.}\ \bibnamefont {Torosov}}\ and\ \bibinfo {author} {\bibfnamefont {N.~V.}\ \bibnamefont {Vitanov}},\ }\bibinfo {title} {High-fidelity error-resilient composite phase gates},\ \href {https://doi.org/10.1103/PhysRevA.90.012341} {\bibfield  {journal} {\bibinfo  {journal} {Phys. Rev. A}\ }\textbf {\bibinfo {volume} {90}},\ \bibinfo {pages} {012341} (\bibinfo {year} {2014})}\BibitemShut {NoStop}%
\bibitem [{\citenamefont {Torosov}\ and\ \citenamefont {Vitanov}(2018)}]{Torosov2018}%
  \BibitemOpen
  \bibfield  {author} {\bibinfo {author} {\bibfnamefont {B.~T.}\ \bibnamefont {Torosov}}\ and\ \bibinfo {author} {\bibfnamefont {N.~V.}\ \bibnamefont {Vitanov}},\ }\bibinfo {title} {Arbitrarily accurate twin composite $\ensuremath{\pi}$-pulse sequences},\ \href {https://doi.org/10.1103/PhysRevA.97.043408} {\bibfield  {journal} {\bibinfo  {journal} {Phys. Rev. A}\ }\textbf {\bibinfo {volume} {97}},\ \bibinfo {pages} {043408} (\bibinfo {year} {2018})}\BibitemShut {NoStop}%
\bibitem [{\citenamefont {Wimperis}(1994)}]{Wimperis1994}%
  \BibitemOpen
  \bibfield  {author} {\bibinfo {author} {\bibfnamefont {S.}~\bibnamefont {Wimperis}},\ }\bibinfo {title} {Broadband, narrowband, and passband composite pulses for use in advanced {NMR} experiments},\ \href {https://doi.org/10.1006/jmra.1994.1159} {\bibfield  {journal} {\bibinfo  {journal} {Journal of Magnetic Resonance, Series A}\ }\textbf {\bibinfo {volume} {109}},\ \bibinfo {pages} {221} (\bibinfo {year} {1994})}\BibitemShut {NoStop}%
\bibitem [{\citenamefont {Gevorgyan}\ and\ \citenamefont {Vitanov}(2021)}]{Gevorgyan2021}%
  \BibitemOpen
  \bibfield  {author} {\bibinfo {author} {\bibfnamefont {H.~L.}\ \bibnamefont {Gevorgyan}}\ and\ \bibinfo {author} {\bibfnamefont {N.~V.}\ \bibnamefont {Vitanov}},\ }\bibinfo {title} {Ultrahigh-fidelity composite rotational quantum gates},\ \href {https://doi.org/10.1103/PhysRevA.104.012609} {\bibfield  {journal} {\bibinfo  {journal} {Phys. Rev. A}\ }\textbf {\bibinfo {volume} {104}},\ \bibinfo {pages} {012609} (\bibinfo {year} {2021})}\BibitemShut {NoStop}%
\bibitem [{\citenamefont {Kukita}\ \emph {et~al.}(2022{\natexlab{a}})\citenamefont {Kukita}, \citenamefont {Kiya},\ and\ \citenamefont {Kondo}}]{KukitaPRA2022}%
  \BibitemOpen
  \bibfield  {author} {\bibinfo {author} {\bibfnamefont {S.}~\bibnamefont {Kukita}}, \bibinfo {author} {\bibfnamefont {H.}~\bibnamefont {Kiya}},\ and\ \bibinfo {author} {\bibfnamefont {Y.}~\bibnamefont {Kondo}},\ }\bibinfo {title} {General off-resonance-error-robust symmetric composite pulses with three elementary operations},\ \href {https://doi.org/10.1103/PhysRevA.106.042613} {\bibfield  {journal} {\bibinfo  {journal} {Phys. Rev. A}\ }\textbf {\bibinfo {volume} {106}},\ \bibinfo {pages} {042613} (\bibinfo {year} {2022}{\natexlab{a}})}\BibitemShut {NoStop}%
\bibitem [{\citenamefont {Kukita}\ \emph {et~al.}(2023)\citenamefont {Kukita}, \citenamefont {Kiya},\ and\ \citenamefont {Kondo}}]{Kukita2023}%
  \BibitemOpen
  \bibfield  {author} {\bibinfo {author} {\bibfnamefont {S.}~\bibnamefont {Kukita}}, \bibinfo {author} {\bibfnamefont {H.}~\bibnamefont {Kiya}},\ and\ \bibinfo {author} {\bibfnamefont {Y.}~\bibnamefont {Kondo}},\ }\bibinfo {title} {Lower bound on operation time of composite quantum gates robust against pulse length error},\ \href {https://doi.org/10.1088/1751-8121/ad0804} {\bibfield  {journal} {\bibinfo  {journal} {Journal of Physics A: Mathematical and Theoretical}\ }\textbf {\bibinfo {volume} {56}},\ \bibinfo {pages} {485305} (\bibinfo {year} {2023})}\BibitemShut {NoStop}%
\bibitem [{\citenamefont {Gevorgyan}\ and\ \citenamefont {Vitanov}(2024)}]{Gevorgyan2023}%
  \BibitemOpen
  \bibfield  {author} {\bibinfo {author} {\bibfnamefont {H.~L.}\ \bibnamefont {Gevorgyan}}\ and\ \bibinfo {author} {\bibfnamefont {N.~V.}\ \bibnamefont {Vitanov}},\ }\bibinfo {title} {Ultrahigh-fidelity composite quantum phase gates},\ \href {https://doi.org/10.1103/PhysRevA.109.052625} {\bibfield  {journal} {\bibinfo  {journal} {Physical Review A}\ }\textbf {\bibinfo {volume} {109}},\ \bibinfo {pages} {052625} (\bibinfo {year} {2024})}\BibitemShut {NoStop}%
\bibitem [{\citenamefont {Fromonteil}\ \emph {et~al.}(2023)\citenamefont {Fromonteil}, \citenamefont {Bluvstein},\ and\ \citenamefont {Pichler}}]{Fromonteil2023}%
  \BibitemOpen
  \bibfield  {author} {\bibinfo {author} {\bibfnamefont {C.}~\bibnamefont {Fromonteil}}, \bibinfo {author} {\bibfnamefont {D.}~\bibnamefont {Bluvstein}},\ and\ \bibinfo {author} {\bibfnamefont {H.}~\bibnamefont {Pichler}},\ }\bibinfo {title} {Protocols for {R}ydberg entangling gates featuring robustness against quasistatic errors},\ \href {https://doi.org/10.1103/PRXQuantum.4.020335} {\bibfield  {journal} {\bibinfo  {journal} {PRX Quantum}\ }\textbf {\bibinfo {volume} {4}},\ \bibinfo {pages} {020335} (\bibinfo {year} {2023})}\BibitemShut {NoStop}%
\bibitem [{\citenamefont {Bando}\ \emph {et~al.}(2013)\citenamefont {Bando}, \citenamefont {Ichikawa}, \citenamefont {Kondo},\ and\ \citenamefont {Nakahara}}]{Bando2013}%
  \BibitemOpen
  \bibfield  {author} {\bibinfo {author} {\bibfnamefont {M.}~\bibnamefont {Bando}}, \bibinfo {author} {\bibfnamefont {T.}~\bibnamefont {Ichikawa}}, \bibinfo {author} {\bibfnamefont {Y.}~\bibnamefont {Kondo}},\ and\ \bibinfo {author} {\bibfnamefont {M.}~\bibnamefont {Nakahara}},\ }\bibinfo {title} {Concatenated composite pulses compensating simultaneous systematic errors},\ \bibfield  {journal} {\bibinfo  {journal} {Journal of the Physical Society of Japan}\ }\textbf {\bibinfo {volume} {82}},\ \href {https://doi.org/10.7566/JPSJ.82.014004} {10.7566/JPSJ.82.014004} (\bibinfo {year} {2013})\BibitemShut {NoStop}%
\bibitem [{\citenamefont {Chen}\ \emph {et~al.}(2023)\citenamefont {Chen}, \citenamefont {Ding}, \citenamefont {Huang},\ and\ \citenamefont {Ye}}]{Chen2021}%
  \BibitemOpen
  \bibfield  {author} {\bibinfo {author} {\bibfnamefont {J.}~\bibnamefont {Chen}}, \bibinfo {author} {\bibfnamefont {D.}~\bibnamefont {Ding}}, \bibinfo {author} {\bibfnamefont {C.}~\bibnamefont {Huang}},\ and\ \bibinfo {author} {\bibfnamefont {Q.}~\bibnamefont {Ye}},\ }\bibinfo {title} {Compiling arbitrary single-qubit gates via the phase shifts of microwave pulses},\ \bibfield  {journal} {\bibinfo  {journal} {Physical Review Research}\ }\textbf {\bibinfo {volume} {5}},\ \href {https://doi.org/10.1103/physrevresearch.5.l022031} {10.1103/physrevresearch.5.l022031} (\bibinfo {year} {2023})\BibitemShut {NoStop}%
\bibitem [{\citenamefont {Low}\ \emph {et~al.}(2014)\citenamefont {Low}, \citenamefont {Yoder},\ and\ \citenamefont {Chuang}}]{Low2013}%
  \BibitemOpen
  \bibfield  {author} {\bibinfo {author} {\bibfnamefont {G.~H.}\ \bibnamefont {Low}}, \bibinfo {author} {\bibfnamefont {T.~J.}\ \bibnamefont {Yoder}},\ and\ \bibinfo {author} {\bibfnamefont {I.~L.}\ \bibnamefont {Chuang}},\ }\bibinfo {title} {Optimal arbitrarily accurate composite pulse sequences},\ \href {https://doi.org/10.1103/PhysRevA.89.022341} {\bibfield  {journal} {\bibinfo  {journal} {Phys. Rev. A}\ }\textbf {\bibinfo {volume} {89}},\ \bibinfo {pages} {022341} (\bibinfo {year} {2014})}\BibitemShut {NoStop}%
\bibitem [{sup()}]{supp}%
  \BibitemOpen
  \href@noop {} {}\bibinfo {note} {See Supplemental Material, which includes Refs.~\cite{SUTER1987509,KukitaPSJ2022,Le2023}}\BibitemShut {NoStop}%
\bibitem [{\citenamefont {Ichikawa}\ \emph {et~al.}(2014)\citenamefont {Ichikawa}, \citenamefont {Filgueiras}, \citenamefont {Bando}, \citenamefont {Kondo}, \citenamefont {Nakahara},\ and\ \citenamefont {Suter}}]{Ichikawa2014}%
  \BibitemOpen
  \bibfield  {author} {\bibinfo {author} {\bibfnamefont {T.}~\bibnamefont {Ichikawa}}, \bibinfo {author} {\bibfnamefont {J.~G.}\ \bibnamefont {Filgueiras}}, \bibinfo {author} {\bibfnamefont {M.}~\bibnamefont {Bando}}, \bibinfo {author} {\bibfnamefont {Y.}~\bibnamefont {Kondo}}, \bibinfo {author} {\bibfnamefont {M.}~\bibnamefont {Nakahara}},\ and\ \bibinfo {author} {\bibfnamefont {D.}~\bibnamefont {Suter}},\ }\bibinfo {title} {Construction of arbitrary robust one-qubit operations using planar geometry},\ \href {https://doi.org/10.1103/PhysRevA.90.052330} {\bibfield  {journal} {\bibinfo  {journal} {Phys. Rev. A}\ }\textbf {\bibinfo {volume} {90}},\ \bibinfo {pages} {052330} (\bibinfo {year} {2014})}\BibitemShut {NoStop}%
\bibitem [{\citenamefont {Jones}(2013)}]{Jones2013}%
  \BibitemOpen
  \bibfield  {author} {\bibinfo {author} {\bibfnamefont {J.~A.}\ \bibnamefont {Jones}},\ }\bibinfo {title} {Designing short robust not gates for quantum computation},\ \href {https://doi.org/10.1103/PhysRevA.87.052317} {\bibfield  {journal} {\bibinfo  {journal} {Phys. Rev. A}\ }\textbf {\bibinfo {volume} {87}},\ \bibinfo {pages} {052317} (\bibinfo {year} {2013})}\BibitemShut {NoStop}%
\bibitem [{\citenamefont {Ma}\ \emph {et~al.}(2023)\citenamefont {Ma}, \citenamefont {Liu}, \citenamefont {Peng}, \citenamefont {Zhang}, \citenamefont {Jandura}, \citenamefont {Claes}, \citenamefont {Burgers}, \citenamefont {Pupillo}, \citenamefont {Puri},\ and\ \citenamefont {Thompson}}]{Ma2023}%
  \BibitemOpen
  \bibfield  {author} {\bibinfo {author} {\bibfnamefont {S.}~\bibnamefont {Ma}}, \bibinfo {author} {\bibfnamefont {G.}~\bibnamefont {Liu}}, \bibinfo {author} {\bibfnamefont {P.}~\bibnamefont {Peng}}, \bibinfo {author} {\bibfnamefont {B.}~\bibnamefont {Zhang}}, \bibinfo {author} {\bibfnamefont {S.}~\bibnamefont {Jandura}}, \bibinfo {author} {\bibfnamefont {J.}~\bibnamefont {Claes}}, \bibinfo {author} {\bibfnamefont {A.~P.}\ \bibnamefont {Burgers}}, \bibinfo {author} {\bibfnamefont {G.}~\bibnamefont {Pupillo}}, \bibinfo {author} {\bibfnamefont {S.}~\bibnamefont {Puri}},\ and\ \bibinfo {author} {\bibfnamefont {J.~D.}\ \bibnamefont {Thompson}},\ }\bibinfo {title} {High-fidelity gates and mid-circuit erasure conversion in an atomic qubit},\ \href {https://doi.org/10.1038/s41586-023-06438-1} {\bibfield  {journal} {\bibinfo  {journal} {Nature}\ }\textbf {\bibinfo {volume} {622}},\ \bibinfo {pages} {279} (\bibinfo {year} {2023})}\BibitemShut {NoStop}%
\bibitem [{\citenamefont {Kang}\ \emph {et~al.}(2023)\citenamefont {Kang}, \citenamefont {Campbell},\ and\ \citenamefont {Brown}}]{Kang2023}%
  \BibitemOpen
  \bibfield  {author} {\bibinfo {author} {\bibfnamefont {M.}~\bibnamefont {Kang}}, \bibinfo {author} {\bibfnamefont {W.~C.}\ \bibnamefont {Campbell}},\ and\ \bibinfo {author} {\bibfnamefont {K.~R.}\ \bibnamefont {Brown}},\ }\bibinfo {title} {Quantum error correction with metastable states of trapped ions using erasure conversion},\ \href {https://doi.org/10.1103/PRXQuantum.4.020358} {\bibfield  {journal} {\bibinfo  {journal} {PRX Quantum}\ }\textbf {\bibinfo {volume} {4}},\ \bibinfo {pages} {020358} (\bibinfo {year} {2023})}\BibitemShut {NoStop}%
\bibitem [{\citenamefont {Jandura}\ \emph {et~al.}(2023)\citenamefont {Jandura}, \citenamefont {Thompson},\ and\ \citenamefont {Pupillo}}]{Jandura2023}%
  \BibitemOpen
  \bibfield  {author} {\bibinfo {author} {\bibfnamefont {S.}~\bibnamefont {Jandura}}, \bibinfo {author} {\bibfnamefont {J.~D.}\ \bibnamefont {Thompson}},\ and\ \bibinfo {author} {\bibfnamefont {G.}~\bibnamefont {Pupillo}},\ }\bibinfo {title} {Optimizing {R}ydberg gates for logical-qubit performance},\ \href {https://doi.org/10.1103/PRXQuantum.4.020336} {\bibfield  {journal} {\bibinfo  {journal} {PRX Quantum}\ }\textbf {\bibinfo {volume} {4}},\ \bibinfo {pages} {020336} (\bibinfo {year} {2023})}\BibitemShut {NoStop}%
\bibitem [{\citenamefont {Shi}\ \emph {et~al.}(2024)\citenamefont {Shi}, \citenamefont {Ding}, \citenamefont {Chen}, \citenamefont {Song}, \citenamefont {Xia}, \citenamefont {Yi},\ and\ \citenamefont {Nori}}]{Shi2023}%
  \BibitemOpen
  \bibfield  {author} {\bibinfo {author} {\bibfnamefont {Z.-C.}\ \bibnamefont {Shi}}, \bibinfo {author} {\bibfnamefont {J.-T.}\ \bibnamefont {Ding}}, \bibinfo {author} {\bibfnamefont {Y.-H.}\ \bibnamefont {Chen}}, \bibinfo {author} {\bibfnamefont {J.}~\bibnamefont {Song}}, \bibinfo {author} {\bibfnamefont {Y.}~\bibnamefont {Xia}}, \bibinfo {author} {\bibfnamefont {X.}~\bibnamefont {Yi}},\ and\ \bibinfo {author} {\bibfnamefont {F.}~\bibnamefont {Nori}},\ }\bibinfo {title} {Supervised learning for robust quantum control in composite-pulse systems},\ \href {https://doi.org/10.1103/PhysRevApplied.21.044012} {\bibfield  {journal} {\bibinfo  {journal} {Physical Review Applied}\ }\textbf {\bibinfo {volume} {21}},\ \bibinfo {pages} {044012} (\bibinfo {year} {2024})}\BibitemShut {NoStop}%
\bibitem [{\citenamefont {McKay}\ \emph {et~al.}(2017)\citenamefont {McKay}, \citenamefont {Wood}, \citenamefont {Sheldon}, \citenamefont {Chow},\ and\ \citenamefont {Gambetta}}]{McKay2016}%
  \BibitemOpen
  \bibfield  {author} {\bibinfo {author} {\bibfnamefont {D.~C.}\ \bibnamefont {McKay}}, \bibinfo {author} {\bibfnamefont {C.~J.}\ \bibnamefont {Wood}}, \bibinfo {author} {\bibfnamefont {S.}~\bibnamefont {Sheldon}}, \bibinfo {author} {\bibfnamefont {J.~M.}\ \bibnamefont {Chow}},\ and\ \bibinfo {author} {\bibfnamefont {J.~M.}\ \bibnamefont {Gambetta}},\ }\bibinfo {title} {Efficient $z$ gates for quantum computing},\ \href {https://doi.org/10.1103/PhysRevA.96.022330} {\bibfield  {journal} {\bibinfo  {journal} {Phys. Rev. A}\ }\textbf {\bibinfo {volume} {96}},\ \bibinfo {pages} {022330} (\bibinfo {year} {2017})}\BibitemShut {NoStop}%
\bibitem [{\citenamefont {Suter}\ and\ \citenamefont {Pines}(1987)}]{SUTER1987509}%
  \BibitemOpen
  \bibfield  {author} {\bibinfo {author} {\bibfnamefont {D.}~\bibnamefont {Suter}}\ and\ \bibinfo {author} {\bibfnamefont {A.}~\bibnamefont {Pines}},\ }\bibinfo {title} {Recursive evaluation of interaction pictures},\ \href {https://doi.org/https://doi.org/10.1016/0022-2364(87)90106-5} {\bibfield  {journal} {\bibinfo  {journal} {Journal of Magnetic Resonance (1969)}\ }\textbf {\bibinfo {volume} {75}},\ \bibinfo {pages} {509} (\bibinfo {year} {1987})}\BibitemShut {NoStop}%
\bibitem [{\citenamefont {Kukita}\ \emph {et~al.}(2022{\natexlab{b}})\citenamefont {Kukita}, \citenamefont {Kiya},\ and\ \citenamefont {Kondo}}]{KukitaPSJ2022}%
  \BibitemOpen
  \bibfield  {author} {\bibinfo {author} {\bibfnamefont {S.}~\bibnamefont {Kukita}}, \bibinfo {author} {\bibfnamefont {H.}~\bibnamefont {Kiya}},\ and\ \bibinfo {author} {\bibfnamefont {Y.}~\bibnamefont {Kondo}},\ }\bibinfo {title} {Short composite quantum gate robust against two common systematic errors},\ \bibfield  {journal} {\bibinfo  {journal} {Journal of the Physical Society of Japan}\ }\textbf {\bibinfo {volume} {91}},\ \href {https://doi.org/10.7566/jpsj.91.104001} {10.7566/jpsj.91.104001} (\bibinfo {year} {2022}{\natexlab{b}})\BibitemShut {NoStop}%
\bibitem [{\citenamefont {Le}\ \emph {et~al.}(2023)\citenamefont {Le}, \citenamefont {Cykiert},\ and\ \citenamefont {Ginossar}}]{Le2023}%
  \BibitemOpen
  \bibfield  {author} {\bibinfo {author} {\bibfnamefont {N.~H.}\ \bibnamefont {Le}}, \bibinfo {author} {\bibfnamefont {M.}~\bibnamefont {Cykiert}},\ and\ \bibinfo {author} {\bibfnamefont {E.}~\bibnamefont {Ginossar}},\ }\bibinfo {title} {Scalable and robust quantum computing on qubit arrays with fixed coupling},\ \href {https://doi.org/10.1038/s41534-022-00668-3} {\bibfield  {journal} {\bibinfo  {journal} {npj Quantum Information}\ }\textbf {\bibinfo {volume} {9}},\ \bibinfo {pages} {1} (\bibinfo {year} {2023})}\BibitemShut {NoStop}%
\end{thebibliography}%


\begin{thebibliography}{10}%
\makeatletter
\providecommand \@ifxundefined [1]{%
 \@ifx{#1\undefined}
}%
\providecommand \@ifnum [1]{%
 \ifnum #1\expandafter \@firstoftwo
 \else \expandafter \@secondoftwo
 \fi
}%
\providecommand \@ifx [1]{%
 \ifx #1\expandafter \@firstoftwo
 \else \expandafter \@secondoftwo
 \fi
}%
\providecommand \natexlab [1]{#1}%
\providecommand \enquote  [1]{``#1''}%
\providecommand \bibnamefont  [1]{#1}%
\providecommand \bibfnamefont [1]{#1}%
\providecommand \citenamefont [1]{#1}%
\providecommand \href@noop [0]{\@secondoftwo}%
\providecommand \href [0]{\begingroup \@sanitize@url \@href}%
\providecommand \@href[1]{\@@startlink{#1}\@@href}%
\providecommand \@@href[1]{\endgroup#1\@@endlink}%
\providecommand \@sanitize@url [0]{\catcode `\\12\catcode `\$12\catcode `\&12\catcode `\#12\catcode `\^12\catcode `\_12\catcode `\%12\relax}%
\providecommand \@@startlink[1]{}%
\providecommand \@@endlink[0]{}%
\providecommand \url  [0]{\begingroup\@sanitize@url \@url }%
\providecommand \@url [1]{\endgroup\@href {#1}{\urlprefix }}%
\providecommand \urlprefix  [0]{URL }%
\providecommand \Eprint [0]{\href }%
\providecommand \doibase [0]{https://doi.org/}%
\providecommand \selectlanguage [0]{\@gobble}%
\providecommand \bibinfo  [0]{\@secondoftwo}%
\providecommand \bibfield  [0]{\@secondoftwo}%
\providecommand \translation [1]{[#1]}%
\providecommand \BibitemOpen [0]{}%
\providecommand \bibitemStop [0]{}%
\providecommand \bibitemNoStop [0]{.\EOS\space}%
\providecommand \EOS [0]{\spacefactor3000\relax}%
\providecommand \BibitemShut  [1]{\csname bibitem#1\endcsname}%
\let\auto@bib@innerbib\@empty
\bibitem [{\citenamefont {Low}\ \emph {et~al.}(2016)\citenamefont {Low}, \citenamefont {Yoder},\ and\ \citenamefont {Chuang}}]{Low2016}%
  \BibitemOpen
  \bibfield  {author} {\bibinfo {author} {\bibfnamefont {G.~H.}\ \bibnamefont {Low}}, \bibinfo {author} {\bibfnamefont {T.~J.}\ \bibnamefont {Yoder}},\ and\ \bibinfo {author} {\bibfnamefont {I.~L.}\ \bibnamefont {Chuang}},\ }\bibinfo {title} {Methodology of resonant equiangular composite quantum gates},\ \href {https://doi.org/10.1103/PhysRevX.6.041067} {\bibfield  {journal} {\bibinfo  {journal} {Physical Review X}\ }\textbf {\bibinfo {volume} {6}},\ \bibinfo {pages} {041067} (\bibinfo {year} {2016})}\BibitemShut {NoStop}%
\bibitem [{\citenamefont {Suter}\ and\ \citenamefont {Pines}(1987)}]{SUTER1987509}%
  \BibitemOpen
  \bibfield  {author} {\bibinfo {author} {\bibfnamefont {D.}~\bibnamefont {Suter}}\ and\ \bibinfo {author} {\bibfnamefont {A.}~\bibnamefont {Pines}},\ }\bibinfo {title} {Recursive evaluation of interaction pictures},\ \href {https://doi.org/https://doi.org/10.1016/0022-2364(87)90106-5} {\bibfield  {journal} {\bibinfo  {journal} {Journal of Magnetic Resonance (1969)}\ }\textbf {\bibinfo {volume} {75}},\ \bibinfo {pages} {509} (\bibinfo {year} {1987})}\BibitemShut {NoStop}%
\bibitem [{\citenamefont {Jones}(2013)}]{Jones2013}%
  \BibitemOpen
  \bibfield  {author} {\bibinfo {author} {\bibfnamefont {J.~A.}\ \bibnamefont {Jones}},\ }\bibinfo {title} {Designing short robust not gates for quantum computation},\ \href {https://doi.org/10.1103/PhysRevA.87.052317} {\bibfield  {journal} {\bibinfo  {journal} {Phys. Rev. A}\ }\textbf {\bibinfo {volume} {87}},\ \bibinfo {pages} {052317} (\bibinfo {year} {2013})}\BibitemShut {NoStop}%
\bibitem [{\citenamefont {Kukita}\ \emph {et~al.}(2022)\citenamefont {Kukita}, \citenamefont {Kiya},\ and\ \citenamefont {Kondo}}]{KukitaPSJ2022}%
  \BibitemOpen
  \bibfield  {author} {\bibinfo {author} {\bibfnamefont {S.}~\bibnamefont {Kukita}}, \bibinfo {author} {\bibfnamefont {H.}~\bibnamefont {Kiya}},\ and\ \bibinfo {author} {\bibfnamefont {Y.}~\bibnamefont {Kondo}},\ }\bibinfo {title} {Short composite quantum gate robust against two common systematic errors},\ \bibfield  {journal} {\bibinfo  {journal} {Journal of the Physical Society of Japan}\ }\textbf {\bibinfo {volume} {91}},\ \href {https://doi.org/10.7566/jpsj.91.104001} {10.7566/jpsj.91.104001} (\bibinfo {year} {2022})\BibitemShut {NoStop}%
\bibitem [{\citenamefont {Bando}\ \emph {et~al.}(2013)\citenamefont {Bando}, \citenamefont {Ichikawa}, \citenamefont {Kondo},\ and\ \citenamefont {Nakahara}}]{Bando2013}%
  \BibitemOpen
  \bibfield  {author} {\bibinfo {author} {\bibfnamefont {M.}~\bibnamefont {Bando}}, \bibinfo {author} {\bibfnamefont {T.}~\bibnamefont {Ichikawa}}, \bibinfo {author} {\bibfnamefont {Y.}~\bibnamefont {Kondo}},\ and\ \bibinfo {author} {\bibfnamefont {M.}~\bibnamefont {Nakahara}},\ }\bibinfo {title} {Concatenated composite pulses compensating simultaneous systematic errors},\ \bibfield  {journal} {\bibinfo  {journal} {Journal of the Physical Society of Japan}\ }\textbf {\bibinfo {volume} {82}},\ \href {https://doi.org/10.7566/JPSJ.82.014004} {10.7566/JPSJ.82.014004} (\bibinfo {year} {2013})\BibitemShut {NoStop}%
\bibitem [{\citenamefont {Le}\ \emph {et~al.}(2023)\citenamefont {Le}, \citenamefont {Cykiert},\ and\ \citenamefont {Ginossar}}]{Le2023}%
  \BibitemOpen
  \bibfield  {author} {\bibinfo {author} {\bibfnamefont {N.~H.}\ \bibnamefont {Le}}, \bibinfo {author} {\bibfnamefont {M.}~\bibnamefont {Cykiert}},\ and\ \bibinfo {author} {\bibfnamefont {E.}~\bibnamefont {Ginossar}},\ }\bibinfo {title} {Scalable and robust quantum computing on qubit arrays with fixed coupling},\ \href {https://doi.org/10.1038/s41534-022-00668-3} {\bibfield  {journal} {\bibinfo  {journal} {npj Quantum Information}\ }\textbf {\bibinfo {volume} {9}},\ \bibinfo {pages} {1} (\bibinfo {year} {2023})}\BibitemShut {NoStop}%
\bibitem [{\citenamefont {Jandura}\ \emph {et~al.}(2023)\citenamefont {Jandura}, \citenamefont {Thompson},\ and\ \citenamefont {Pupillo}}]{Jandura2023}%
  \BibitemOpen
  \bibfield  {author} {\bibinfo {author} {\bibfnamefont {S.}~\bibnamefont {Jandura}}, \bibinfo {author} {\bibfnamefont {J.~D.}\ \bibnamefont {Thompson}},\ and\ \bibinfo {author} {\bibfnamefont {G.}~\bibnamefont {Pupillo}},\ }\bibinfo {title} {Optimizing {R}ydberg gates for logical-qubit performance},\ \href {https://doi.org/10.1103/PRXQuantum.4.020336} {\bibfield  {journal} {\bibinfo  {journal} {PRX Quantum}\ }\textbf {\bibinfo {volume} {4}},\ \bibinfo {pages} {020336} (\bibinfo {year} {2023})}\BibitemShut {NoStop}%
\bibitem [{\citenamefont {Cummins}\ \emph {et~al.}(2003)\citenamefont {Cummins}, \citenamefont {Llewellyn},\ and\ \citenamefont {Jones}}]{Cummins2003}%
  \BibitemOpen
  \bibfield  {author} {\bibinfo {author} {\bibfnamefont {H.~K.}\ \bibnamefont {Cummins}}, \bibinfo {author} {\bibfnamefont {G.}~\bibnamefont {Llewellyn}},\ and\ \bibinfo {author} {\bibfnamefont {J.~A.}\ \bibnamefont {Jones}},\ }\bibinfo {title} {Tackling systematic errors in quantum logic gates with composite rotations},\ \href {https://doi.org/10.1103/PhysRevA.67.042308} {\bibfield  {journal} {\bibinfo  {journal} {Phys. Rev. A}\ }\textbf {\bibinfo {volume} {67}},\ \bibinfo {pages} {042308} (\bibinfo {year} {2003})}\BibitemShut {NoStop}%
\bibitem [{\citenamefont {Wimperis}(1994)}]{Wimperis1994}%
  \BibitemOpen
  \bibfield  {author} {\bibinfo {author} {\bibfnamefont {S.}~\bibnamefont {Wimperis}},\ }\bibinfo {title} {Broadband, narrowband, and passband composite pulses for use in advanced {NMR} experiments},\ \href {https://doi.org/10.1006/jmra.1994.1159} {\bibfield  {journal} {\bibinfo  {journal} {Journal of Magnetic Resonance, Series A}\ }\textbf {\bibinfo {volume} {109}},\ \bibinfo {pages} {221} (\bibinfo {year} {1994})}\BibitemShut {NoStop}%
\bibitem [{\citenamefont {Cummins}\ and\ \citenamefont {Jones}(2000)}]{Cummins2000}%
  \BibitemOpen
  \bibfield  {author} {\bibinfo {author} {\bibfnamefont {H.~K.}\ \bibnamefont {Cummins}}\ and\ \bibinfo {author} {\bibfnamefont {J.~A.}\ \bibnamefont {Jones}},\ }\bibinfo {title} {Use of composite rotations to correct systematic errors in {NMR} quantum computation},\ \href {https://doi.org/10.1088/1367-2630/2/1/006} {\bibfield  {journal} {\bibinfo  {journal} {New Journal of Physics}\ }\textbf {\bibinfo {volume} {2}},\ \bibinfo {pages} {006} (\bibinfo {year} {2000})}\BibitemShut {NoStop}%
\end{thebibliography}%

\end{document}



\title{Supplemental Materials for `Robust and Parallel Control of Many Qubits'}
\author{Wenjie Gong}
\author{Soonwon Choi}%
\affiliation{%
Center for Theoretical Physics, Massachusetts Institute of Technology, Cambridge, MA 02139, USA
}%

\date{\today}
\maketitle
\tableofcontents
\section{Optimal scaling of the number of pulses in high order composite pulses}\label{app:optcp}
In this section, we show that any generic composite pulse sequence that suppresses an imperfection $\epsilon$ to order~$\sim O(\epsilon^{n+1})$ has an optimal number of pulses scaling linearly in $n$. This can be straightforwardly seen through parameter counting. 

To be specific, we carry out this counting for a generic pulse sequence $\mathcal{P}_1 \mathcal{P}_2 \cdots \mathcal{P}_k$ =  $\mathcal{U(\epsilon)}$= $\mathcal{U}(0)+O(\epsilon^{n+1})$ , where $\mathcal{P}_i\in$ SU(2), so $\mathcal{U(\epsilon)} \in$ SU(2). Thus, $\mathcal{U(\epsilon)}$ takes the form
\begin{align}
\mathcal{U(\epsilon)} = \begin{pmatrix}
a(\epsilon) & b(\epsilon)\\
-b(\epsilon)^* & a(\epsilon)^*
\end{pmatrix}.
\end{align}
In the absence of the error ($\epsilon=0$), this unitary implements our target unitary, enforced by $a(0) = a_0$, $b(0) = b_0$. For robustness up to the $n$-th order, we have that $\frac{d^{n'} a}{d^{n'}\epsilon}|_{\epsilon = 0} = 0, \frac{d^{n'} b}{d^{n'}\epsilon}|_{\epsilon = 0} = 0$ for all $1 < n' \leq n$. Naively, setting the zero-th order terms of $a(\epsilon)$, $b(\epsilon)$ equal to their target values and their derivatives with respect to $\epsilon$ up to order $n$ to zero results in $2(n+1)$ complex constraints, or $4(n+1)$ real equations, that must be fulfilled by the pulses $\{\mathcal{P}_i\}$. However, as $\mathcal{U}$ is unitary, we must additionally have $|a(\epsilon)|^2 + |b(\epsilon)|^2 = 1$. Intuitively, we expect that this will reduce the number of equations by 1 at each order, leading to $3(n+1)$ total equations. As there are two free parameters, $\theta$ and $\phi$, for every pulse $\mathcal{P}_i$, we then require at least $\lceil \frac{3(n+1)}{2} \rceil \sim O(n)$ pulses for an error compensating pulse sequence of $O(\epsilon^{n+1})$ in general. 

We now prove our intuition regarding the effect of unitarity more rigorously. 
We write $a(\epsilon)$, $b(\epsilon)$ as 
\begin{align}
    a(\epsilon) = \sum_{i = 0} a_i \epsilon^{i}; \;\; b(\epsilon) = \sum_{i = 0} b_i \epsilon^{i},
\end{align}
such that
\begin{align}
    |a(\epsilon)|^2 &= \frac{1}{2}\sum_{i = 0} \sum_{j = 0}^{i} \bigg(a_ja^*_{i-j} + a^*_{i-j}a_j\bigg)\epsilon^{i};  \\ |b(\epsilon)|^2 &= \frac{1}{2}\sum_{i = 0} \sum_{j = 0}^{i} \bigg(b_j b^*_{i-j} + b^*_{i-j} b_j\bigg)\epsilon^{i}.
\end{align}

Thus, $|a(\epsilon)|^2 + |b(\epsilon)|^2 = 1$ implies that
\begin{align}
|a_0|^2 + |b_0|^2 &= 1;\\
\sum_{j = 0}^{i} \bigg(a_ja^*_{i-j} + a^*_{i-j}a_j + b_j b^*_{i-j} + b^*_{i-j} b_j\bigg) &= 0\text{ for $i > 0$}.
\end{align}
We rewrite these conditions as
\begin{align}
    \re[a_0]^2 +  \im[a_0]^2 +  \re[b_0]^2 +  \im[b_0]^2 &= 1;\\
    \sum_{j = 0}^{i} \bigg(2\re(a_j)\re(a_{i-j}) + 2\im(a_{i-j})\im(a_j)
    + 2\re(b_j)\re(b_{i-j}) + 2\im(b_{i-j})\im(b_j)\bigg) &= 0\text{ for $i > 0$}. 
\end{align}

Note that for $i = 0$, if three of the four $\re(a_0), \im(a_0)$, $\re(b_0)$, $\im(b_0)$ are equivalent to their target values, then the last constraint will be automatically satisfied by unitarity. Through an inductive argument, if $\re(a_i), \im(a_i), \re(b_i), \im(b_i)$ are equal to their target values for $i=0$, and zero for all $ 0<i < n$, and if three of the four $\re(a_n), \im(a_n)$, $\re(b_n)$, $\im(b_n)$ are equal to zero, then unitarity guarantees that the last will also be zero. Thus, we only have 3 effective equations for each $n$, leading to $3(n+1)$ constraints.

\section{Proof of Theorem 1}\label{app:thmtrs}
Here, we prove Theorem 1 in the main text, copied below for convenience: 
\begin{theorem}\label{thm:trs}
Under amplitude error in the global drive,  a given time-reversal symmetric composite sequence 
\begin{align*}
    [{\theta_1}]^\epsilon_{\phi_1}... [{\theta_k}]^\epsilon_{\phi_k}[{\theta_k}]^\epsilon_{\phi_k}... [{\theta_1}]^\epsilon_{\phi_1} = [{\theta}]_{\phi} + O(\epsilon^{n+1})
\end{align*}
implies the class of composite sequences 
\begin{align*}
Z(\alpha) \; [{\theta_1}]^\epsilon_{\phi_1}... [{\theta_k}]^\epsilon_{\phi_k} \; Z(\beta) \; [{\theta_k}]^\epsilon_{\phi_k}...  [{\theta_1}]^\epsilon_{\phi_1} \; Z(\gamma)= Z(\alpha)\; [\theta/2]_{\phi} \; Z(\beta) \; [\theta/2]_{\phi} \; Z(\gamma) + O(\epsilon^{n+1}).
\end{align*}
\end{theorem}

This Theorem implies that, given one composite sequence, one can readily produce a class of composite sequences only by either inserting $Z$ rotations or shifting the phase angles of each pulse, both of which we assume can be done with high precision.

\textbf{Proof: } First, we note that as amplitude error manifests as $\theta\to\theta(1+\epsilon)$ in the global pulses, we can use the identity 
\begin{align}
    Z(\phi) [\theta]^\epsilon_{0} Z(-\phi) = [\theta]^\epsilon_{\phi}.
\end{align}
Thus, WLOG, it suffices to show a special case when $\phi = 0$. That is, 
\begin{align}
  [{\theta_1}]^\epsilon_{\phi_1}... [{\theta_k}]^\epsilon_{\phi_k}[{\theta_k}]^\epsilon_{\phi_k}... [{\theta_1}]^\epsilon_{\phi_1} &= [{\theta}]_{0} + O(\epsilon^{n+1}) \label{eq:trs0}\\
\implies Z(\alpha) \; [{\theta_1}]^\epsilon_{\phi_1}... [{\theta_k}]^\epsilon_{\phi_k} \; Z(\beta) \; [{\theta_k}]^\epsilon_{\phi_k}...  [{\theta_1}]^\epsilon_{\phi_1} \; Z(\gamma) &= Z(\alpha)\; [\theta/2]_{0} \; Z(\beta) \; [\theta/2]_{0} \; Z(\gamma) + O(\epsilon^{n+1}).\label{eq:trs1}
\end{align}
Theorem \ref{thm:trs} then follows by a rotation of our coordinate system $\alpha\to \alpha + \phi$, $\gamma \to \alpha - \phi$, and redefining $\phi_i \to \phi_i - \phi$. Now, assuming Eq. \ref{eq:trs0}, we can write
\begin{align}
    [{\theta_1}]^\epsilon_{\phi_1}... [{\theta_k}]^\epsilon_{\phi_k}Z(\beta)[{\theta_k}]^\epsilon_{\phi_k}... [{\theta_1}]^\epsilon_{\phi_1} &=  [{\theta_1}]^\epsilon_{\phi_1}... [{\theta_k}]^\epsilon_{\phi_k}\big(\cos(\frac{\beta}{2}) - i\sin(\frac{\beta}{2})\hat \sigma_z\big)[{\theta_k}]^\epsilon_{\phi_k}... [{\theta_1}]^\epsilon_{\phi_1} \\
    &= \cos(\frac{\beta}{2})\underbrace{[{\theta_1}]^\epsilon_{\phi_1}... [{\theta_k}]^\epsilon_{\phi_k}[{\theta_k}]^\epsilon_{\phi_k}... [{\theta_1}]^\epsilon_{\phi_1} }_{[\theta]_{0} + O(\epsilon^{n+1})}  \notag\\
    &- i\sin(\frac{\beta}{2})\hat \sigma_z \underbrace{[{\theta_1}]^\epsilon_{\phi_1+\pi}... [{\theta_k}]^\epsilon_{\phi_k+\pi}[{\theta_k}]^\epsilon_{\phi_k}... [{\theta_1}]^\epsilon_{\phi_1} }_{=\id},
\end{align}
where we have used $[\theta]^\epsilon_\phi \hat \sigma_{z} = \hat \sigma_{z} [\theta]^\epsilon_{\phi+\pi}$ and that $[\theta_k]^{\epsilon}_{\phi_k+\pi}[\theta_k]^{\epsilon}_{\phi_k} = \id$.
Similarly, 
\begin{align*}
    [\theta/2]_{0}Z(\beta)[\theta/2]_{0} = \cos(\frac{\beta}{2})[{\theta}]_{0} - i\sin(\frac{\beta}{2}) \hat \sigma_{z}.
\end{align*}
By equating two results, we obtain
\begin{align}\label{eq:trs2}
  [{\theta_1}]^\epsilon_{\phi_1}... [{\theta_k}]^\epsilon_{\phi_k}Z(\beta)[{\theta_k}]^\epsilon_{\phi_k}... [{\theta_1}]^\epsilon_{\phi_1}
    =\bigg[{\frac{\theta}{2}}\bigg]_{0}Z(\beta)\bigg[{\frac{\theta}{2}}\bigg]_{0} + O(\epsilon^{n+1}).
\end{align}
Multiplying Eq. \ref{eq:trs2} on the left by $Z(\alpha)$ and on the right by $Z(\gamma)$  yields Eq. \ref{eq:trs1} as desired. $\square$

\section{Robust pulse sequences}
\begin{figure}
    \centering
    \includegraphics{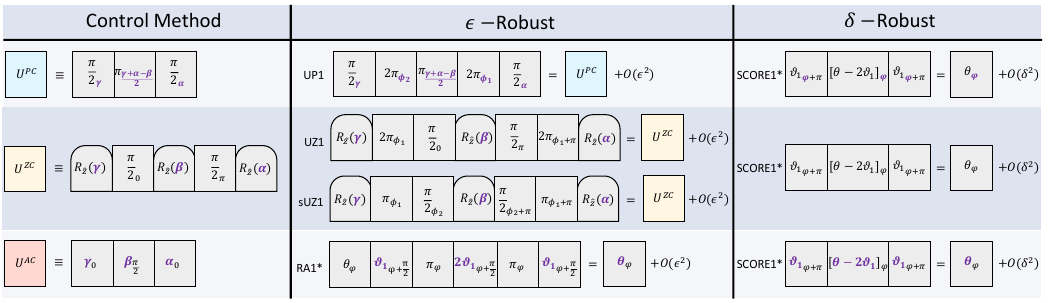}
    \caption{The first column lists the basic sequences for realizing parallel, site-dependent unitaries under each of PC, ZC, and AC, and the next columns provide their robust implementations. Solutions for the robust pulse parameters are given in the main text. Locally tunable parameters are indicated in purple for each case.  For ZC, robust sequences are given for the regime $\epsilon, \delta \gg \epsilon_s$, where local $Z$ rotations are approximately perfect. Robust sequences for when $\epsilon, \delta \approx \epsilon_s$ are found numerically (see Sec. \ref{app:num}). For AC, shorter $\epsilon$-robust sequences for arbitrary unitaries can also be found numerically. For all cases, doubly-robust sequences are obtained by concatenation (see Sec. \ref{app:concat}) or numerically. * indicates that the corresponding robust unitaries are achieved by directly replacing each pulse in the corresponding basic sequence with the robust pulse.}
    \label{fig:pultab}
\end{figure}

Here, we provide more details on the robust pulse sequences introduced in the main text; a concise diagrammatic table of these sequences is shown in Fig. \ref{fig:pultab}. Specifically, we present the general closed-form solution to our $\epsilon$-robust pulse primitive UP1 for arbitrary unitaries under phase control. We further discuss the derivation of our $\epsilon$-robust composite pulse RA1 for arbitrary rotations under amplitude control. Note that rotations are pulses $\theta_\phi$, while unitaries refer to any SU(2) operation. 
We then give higher-order extensions of the first-order sequences in the main text. Finally, we demonstrate the concatenation procedure explicitly to construct one of our $\epsilon, \delta$-robust pulse sequences.

For the discussion in this section, it is convenient to first recall our basic sequences for achieving parallel, arbitrary unitaries under each of phase control (PC), Z control (ZC), or amplitude control (AC) (Fig. \ref{fig:pultab}):
\begin{align}
    U^{PC}(\alpha_i, \beta_i, \gamma_i)&\equiv [\pi/2]_{\alpha_i} [\pi]_{\frac{\gamma_i +\alpha_i - \beta_i}{2}}[\pi/2]_{\gamma_i}, \label{eq:arbupcap} \\
    U^{ZC}(\alpha_i, \beta_i, \gamma_i) &\equiv Z(\alpha_i) [\pi/2]_\pi Z(\beta_i) [\pi/2]_0  Z(\gamma_i), \label{eq:arbuzcap} \\
    U^{AC}(\alpha_i, \beta_i, \gamma_i) &\equiv {[\alpha_i]}_0{[\beta_i]}_{\frac{\pi}{2}}{[\gamma_i]}_0,\label{eq:arbuacap}
\end{align}
where 
\begin{align}
    U^{PC}(\alpha_i, \beta_i, \gamma_i) = Z(\alpha_i)  Y(\beta_i)Z(2\pi - \gamma_i);\;\; U^{ZC}(\alpha_i, \beta_i, \gamma_i) = Z(\alpha_i) Y(\beta_i) Z(\gamma_i);\;\; U^{AC}(\alpha_i, \beta_i, \gamma_i) = X(\alpha_i) Y(\beta_i)X(\gamma_i).
    \notag
\end{align} Tensor products over all qubits $i$ are omitted for clarity.

\subsection{UP1 closed-form solution}
Recall the form of UP1 (Eq. \ref{eq:up1}
in the main text), 
\begin{align}\label{eq:up1ap}
    \text{UP1} (\alpha, \beta, \gamma) \equiv [\pi/2]_{\alpha} [2\pi]_{\phi_1}  [\pi]_{\frac{\gamma + \alpha-\beta}{2}} [2\pi]_{\phi_{2}}[\pi/2]_{\gamma}.
\end{align}
We want to solve for $\phi_1$, $\phi_2$ such that $\text{UP1} (\alpha, \beta, \gamma;\epsilon) = U^{PC}(\alpha, \beta, \gamma) + O(\epsilon^{2})$. Thus, enforcing UP1$(\alpha, \beta, \gamma; \epsilon=0)$ = $ U^{PC}(\alpha, \beta, \gamma)$ and $\frac{d}{d\epsilon}\text{UP1} (\alpha, \beta, \gamma;\epsilon)|_{\epsilon =0} = 0$ straightfowardly yields the following system of equations for $\phi_1, \phi_2$:
\begin{equation}
\begin{aligned}
    \sin(\tilde \phi_1)+ \sin(\tilde \phi_2) &= -\frac{1}{2}\sin(\frac{\alpha-\gamma}{2}) \\
    \cos(\tilde \phi_1)+ \cos(\tilde \phi_2) &= -\frac{1}{2}\bigg(\cos(\frac{\beta}{2}) + \cos(\frac{\alpha-\gamma}{2})\bigg)\\
    \tilde \phi_1 &= - \phi_1 -\frac{\beta}{2} + \alpha\\
    \tilde \phi_2 &=  \phi_2 +\frac{\beta}{2} -\gamma.
\end{aligned}\label{eq:up1eqs}
\end{equation}
We can find $\phi_1, \phi_2$ in closed form by making a Weirstrass substitution, $\tilde t_i = \tan(\frac{\tilde \phi_i}{2})$. This then turns Eq. \ref{eq:up1eqs} into a polynomial equation.
Letting $a \equiv \sin(\frac{\alpha-\gamma}{2})$, $b \equiv \cos(\frac{\beta}{2})$, and $c \equiv \cos(\frac{\alpha-\gamma}{2})$, we obtain the following solutions:
\begin{equation}
    \begin{aligned}
\tilde t_1 &= -\frac{4a +\sqrt{(a^2 + (b+c)^2)(16-a^2-(b+c)^2)}}{a^2 + (b+c)(b+c-4)}\\
\tilde t_2 &= \frac{-4a +\sqrt{(a^2 + (b+c)^2)(16-a^2-(b+c)^2)}}{a^2 + (b+c)(b+c-4)}.
\end{aligned}
\end{equation}

The same solution with $\tilde t_1 \leftrightarrow \tilde t_2$ also works due to the symmetry of $\tilde \phi_1$ and $\tilde \phi_2$. Which one performs best, i.e. has lowest next-leading order error, depends on the specific parameter values. Moreover, we note that a solution always exists if $\alpha \neq \gamma$, since the term under the square root is always positive and the denominator is non-zero. For the case when $\alpha = \gamma$, we have the simpler solution
\begin{align}
    \phi_1 = \phi_2 = -\frac{\beta}{2} +\gamma \pm \cos^{-1}(-\frac{1}{2}\cos(\frac{\beta}{4})^2), 
\end{align}
where the sign of the $\cos^{-1}$ term can also be chosen by minimizing the next-leading order error. 

\subsection{RA1 derivation}\label{app:ra1}
For a composite pulse to be compatible with parallelization under AC, only the pulse areas of the sequence can vary as a function of the target operation; the phases must stay constant. As the opposite is generally true for $\epsilon$-robust sequences---for example, BB1 and SCROFULOUS both involve pulses with fixed pulse areas and varying phases---designing an $\epsilon$-robust composite pulse for AC may initally seem difficult to approach. We tackle this by first relating our problem to traditional composite pulses with tunable phases in the $x-y$ plane.

A rotation about some axis with phase $\phi$ in the $x-y$ plane can be equivalently written as
\begin{align}
     [\theta]_{\phi} = Z(\phi) [\theta]_{0}Z(-\phi).
\end{align}
Composite pulses can therefore be formulated as alternating rotations about the $\hat x$ and $\hat z$ axes; indeed, this is the basis of quantum signal processing \cite{Low2016}. 

As the form of the Hamiltonian limits pulses to SU(2) rotations about axes in the $x-y$ plane, we do not have access to individual $Z$ rotations under AC. However, by a rotation of our coordinate system, robust composite operations should similarly be realizable for alternating rotations about the $\hat x$ and $\hat y$ axes. For an $\epsilon$-robust rotation $[\theta]_0$ under AC, we can thus make the initial pulse primitive ansatz
\begin{align}\label{eq:2piansatz}
    Y(\vartheta_1; \epsilon) [2 \pi]^\epsilon_0 Y(-2\vartheta_1; \epsilon)  [2\pi]^\epsilon_0  Y(\vartheta_1; \epsilon) [\theta]^\epsilon_0 = [\theta]_0 + O(\epsilon^2),
\end{align}
where we note that $Y(\vartheta_1; \epsilon) [2 \pi]_0^{\epsilon} Y(-\vartheta_1; \epsilon) = [{\vartheta_1}]_{\frac{\pi}{2}}^{\epsilon} [2\pi]_0^{\epsilon}{[-\vartheta_1]}_{\frac{\pi}{2}}^{\epsilon}$ is just a faulty $2\pi(1+\epsilon)$ rotation about the axis $\cos(\vartheta_1(1+\epsilon))\hat x + \sin(\vartheta_1 (1+\epsilon))\hat z$. Adding such $2\pi$ pulses in  Eq. \ref{eq:2piansatz} thus hearkens back to our technique for constructing $\epsilon$-robust pulse primitives using $2\pi$ pulses. 

We can indeed find a valid solution $\vartheta_1$ which satisfies Eq. \ref{eq:2piansatz}, and this inspires us to further improve the ansatz by replacing $[2\pi]_0$ with $[\pi]_0$ in Eq. \ref{eq:2piansatz}. In the presence of error, we can write $[\pi]_0^{\epsilon} = [\pi]_0 [\pi \epsilon]_0$. From here, the two perfect $[\pi]_0$ pulses can be commuted out using the identity
\begin{align}
    [\theta]_\beta [\pi]_\alpha = [\pi]_\alpha [\theta]_{2\alpha-\beta},
\end{align}
a common technique in NMR analysis \cite{SUTER1987509, Jones2013}.
This gives a shorter ansatz
\begin{align}\label{eq:ra1}
 \text{RA1}(\theta; \epsilon) \equiv [{\vartheta_1}]^\epsilon_{\frac{\pi}{2}}  [\pi]^\epsilon_0 [{2\vartheta_1}]^\epsilon_{\frac{\pi}{2}} [\pi]^\epsilon_0  [{\vartheta_1}]_{\frac{\pi}{2}}^\epsilon [\theta]_0^\epsilon = [\theta]_0 + O(\epsilon^2)
\end{align}
with solution
\begin{align}\label{eq:ra1sol}
    \vartheta_1 = \cos^{-1}(-\frac{\theta}{2\pi}),
\end{align}
which is indeed our RA1 from the main text. Notably, the first part of RA1 is already robust to $\delta$ error to second order, 
\begin{align}
\underbrace{[{\vartheta_1}]^\delta_{\frac{\pi}{2}}  [\pi]^\delta_0 [{2\vartheta_1}]^\delta_{\frac{\pi}{2}} [\pi]^\delta_0  [{\vartheta_1}]_{\frac{\pi}{2}}^\delta }_{O(\delta^2)} [\theta]_0^\delta. 
\end{align} 
Thus, RA1 is also amenable to concatenation; by simply replacing the first $[\theta]_0$ with a composite SCORE1 pulse, the entire sequence will compensate against both $\epsilon$ and $\delta$. 

RA1 can also be related to $\epsilon_s$-compensation in the regime $\epsilon_s \gg \epsilon, \delta$ under ZC. Using the identity $Y(-\frac{\pi}{2})[\theta]_0 Y(\frac{\pi}{2}) = Z(\theta)$, a rotation of Eq. \ref{eq:ra1} yields an $\epsilon_s$-robust composite $Z$ pulse,
\begin{align}\label{eq:rz1}
    \text{RZ1}(\theta; \epsilon_s) &\equiv
     \big[\frac{\pi}{2}\big]_{\pi} Z{(\vartheta_1;\epsilon_s)} \big[\frac{\pi}{2}\big]_{0} Z(\pi;\epsilon_s) \big[\frac{\pi}{2}\big]_{\pi} Z({2\vartheta_1};\epsilon_s) \big[\frac{\pi}{2}\big]_{0} Z(\pi;\epsilon_s) \big[\frac{\pi}{2}\big]_{\pi}Z({\vartheta_1};\epsilon_s) \big[\frac{\pi}{2}\big]_{0} Z(\theta;\epsilon_s)  \\
    &= Z(\theta) + O(\epsilon_s^2). \notag
\end{align}

\subsection{Higher-order extensions}
Here, we give extensions of the pulse sequences introduced in the main text for higher-order compensation. 

For phase control (PC), our $\epsilon$-robust UP1 (Eq. \ref{eq:up1} in the main text) can be written generally as
\begin{align}
   \text{UP}\textit{n}(\alpha\,\beta, \gamma) \equiv  [\pi/2]_{\alpha} [2\pi]_{\phi_1} ... [2\pi]_{\phi_{n}} [\pi]_{\frac{\gamma + \alpha-\beta}{2}} [2\pi]_{\phi_{n+1}}...[2\pi]_{\phi_{2n}} [\pi/2]_{\gamma} \label{eq:upn}
\end{align}
with $\text{UP}\textit{n}(\alpha\,\beta, \gamma; \epsilon) = U^{PC}(\alpha, \beta, \gamma) + O(\epsilon^{n+1})$ for the appropriate $\vec{\phi}$. Valid solutions $\vec{\phi}$, if they exist, depend on $\alpha, \beta, \gamma$ and must be found numerically in general. We include $\vec{\phi}$ up to $n=2$ in our numerical testing in Section \ref{app:numtest}. 

For Z control (ZC), our $\epsilon$-robust UZ1 (Eq. \ref{eq:uz1} in the main text) is constructed generally as
\begin{align}
    \text{UZ}\textit{n}(\alpha\,\beta, \gamma) &\equiv Z(\alpha) [2\pi]_{\phi_1 + \pi} ... [2\pi]_{\phi_n + \pi} [\pi/2]_\pi Z(\beta)[\pi/2]_0[2\pi]_{\phi_{n}}...[2\pi]_{\phi_{1}}Z(\gamma)  \label{eq:uzn}
\end{align}
with $\text{UZ}\textit{n}(\alpha\,\beta, \gamma;\epsilon) = U^{ZC}(\alpha, \beta, \gamma) + O(\epsilon^{n+1})$ for the appropriate $\vec{\phi}$. Notably, due to our Theorem \ref{thm:trs}, $\vec{\phi}$ here does not depend on $\alpha, \beta, \gamma$, and we list $\vec{\phi}$ up to $n=5$ in Table \ref{tab:uzn}. Similarly, our $\epsilon$-robust sUZ1 (Eq. \ref{eq:suz1} in the main text) is written generally as 
\begin{align}
    \text{sUZ}\textit{n}(\alpha\,\beta, \gamma) &\equiv Z(\alpha) [\pi]_{\phi_1 + \pi} ... [\pi]_{\phi_n + \pi} [\pi/2]_{\phi_{n+1} + \pi} Z(\beta) [\pi/2]_{\phi_{n+1}}[\pi]_{\phi_{n}} ... [\pi]_{\phi_{1}}Z(\gamma)  \label{eq:suzn}
\end{align}
with $\text{sUZ}\textit{n}(\alpha\,\beta, \gamma;\epsilon) = U^{ZC}(\alpha, \beta, \gamma) + O(\epsilon^{n+1})$ for the appropriate $\vec{\phi}$. Again,  $\vec{\phi}$ is universal at each order for all $\alpha, \beta, \gamma$, and we give $\vec{\phi}$ up to $n=5$ in Table \ref{tab:suzn}. 

For AC, a second-order version of RA1 can constructed following a similar procedure as Section \ref{app:ra1},
\begin{align}
     \text{RA2}(\theta; \epsilon)&\equiv[{\vartheta_1}]_{\frac{\pi}{2}}^\epsilon  [\pi]_0^\epsilon {[\vartheta_1-\vartheta_2]}_{\frac{\pi}{2}}^\epsilon [\pi]_0^\epsilon   {[\vartheta_2-\vartheta_3]}_{\frac{\pi}{2}}^\epsilon [\pi]_0^\epsilon [{2\vartheta_3}]_{\frac{\pi}{2}}^\epsilon [\pi]_0^\epsilon  {[\vartheta_2-\vartheta_3]}_{\frac{\pi}{2}}^\epsilon [\pi]_0^\epsilon {[\vartheta_1-\vartheta_2]}_{\frac{\pi}{2}}^\epsilon [\pi]_0^\epsilon [{\vartheta_1}]_{\frac{\pi}{2}}^\epsilon [{\theta}]_0^\epsilon    \label{eq:ra2}\\  &= \theta_0+ O(\epsilon^3) , \notag
\end{align}
where $\vec{\vartheta}$ satisfies the following system of equations,
\begin{align}
    \theta + 2\pi(\cos(\vartheta_1) +\cos(\vartheta_2) +\cos(\vartheta_3)) &= 0\\
    \vartheta_1 \sin(\vartheta_1) + \vartheta_2 \sin(\vartheta_2) + \vartheta_3 \sin(\vartheta_3) &= 0\\
    \cos(\vartheta_1) \sin(\vartheta_1) + \cos(\vartheta_2)\big(2\sin(\vartheta_1)+ \sin(\vartheta_2)\big)
    + \cos(\vartheta_3)\big(2(\sin(\vartheta_1)+ \sin(\vartheta_2)) + \sin(\vartheta_3)\big) &= 0.
\end{align}
Some solutions for various $\vec{\vartheta}$ are given in Table \ref{tab:ra2}. However, due to its length, we do not expect that this sequence will be practical to implement in the near term. 

Our $\delta$-robust SCORE1 (Eq. \ref{eq:score1} in the main text) is constructed generally as
\begin{align}
\text{SCORE}\textit{n}(\theta, \phi) &\equiv [\vartheta_1]_{\phi}...[\vartheta_n]_{\phi + \pi}... [\Theta]_{\phi}[\vartheta_n]_{\phi+\pi} ... [\vartheta_1]_\phi  .\label{eq:scoren}
\end{align}
Here, $\vec{\vartheta}$ is appropriately selected such that $\text{SCORE}\textit{n}(\theta, \phi; \delta) = [\theta]_\phi + O(\delta^{n+1})$, $\Theta =\theta + \sum_{k = 1}^{n}(-1)^{n-k}2\vartheta_k$, and the global phases alternate between $\phi$ and $\phi+\pi$. $\vec{\vartheta}$ depends on $\theta$, and solutions for various $\theta$ are listed up to $n = 4$ in Table \ref{tab:scoren}. Intuitively, SCORE\textit{n} is easily seen to equate to $[\theta]_\phi$ at $\delta\,=\,0$; when $\delta\,\neq\,0$, errors accumulated in additional $\vartheta_i$ rotations cancel with error in the central pulse. This type of structure is coined switchback in Ref. \cite{KukitaPSJ2022}, hence inspiring our name Switchback for Compensating Off Resonance Error (SCORE).

\subsection{Concatenation procedure}\label{app:concat}
Here, we describe the concatenation procedure \cite{Bando2013} for constructing $\epsilon, \delta$-robust pulse sequences from our $\epsilon$- and $\delta$-robust pulse primitives. As described in the main text, concatenation involves replacing each pulse in an $\epsilon$-robust pulse sequence with a $\delta$-robust composite pulse, or vice versa, in order to form an overall sequence which simultaneously compensates against both errors. 
In this work, we choose $\epsilon$-robust outer sequences and $\delta$-robust inner sequences. An appropriate $\delta$-robust inner composite pulse $\mathcal{C_\delta}$ must be residual error preserving (REP) in $\epsilon$, defined by enforcing that the first-order error of $\mathcal{C_\delta}$ and an elementary pulse $[\theta]^\epsilon_{\phi}$ are equivalent:  $\frac{d}{d\epsilon}\mathcal{C_\delta(\epsilon)}|_{\epsilon=0}\,=\,\frac{d}{d\epsilon}[\theta]^\epsilon_\phi|_{\epsilon=0}$. 

Specifically, we use SCORE\textit{n} as our inner $\delta$-robust composite pulse for all individual control schemes,
\begin{align}
\text{SCORE}\textit{n}(\theta, \phi; \delta) \equiv [\vartheta_1]^\delta_{\phi}...[\vartheta_n]^\delta_{\phi+\pi} [\Theta]^\delta_{\phi}[\vartheta_n]^\delta_{\phi+\pi} ... [\vartheta_1]^\delta_\phi =[\theta]_\phi + O(\delta^{n+1}),
\end{align}
with $\Theta =\theta + \sum_{k = 1}^{n}(-1)^{n-k}2\vartheta_k$. SCORE1 is residual-error preserving in $\epsilon$ ($\epsilon$-REP), namely: $\frac{d}{d\epsilon}\text{SCORE1}(\theta, \phi; \epsilon)|_{\epsilon=0}$ = $\frac{d}{d\epsilon}[\theta]^\epsilon_\phi|_{\epsilon=0}$. This ensures that replacing each $[\theta]_\phi$ in an $\epsilon$-robust primitive with a SCORE1 pulse will not spoil the outer sequence's $\epsilon$-compensating structure. Moreover, we note that as $[2\pi]^\delta_\phi = [2\pi]_\phi + O(\delta^2)$, $2\pi$ pulses in any outer sequence (e.g. UP1 or UZ1 in the main text) do not need to be replaced.

We now explicitly show how our SCORE1inUP1 is obtained by concatenating SCORE1 into UP1. Recall the form of UP\textit{n},
\begin{align}
    \text{UP}\textit{n}(\alpha, \beta, \gamma; \epsilon) \equiv
    [\pi/2]^\epsilon_{\alpha} [2\pi]^\epsilon_{\phi_1} ... [2\pi]^\epsilon_{\phi_{n}} [\pi]^\epsilon_{\frac{\gamma + \alpha-\beta}{2}} [2\pi]^\epsilon_{\phi_{n+1}}...  [2\pi]^\epsilon_{\phi_{2n}} [\pi/2]^\epsilon_{\gamma} = U^{PC}(\alpha, \beta,\gamma) + O(\epsilon^{n+1}),
\end{align}
where $U^{PC}(\alpha, \beta,\gamma)\equiv Z(\alpha)  Y(\beta)Z(2\pi - \gamma)$ as defined in the main text. Conducting the replacement procedure as outlined above, we find
\begin{align}
    \text{SCORE1inUP1}(\alpha, \beta, \gamma; \epsilon, \delta) &\equiv \text{SCORE1}(\pi/2, \alpha;\epsilon, \delta) [2\pi]^{\epsilon, \delta}_{\phi_1} \text{SCORE1}(\pi, \frac{\gamma + \alpha-\beta}{2}; \epsilon, \delta)[2\pi]^{\epsilon, \delta}_{\phi_{2}}\text{SCORE1}(\pi/2, \gamma;\epsilon, \delta) \notag \\
    &= U^{PC}(\alpha, \beta,\gamma) + O(\epsilon^2, \delta^2, \epsilon\delta),
\end{align}
thus yielding a composite pulse robust against both $\epsilon$ and $\delta$ error which achieves arbitrary, parallel unitaries under PC.

\section{Numerical search method}\label{app:num}
Here, we elaborate on our numerical method to find robust pulse sequences for arbitrary unitaries under both AC and ZC.

For AC, we consider generic pulse sequences of the form
\begin{align}\label{eq:numpulgenac}
[{\theta_1}]^{\epsilon, \delta}_{\phi_1} \cdots  [{\theta_k}]^{\epsilon, \delta}_{\phi_k} = \mathcal{U}(\epsilon, \delta)
\end{align}
where we express a SU(2) pulse under both amplitude ($\epsilon$) error and off-resonance ($\delta$) error as $[{\theta_\phi}]^{\epsilon, \delta} \equiv \exp(-i\frac{\theta(1+\epsilon)}{2}(\hat \sigma_{\phi} + \delta \hat \sigma_z))$. 

For ZC, we consider sequences of alternating rotations about $\hat z$ and in the $x-y$ plane,
\begin{align}\label{eq:numpulgenzc}
Z(\theta_1; \epsilon_s) \; [{\theta_2}]_{\phi_2}^{\epsilon, \delta, \epsilon_s} \; Z(\theta_3; \epsilon_s) \cdots  [{\theta_k}]_{\phi_k}^{\epsilon, \delta, \epsilon_s} = \mathcal{U}(\epsilon, \delta, \epsilon_s).
\end{align}
Stark error ($\epsilon_s$ error) manifests as $Z(\theta_1; \epsilon_s) \equiv Z(\theta(1+\epsilon_s))$, and we define our $x-y$ rotations $[{\theta_\phi}]^{\epsilon, \delta, \epsilon_s}$ as a local $Z$ conjugated by global pulses, 
\begin{align}
    [{\theta_\phi}]^{\epsilon, \delta, \epsilon_s} \equiv [{\pi/2}]_{\phi-\frac{\pi}{2}}^{\epsilon, \delta} \; Z(\theta; \epsilon_s)\; [{\pi/2}]_{\phi+\frac{\pi}{2}}^{\epsilon, \delta}.
\end{align} Although more generic sequence forms exist, we expect that alternating rotations about perpendicular axes can generate the appropriate solutions---indeed, traditional composite pulses can be formulated as alternating $X$ and $Z$ rotations.

For both AC and ZC, given a target $\mathcal{U}$ and fixed global phases $\vec{\phi}$ we numerically find the pulse areas $\vec{\theta}$ leading to $\mathcal{U}(\vec{\zeta})$ robust against multiple errors $\vec{\zeta}$. This is conducted with the procedure given below, similar to Ref. \cite{Le2023}. Our ansatzes for the global $\vec{\phi}$, as well as examples of specific $\vec{\theta}$ obtained for different unitaries, are listed in Tables ~\ref{tab:numpulzc}-\ref{tab:numpulac}. 

 \begin{enumerate}
     \item Starting from an appropriate initial condition $\vec{\theta}_0$, fixed global phases $\vec{\phi}$, target $\mathcal{U}$, and region over which to achieve robustness $\vec{\zeta} \in \mathrm{Z} \equiv (\epsilon \in [-\epsilon_0, \epsilon_0], \delta \in[-\delta_0, \delta_0], \epsilon_s \in [-{\epsilon_s}_0, {\epsilon_s}_0])$, we first maximize the fidelity over $\vec{\theta}$ 
     at both the center and extreme points $\vec{\zeta}_i$ of the hypercube  $\mathrm{Z}$ with gradient ascent. The function to be maximized is 
     \begin{align}
         \mathcal{F}(\vec{0})+ \frac{1}{N}\sum_{\vec{\zeta}_i} \mathcal{F}(\vec{\zeta}_i), \label{eq:fiderrap}\\
         \mathcal{F}(\vec{\zeta}) = \frac{1}{2}|\Tr(\mathcal{U}(\vec{\zeta})^\dagger \mathcal{U})|,\label{eq:fidap}
     \end{align}
     where $N$ is the number of extreme points $\vec{\zeta}_i$. 
     After this step, we are only ensured high $\mathcal{F}$ at $\vec{\zeta} = \vec{0}$ or  $\vec{\zeta}_i$; points within the hypercube may still have low fidelity.
\item Starting from the result from Step 1, we again maximize the fidelity $\mathcal{F}(\vec{0})$ over $\vec{\theta}$. As we use gradient ascent, the Jacobian is zero and the Hessian is negative semi-definite. Thus, the fidelity $\mathcal{F}$ can be approximated as a paraboloid over $\mathrm{Z}$ with global maximum at $\vec{\zeta} = \vec{0}$ and minimum at the extreme points. As this step is generally a perturbation of the solution from Step 1, the fidelity at these extreme points, or global minima, are still high; this is confirmed in our numerical testing. Moreover, as the final $\vec{\theta}$ maximizes $\mathcal{F}(\vec{0})$, we ensure that the composite pulse $\mathcal{U}(\vec{\zeta})$ does indeed precisely achieve the target $\mathcal{U}$ at zero error.
 \end{enumerate}
In general, this algorithm is repeated for a large number of initial conditions $\vec{\theta_0}$, and the solution with the best performance, i.e. has the highest Eq. \ref{eq:fiderrap}, is chosen. Computations are efficiently carried out by storing SU(2) rotations  $a\id - i (b \hat \sigma_x + c \hat \sigma_y + d \hat \sigma_z)$ as quaternions $(a, b, c, d)$.

For Eq. \ref{eq:numpulgenac}, the Jacobian $\frac{\partial}{\partial \theta_i} \mathcal{U}(\epsilon, \delta)$ can be simply expressed as
\begin{align}
    \frac{\partial}{\partial \theta_i} \mathcal{U}(\epsilon, \delta) &= \frac{\pi}{2\alpha}[{\theta_1}]_{\phi_1}^{\epsilon, \delta} \dots  {[\theta_i + \alpha]}_{\phi_i}^{\epsilon, \delta} \dots [{\theta_k}]_{\phi_k}^{\epsilon, \delta},\\
    \alpha &= \frac{\pi}{(1+\epsilon)\sqrt{1+\delta^2}}.
\end{align}
Similarly, for Eq. \ref{eq:numpulgenzc},
\begin{align}
    \frac{\partial}{\partial \theta_i} \mathcal{U}(\epsilon, \delta, \epsilon_s) &= \frac{\pi}{2\alpha_s} Z(\theta_1; \epsilon_s) \;  [{\theta_2}]_{\phi_2}^{\epsilon, \delta, \epsilon_s} \cdots Z(\theta_i + \alpha_s; \epsilon_s) \cdots  [{\theta_k}]_{\phi_k}^{\epsilon, \delta, \epsilon_s},\\
    \alpha_s &= \frac{\pi}{(1+\epsilon_s)}.
\end{align}
$\frac{\partial}{\partial \theta_i}\mathcal{F}(\vec{\zeta})$  then follows straightforwardly for both cases using Eq. \ref{eq:fidap}. All optimization is conducted using gradient ascent with the LBFGS Hessian approximation implemented in the Python method \texttt{scipy.optimize.minimize}, where we constrain $\theta_i \in [0, 3\pi]$. 

To find robust sequences $\mathcal{U}(\vec{\zeta})$, an alternative measure to optimize could be the minimization of
\begin{align}
    1 - \mathcal{F}(\vec{0}) + \frac{\partial}{\partial \vec{\zeta}}\mathcal{U}(\vec{\zeta}),
\end{align}
which directly enforces first-order error compensation \cite{Jandura2023}. However, we find that pulse sequences found using the method presented in this section perform better under numerical testing.

\section{Numerical testing}\label{app:numtest}
Here, we first show how we calculate an analytical expression for the gate fidelity $\mathcal{F}$ defined in the main text. For each of PC, ZC, and AC, we then list all of the composite pulse sequences tested, providing more exhaustive phase diagrams for different parameter regimes. Numerical parameters of the simulated pulse sequences---including our numerical search results for AC and ZC--- will also be provided.

\subsection{Calculating fidelity}
As discussed in the main text, we develop the fidelity measure
\begin{align}\label{eq:fidgateap}
    \mathcal{F} = \mathbb{E}_{\ket{\psi}\sim \text{Haar}} \big[ e^{-\gamma t} |\braket{\psi|\mathcal{U}^\dagger \mathcal{U}(\epsilon, \delta, \epsilon_s)|\psi}|^2 + (1- e^{-\gamma t})\Tr{(\frac{\id}{2}\mathcal{U}\ket{\psi}\bra{\psi}\mathcal{U}^\dagger)}\big]
\end{align} in order to assess the performance of a pulse sequence $\mathcal{U}(\epsilon, \delta, \epsilon_s)$.

In constructing $\mathcal{F}$, we consider the squared overlap between a target final state $\mathcal{U}\ket{\psi}$ and implemented final state $\mathcal{U}(\epsilon, \delta,\epsilon_s)\ket{\psi}$.  Moreover, we also account for the effect of Markovian decoherence $\gamma$, in which the state $\mathcal{U}(\epsilon, \delta,\epsilon_s)\ket{\psi}$ decoheres into the maximally mixed state $\frac{\id}{2}$ with probability $1-e^{-\gamma t}$. The initial state $\ket{\psi}$ is averaged over the Haar measure. Now, Eq. \ref{eq:fidgateap} simplifies to 
\begin{align}
    \mathcal{F} &= e^{-\gamma t}\mathbb{E}_{\ket{\psi}\sim \text{Haar}} \big[  |\braket{\psi|\mathcal{U}^\dagger \mathcal{U}(\epsilon, \delta, \epsilon_s)|\psi}|^2\big] + \frac{1- e^{-\gamma t}}{2}.
\end{align}
The average over the Haar measure can be performed by rewriting
\begin{eqnarray}
|\braket{\psi|\mathcal{U}^\dagger \mathcal{U}(\epsilon, \delta, \epsilon_s)|\psi}|^2 = \Tr\big[(\ketbra{\psi}{\psi}\otimes\ketbra{\psi}{\psi}) (\mathcal{U}^\dagger \mathcal{U}(\epsilon, \delta)\otimes \mathcal{U}(\epsilon, \delta)^\dagger\mathcal{U} )\big]
\end{eqnarray}
and using the fact that
\begin{align}
    \mathbb{E}_{\ket{\psi}\sim \text{Haar}} [\ketbra{\psi}{\psi}\otimes\ketbra{\psi}{\psi}] = \frac{\id + \text{SWAP}}{d(d+1)},
\end{align}
where $d$ is the dimension of $\ket{\psi}$, $\id$ is the identity over the doubled Hilbert space, and SWAP is the swap operator over the two identical Hilbert spaces. Following the above, we straightforwardly compute
\begin{align}
      \mathbb{E}_{\ket{\psi}\sim \text{Harr}} \big[  |\braket{\psi|\mathcal{U}^\dagger \mathcal{U}(\epsilon, \delta, \epsilon_s)|\psi}|^2\big] = \frac{|\Tr(\mathcal{U}^\dagger \mathcal{U}(\epsilon, \delta))|^2 + d}{d(d+1)},
\end{align}
where we have used the identity
\begin{align}
    \Tr(\text{SWAP} (A\otimes B)) = \Tr(AB). 
\end{align}
The full simplified Eq. \ref{eq:fidgateap} is then
\begin{align}
    \mathcal{F} = e^{-\gamma t}\frac{|\Tr(\mathcal{U}^\dagger \mathcal{U}(\epsilon, \delta))|^2 + 2}{6} + \frac{1- e^{-\gamma t}}{2},
\end{align}
where we have $d = 2$ as the dimension of our system.
\subsection{Results}
\begin{figure}
    \centering
    \includegraphics{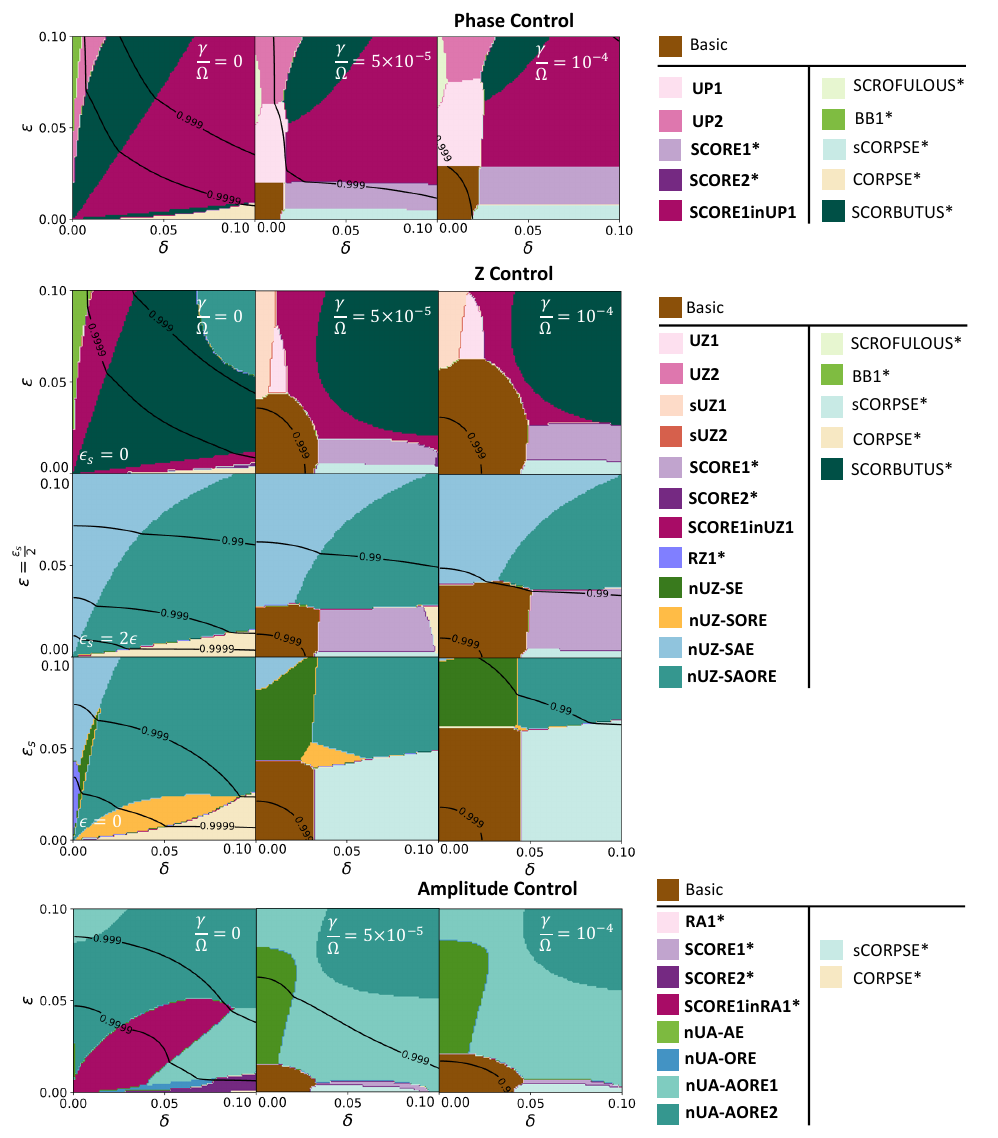}
    \caption{Phase diagrams of our robust pulse sequences under PC, ZC, and AC. Phases correspond to pulse sequences with $\max \mathcal{F}$ at each point $(\epsilon, \delta, \gamma)$, with $\mathcal{F}$ averaged over $\mathcal{U} \in [H, Z(\frac{\pi}{4}), X(\frac{\pi}{2}), Y(\frac{\pi}{2})]$. For AC and ZC, the total pulse area $T$ for each pulse sequence is taken to be the maximum of those out of the target $\mathcal{U}$, and local $Z$ gates are assumed to be five times slower than global pulses. Pulse sequences introduced in this work are bolded; an asterisk indicates direct replacement.}
    \label{fig:phcomp}
\end{figure}

Using our gate fidelity $\mathcal{F}$ (Eq. \ref{eq:fidgateap}), we analyze the performance of our robust sequences against existing composite sequences in the presence of error for PC, ZC, and AC.  A comprehensive list of all relevant sequences and their properties is given in Table \ref{tab:pulcomp}. To compare amongst these sequences, we compute $\mathcal{F}$ averaged across $\mathcal{U} \in [H, Z(\frac{\pi}{4}), X(\frac{\pi}{2}), Y(\frac{\pi}{2})]$. We then construct phase diagrams illustrating the sequence with highest $\mathcal{F}$ at every point $(\epsilon, \delta, \gamma)$ (Fig. \ref{fig:phcomp}). To build these phase diagrams, we also average $\mathcal{F}$ over positive and negative values of each quasistatic error; the nominal amount of error can then be interpreted as the standard deviation of fluctuations in the imperfect parameter.

\paragraph{PC sequences}
Under PC, our robust composite sequences for arbitrary, parallel unitaries $\bigotimes_i U^{PC}(\alpha_i, \beta_i, \gamma_i)$ are based upon the basic sequence Eq. \ref{eq:arbupcap}.
Specifically, our $\epsilon$-robust unitaries are given by UP\textit{n}, which consists of Eq. \ref{eq:arbupcap} with additional $2\pi$ pulses; our $\delta$-robust unitaries are achieved by direct replacement of each pulse in Eq. \ref{eq:arbupcap} with a SCORE\textit{n} composite pulse; and a $\epsilon, \delta$-compensating sequence SCORE1inUP1 is formed by concatenating SCORE1 into UP1. The phases of UP1, 2 used to implement each $\mathcal{U} \in [H, Z(\frac{\pi}{4}), X(\frac{\pi}{2}), Y(\frac{\pi}{2})]$ are listed in Table \ref{tab:up12}. Parameters for SCORE1, 2 can be found through Table \ref{tab:scoren}.

\paragraph{ZC sequences}
Under ZC, we realize arbitrary, parallel unitaries $\bigotimes_i U^{ZC}(\alpha_i, \beta_i, \gamma_i)$ through the basic sequence Eq. \ref{eq:arbuzcap}. Here, the local $Z$ gates may also contain a multiplicative error $\epsilon_s$. Thus, we focus on three regimes: $\epsilon_s \ll \epsilon, \delta$; $\epsilon_s \gg \epsilon, \delta$, and $\epsilon_s \approx \epsilon, \delta$. When $\epsilon_s \ll \epsilon, \delta$, $\epsilon$-robust unitaries are achieved by UZ\textit{n} and sUZ\textit{n}, while $\delta$-robust unitaries are implemented by directly replacing the $\frac{\pi}{2}$ pulses in Eq. \ref{eq:arbuzcap} with SCORE\textit{n} composite pulses. When $\epsilon_s \gg \epsilon, \delta$, we can form $\epsilon_s$-robust unitaries in analogy with AC by direct replacement with RZ1 (Eq. \ref{eq:rz1}). Moreover, as $\epsilon_s$ error becomes prominent, we also find shorter $\epsilon_s$-;  $\epsilon_s, \epsilon$-;  $\epsilon_s, \delta$-; and $\epsilon_s, \epsilon, \delta$-compensating sequences using our numerical method. The parameters of our numerical sequences for each of $\mathcal{U} \in [H, Z(\frac{\pi}{4}), X(\frac{\pi}{2}), Y(\frac{\pi}{2})]$  are given in Table \ref{tab:numpulzc}. SCORE1, 2 values are
in Table \ref{tab:scoren}, UZ1, 2 in Table \ref{tab:uzn}, and sUZ1, 2 in Table \ref{tab:suzn}, while RZ1 has closed form solution Eq. \ref{eq:ra1sol}. 

When computing $\mathcal{F}$, we assume that $Z$ operations are five times slower than global operations. Moreover, the total pulse area $T$ of each sequence also varies based on the target unitary $\mathcal{U}$. However, as we intend these operations to be performed in parallel, the actual time $t$ will be determined by the longest pulse area of the simultaneously performed sequences. Different $Z$ rotation amplitudes for a fixed time $t$ are then achieved by tuning the Stark shift $\frac{\Omega_i^2}{2\Delta}$ at each qubit $i$. For our numerical simulations, we thus take $T$ to be the maximum pulse area required of the target $\mathcal{U}$ for each pulse sequence. 

\paragraph{AC sequences}
Under AC, arbitrary, parallel unitaries $\bigotimes_i U(\alpha_i, \beta_i, \gamma_i)$ are implemented through the basic sequence Eq. \ref{eq:arbuacap}. While $\epsilon$-robust unitaries can be constructed by directly replacing each pulse in Eq. \ref{eq:arbuacap} with RA1 (Eq. \ref{eq:ra1}), and $\delta$-robust unitaries can be similarly built by direct replacement with SCORE\textit{n}, we additionally find $\epsilon$-, $\delta$-, and $\epsilon, \delta$-robust sequences to achieve $\mathcal{U} \in [H, Z(\frac{\pi}{4}), X(\frac{\pi}{2}), Y(\frac{\pi}{2})]$ under AC using our numerical method. For $\epsilon$- and $\epsilon, \delta$-compensation, our numerical solutions show significant improvement in both the number of pulses and total pulse area over our analytical sequence forms. The specific parameters of our numerical solutions are given in Table \ref{tab:numpulac}. Values for SCORE1, 2 can be found in Table \ref{tab:scoren}, and RA1 has closed form solution given by Eq. \ref{eq:ra1sol}. Like ZC, the total length in pulse area $T$ of each composite sequence varies depending on the target $\mathcal{U}$. We similarly take $T$ for each pulse sequence to be the longest of those among the target ensemble $\mathcal{U}$.
\begin{table}[h]
    \centering
    \begin{tabular}{c c c c c}
        
         Pulse & $k$ & $T$ & Rob. &$n$ \\ \hline \hline
          \multicolumn{5}{c}{Phase Control} \\ \hline \hline
         Basic (Eq. \ref{eq:arbupcap})  & 3 & 2 & --- & ---\\ \hline
         \multirow{2}{*}{\textbf{UP\textit{n}}} & 5 & 6 & \multirow{2}{*}{$\epsilon$ err.} & $\epsilon^1$\\
         & 7 & 10 &  & $\epsilon^2$\\ \hline
         SCROFULOUS (\cite{Cummins2003}) & 9 & 7.6 &$\epsilon$ err.  & $\epsilon^1$\\ \hline
        BB1 (\cite{Wimperis1994}) & 12 & 14 & $\epsilon$ err.  & $\epsilon^2$
         \\ \hline
         \multirow{2}{*}{\textbf{SCORE\textit{n}*}} & 9 & 8.4 &\multirow{2}{*}{$\delta$ err.}  & $\delta^1$\\
          & 15 & 12.3 &   & $\delta^2$
         \\ \hline
          CORPSE (\cite{Cummins2000}) & 9 & 12.4 & $\delta$ err. & $\delta^1$ \\ \hline
         Short (s) CORPSE* (\cite{Cummins2003}) & 9 & 6.4 & $\delta$ err. &  $\delta^1$ \\ \hline
          \textbf{SCORE1inUP1} & 11 & 12.4 &  $\epsilon$, $\delta$ err. & $\epsilon^1, \delta^1$ \\ \hline
         SCORBUTUS* (\cite{KukitaPSJ2022}) & 15 & 13.7 & $\epsilon$, $\delta$ err. & $\epsilon^1, \delta^1$ \\ \hline \hline
         \multicolumn{5}{c}{Z Control}\\ \hline \hline
         Basic (Eq. \ref{eq:arbuzcap})& 5 & 2.5 & --- & ---\\ \hline
         \multirow{2}{*}{\textbf{UZ\textit{n}}}  & 7 & 17.5 & \multirow{2}{*}{$\epsilon$ err.} & $\epsilon^1$\\
          & 9 & 21.5 &  & $\epsilon^2$\\\hline
          \multirow{2}{*}{\textbf{sUZ\textit{n}}}  & 7 & 15.5 & \multirow{2}{*}{$\epsilon$ err.} & $\epsilon^1$\\
          & 9 & 17.5 & & $\epsilon^2$\\\hline
         SCROFULOUS (\cite{Cummins2003}) & 9 & 17.1 & $\epsilon$ err.  & $\epsilon^1$\\ \hline
         BB1 (\cite{Wimperis1994}) & 11 & 21.5 & $\epsilon$ err.  & $\epsilon^2$\\ \hline
          \multirow{2}{*}{\textbf{SCORE\textit{n}*}} & 9 & 18.6 & \multirow{2}{*}{$\delta$ err.} & $\delta^1$\\
          & 13 & 20.6 &  & $\delta^2$
         \\ \hline
         CORPSE* (\cite{Cummins2000}) & 9 & 20.6 & $\delta$ err. &  $\delta^1$ \\ \hline
          sCORPSE* (\cite{Cummins2003}) & 9 & 16.6 & $\delta$ err. &  $\delta^1$ \\ \hline
         \textbf{SCORE1inUZ1} & 11 & 22.6 & $\epsilon$, $\delta$ err. & $\epsilon^1, \delta^1$ \\ \hline
         SCORBUTUS* (\cite{KukitaPSJ2022})
         & 13 & 21.2 & $\epsilon$, $\delta$ err. & $\epsilon^1, \delta^1$ \\ 
         \hline 
         \textbf{RZ1*} (Eq. \ref{eq:rz1})& 34 & 90.1 & $\epsilon_s$ err. & $\epsilon_s^1$ \\ \hline
         \textbf{nUZ-SE} & 12 & 41.4 & $\epsilon_s$ err. & ---\\ \hline
         \textbf{nUZ-SORE} & 13 & 43.8 & $\epsilon_s$, $\delta$ err. & ---\\ \hline
         \textbf{nUZ-SAE} & 13 & 44.6 & $\epsilon_s$, $\epsilon$ err. & ---\\ \hline
         \textbf{nUZ-SAORE} & 16 & 51.0 & $\epsilon_s$, $\epsilon$, $\delta$ err. & ---\\ \hline \hline
         \multicolumn{5}{c}{Amplitude Control}\\ \hline \hline
         Basic (Eq. \ref{eq:arbuacap})& 3 & 2.25 & --- & ---\\ \hline
         \textbf{nUA-AE} & 6 & 8.5  & $\epsilon$ err. & ---\\\hline
         \textbf{RA1*} (Eq. \ref{eq:ra1})& 18 & 15.8 & $\epsilon$ err. & $\epsilon^1$ \\ \hline
          \textbf{nUA-ORE}  & 6 & 12.1 & $\delta$ err. & ---\\\hline
           \multirow{2}{*}{\textbf{SCORE\textit{n}*} } & 9 & 8.6 & \multirow{2}{*}{$\delta$ err.} & $\delta^1$\\
          & 15 & 13.2 &  & $\delta^2$
         \\ \hline
          CORPSE* (\cite{Cummins2000}) & 9 & 13.1 &  $\delta$ err. &  $\delta^1$ \\ \hline
           sCORPSE* (\cite{Cummins2003}) & 9 & 7.1 &  $\delta$ err. &  $\delta^1$ \\ \hline
         \textbf{nUA-AORE1}   & 8 & 9.1 & $\epsilon$, $\delta$ err.  &  ---  \\\hline
         \textbf{nUA-AORE2}  & 8 & 11.5 & $\epsilon$, $\delta$ err.  &  --- \\\hline 
         \textbf{SCORE1inRA1*} & 24 & 22.1 & $\epsilon$, $\delta$ err. & $\epsilon^1$, $\delta^1$ \\ 
    \end{tabular}
    \caption{Relevant pulse sequences and their properties for realizing robust, parallel, arbitrary single-qubit unitaries under PC, ZC, and AC. $k$ is the number of pulses, $T$ the total pulse area in units of $\pi$, and $n$ the highest order of error compensation. Sequence names which start with `n' indicate numerical solutions, with AE short for amplitude error, ORE for off-resonance error, and SE for Stark error. For ZC and AC, $T$ is the maximum pulse area of the target unitaries $\mathcal{U} \in [H, Z(\frac{\pi}{4}), X(\frac{\pi}{2}), Y(\frac{\pi}{2})]$. Pulse sequences introduced in our work are bolded. An asterisk indicates that robust arbitrary unitaries were achieved by directly replacing of each pulse the basic sequence. }
    \label{tab:pulcomp}
\end{table}

\begin{table}
    \centering
    \begin{tabular}{c|c|l|l}
    \multirow{2}{*}{$\mathcal{U}$} & $\alpha, \beta, \gamma$ &  $n = 1$ &  $n = 2$ \\
    &(in Eq. \ref{eq:arbupcap}) & $\phi_1, \phi_2$ & $\phi_1, \phi_2, \phi_3, \phi_4$  \\\hline \hline
    $H$ & 0, $\frac{1}{2}$, 1 & 0.65314, 1.04497 & 0.05665, 1.59435, 1.59435, 1.62537\\
    $Z(\frac{\pi}{4})$  & $\frac{1}{4}$, 0, 0 &0.15641, 1.03141 & 1.52435, 1.79167, 1.79167, 1.39935\\
    $Y(\frac{\pi}{2})$ & 0, $\frac{1}{2}$, 0&1.27333, 1.27333 & 0.38338, 0.09060, 0.09060, 0.38338\\
    $X(\frac{\pi}{2})$& $\frac{3}{2}$, $\frac{1}{2}$, $\frac{3}{2}$ & 0.77333, 0.77333 & 1.88338, 1.59060, 1.59060, 1.88338
\end{tabular}
    \caption{Phase values for UP1 and UP2 for composite $\mathcal{U} \in [H, Z(\frac{\pi}{4}), X(\frac{\pi}{2}), Y(\frac{\pi}{2})]$, given in units of $\pi$.}
    \label{tab:up12}
\end{table}

\begin{table}
    \centering
    \begin{tabular}{c|c|l}
    \multirow{2}{*}{$\mathcal{U}$} & $\alpha, \beta, \gamma$ &   \multirow{2}{*}{$\vec{\theta}$} \\
    &(in Eq. \ref{eq:arbuzcap})  & \\\hline \hline
    \multicolumn{3}{c}{nUZ-SE, $\vec{\phi} = (-, \frac{\pi}{2}, -, \frac{\pi}{2}, -, \frac{\pi}{2})$} \\
    \hline \hline
     $H$ & 0, $\frac{1}{2}$, 1 & 1.08366, 0.53716, 1.20067, 1.14080, 1.20596, 0.07654\\
     $Z(\frac{\pi}{4})$ & $\frac{1}{8}$, 0,  $\frac{1}{8}$ &1.25473, 1.70713, 1.00383, 0.83698, 2.00768, 0.87013\\
      $Y(\frac{\pi}{2})$ & 0, $\frac{1}{2}$, 0 & 0.58353, 0.99597, 1.16713, 0.99597, 0.58359, 0.49791 \\
       $X(\frac{\pi}{2})$& $\frac{3}{2}$, $\frac{1}{2}$, $\frac{1}{2}$ & 0.02064, 0.62665, 1.13836, 1.13812, 0.62691, 0.53602 \\\hline \hline
    \multicolumn{3}{c}{nUZ-SORE, $\vec{\phi} = (-, \frac{\pi}{2}, -, \frac{3\pi}{2}, -, \frac{\pi}{2}, -)$} \\
    \hline \hline
     $H$ & \multirow{4}{*}{"} &  0.84840, 0.55252, 1.59811, 1.33208, 0, 2.17230, 1.61088 \\
     $Z(\frac{\pi}{4})$ &  &0.13063, 0.47803, 0.47561, 2.5334 , 0.47554, 0.47811, 1.65445\\
      $Y(\frac{\pi}{2})$ & & 0.37560, 0.57247, 0.63551, 2.66431, 0.18085, 2.32658, 1.41004  \\
       $X(\frac{\pi}{2})$& &0.46315, 0.14846, 2.29735, 0.66463, 0.54004, 2.22517, 1.10436 \\ \hline \hline
       \multicolumn{3}{c}{nUZ-SAE, $\vec{\phi} = (-, \frac{\pi}{2}, -, \frac{\pi}{2}, -, \frac{\pi}{2}, -)$} \\
    \hline \hline
     $H$ & \multirow{4}{*}{"} &  0.03636, 2.26328, 0.74548, 1.29916, 1.86800, 0.49580, 0.19747\\
     $Z(\frac{\pi}{4})$ &  &0.53347, 1.10347, 0.88905, 1.99874, 1.10963, 0.89535, 1.71515\\
      $Y(\frac{\pi}{2})$ & & 1.76768, 0.52059, 1.57813, 0.85407, 2.13223, 0.90779, 0.55527\\
       $X(\frac{\pi}{2})$& &  0.60962, 0.58723, 0.87350, 1.69418, 1.15033, 0.64279, 1.43546 \\
         \hline \hline
       \multicolumn{3}{c}{nUA-SAORE, $\vec{\phi} = (-, \frac{\pi}{2}, -, \frac{3\pi}{2}, -, \frac{\pi}{2}, -, \frac{3\pi}{2})$} \\
    \hline \hline
     $H$ & \multirow{4}{*}{"} &  0.99248, 0.49112, 0.22071, 1.73502, 0, 0.74678, 0.22060,
         0.00018\\
     $Z(\frac{\pi}{4})$ &  &0.49956, 1.49761, 1.65139, 0.64947, 0.26679, 1.50275, 1.65547,
         0.67456\\
      $Y(\frac{\pi}{2})$ & & 0.27126, 0.87319, 1.67632, 1.77268, 0.07852, 0.87415, 1.49395,
         1.35859\\
       $X(\frac{\pi}{2})$& &  0.66534, 0.33698, 1.64227, 0.86345, 2.55432, 0.89766, 1.36726,
         1.07988\\
\end{tabular}
    \caption{Numerically determined pulse areas for nUZ-SE, nUZ-SORE, nUZ-SAE, and nUZ-SAORE for robust $\mathcal{U} \in [H, Z(\frac{\pi}{4}), X(\frac{\pi}{2}), Y(\frac{\pi}{2})]$, given in units of $\pi$. Pulse sequences all take the form of Eq. \ref{eq:numpulgenzc}, with $\vec{\phi}$ listed in the table. A $-$ in $\vec{\phi}$ indicates a $Z$ rotation, while rotations with $\phi_i$ are implemented by a local $Z$ conjugated by global pulses, $[\frac{\pi}{2}]_{\phi_i-\frac{\pi}{2}} Z(\theta_i) [\frac{\pi}{2}]_{\phi_i+\frac{\pi}{2}}$. Thus, each $x-y$ rotation actually requires three physical pulses. }
    \label{tab:numpulzc}
\end{table}

\begin{table}
    \centering
    \begin{tabular}{c|c|l}
    \multirow{2}{*}{$\mathcal{U}$} & $\alpha, \beta, \gamma$ &   \multirow{2}{*}{$\vec{\theta}$} \\
    &(in Eq. \ref{eq:arbuacap})  & \\\hline \hline
    \multicolumn{3}{c}{nUA-AE, $\vec{\phi} = (0, \frac{\pi}{2}, 0, \frac{\pi}{2}, 0, \frac{\pi}{2})$} \\
    \hline \hline
     $H$ & 1, $\frac{1}{2}$, 0 & 1.51178, 0.89113, 1.50143, 0.53102, 0.89067, 0.02874\\
     $Z(\frac{\pi}{4})$ & $\frac{1}{2}$, $\frac{1}{4}$, $\frac{1}{2}$ &1.47015, 1.92121, 0.78077, 1.22039, 2.29714, 0.84557\\
      $Y(\frac{\pi}{2})$ & 0, $\frac{1}{2}$, 0 & 0.58208, 0.99784, 1.16413, 0.99784, 0.58205, 0.49890 \\
       $X(\frac{\pi}{2})$& $\frac{1}{4}$, 0, $\frac{1}{4}$ & 0.49866, 0.58229, 0.99737, 1.16431, 0.99737, 0.58201 \\\hline \hline
    \multicolumn{3}{c}{nUA-ORE, $\vec{\phi} = (0, \pi, \frac{\pi}{2}, \frac{3\pi}{2}, 0, \pi)$} \\
    \hline \hline
     $H$ & \multirow{4}{*}{"} &  1.47516, 2.47516, 1.83179, 1.33179, 0.11990, 0.11990  \\
     $Z(\frac{\pi}{4})$ &  &2.2645 , 1.76446, 2.32205, 2.07205, 1.57995, 2.07998\\
      $Y(\frac{\pi}{2})$ & & 1.06356, 0.06356, 1.75095, 2.25095, 1.06692, 2.06692  \\
       $X(\frac{\pi}{2})$& & 0.15765, 1.71676, 2.01923, 2.01923, 2.09754, 0.03842  \\ \hline \hline
       \multicolumn{3}{c}{nUA-AORE1, $\vec{\phi} = (0, \frac{\pi}{2}, \frac{3\pi}{2}, \pi, \frac{\pi}{2}, 0, \pi, \frac{3\pi}{2})$} \\
    \hline \hline
     $H$ & \multirow{4}{*}{"} &  1.55547, 1.52935, 1.86082, 1.82989, 0.37897, 0.00049, 0.63237,
         1.34055  \\
     $Z(\frac{\pi}{4})$ &  &0.54062, 1.87900  , 1.69114, 2.14132, 0.06719, 1.60977, 0,
         0\\
      $Y(\frac{\pi}{2})$ & & 0.23369, 1.80984, 1.34955, 0.45633, 0.23642, 1.73862, 1.29756,
         0.00004 \\
       $X(\frac{\pi}{2})$& &  0.50000, 0.55366, 1.96769, 0.05042, 3.00000, 0.28430, 0.33472,
         1.58596 \\
         \hline \hline
       \multicolumn{3}{c}{nUA-AORE2, $\vec{\phi} = (0, \frac{\pi}{2}, 0, \frac{\pi}{2}, \frac{3\pi}{2}, \pi, \frac{3\pi}{2}, \pi)$} \\
    \hline \hline
     $H$ & \multirow{4}{*}{"} &  0.46187, 1.35972, 0.51422, 0.59744, 0.16868, 1.22126, 0.04752,
         0.14264\\
     $Z(\frac{\pi}{4})$ &  &2.23936, 1.11608, 0.37698, 1.02046, 2.53839, 0.86708, 2.38421,
         0.91721 \\
      $Y(\frac{\pi}{2})$ & & 1.31437, 0.67537, 1.44474, 1.57055, 1.44498, 1.07006, 1.2211 ,
         0.73226 \\
       $X(\frac{\pi}{2})$& &  1.62544, 0.42495, 1.33176, 0.07258, 0.67993, 0.47797, 1.31842,
         2.70594\\
\end{tabular}
    \caption{Numerically determined pulse areas for nUA-AE, nUA-ORE, nUA-AORE1, and nUA-AORE2 for robust $\mathcal{U} \in [H, Z(\frac{\pi}{4}), X(\frac{\pi}{2}), Y(\frac{\pi}{2})]$, given in units of $\pi$. Pulse sequences all take the form of Eq. \ref{eq:numpulgenac} with global phases listed in the table.}
    \label{tab:numpulac}
\end{table}


\begin{table}
    \centering
    \begin{tabular}{c|llll}
    \multirow{2}{*}{$\theta$} & $n = 1$ &  $n = 2$ &  $n = 3$ &  $n = 4$\\
    & $\vartheta_1$ & $\vartheta_1$, $\vartheta_2$ & $\vartheta_1$, $\vartheta_2$, $\vartheta_3$ & $\vartheta_1$, $\vartheta_2$, $\vartheta_3$, $\vartheta_4$ \\\hline \hline
    $\frac{1}{4}$ & 0.81372 & 0.06352, 0.93832 & 1.37501, 1.99971, 0.560983 & 1.34820, 1.32669, 1.77042, 2.16800\\
    $\frac{1}{3}$ & 0.75290 & 0.08577, 0.91864 & 0.77637, 0.31366, 1.82680 & 1.41901, 1.35864, 1.77664, 2.13759\\
    $\frac{1}{2}$ & 0.63497 & 0.13343, 0.88199 & 0.72996, 0.23033, 1.73323  & 1.55280, 1.42267, 1.78586, 2.07559\\
    $\frac{2}{3}$ & 0.52412 & 0.18738, 0.85099 & 0.68181, 0.18894, 1.66825  & 1.67478, 1.47865, 1.78919, 2.02043\\
    $\frac{3}{4}$ & 0.47215 & 0.21761, 0.83860 & 0.65222, 0.17673, 1.64672 & 1.73053, 1.49972, 1.78853, 1.99939\\
    $1$ & $\frac{1}{3}$ &0.32640, 0.82043 & 0.54046, 0.16221, 1.62138 & 1.87342, 1.52524, 1.78436, 1.97330
\end{tabular}
    \caption{Angle values for SCORE\textit{n}, given in units of $\pi$.}
    \label{tab:scoren}
\end{table}

\begin{table}[h]
    \centering
    \begin{tabular}{ccl}
    Pulse  &$O(\epsilon^n)$ & $\phi_1, \phi_2, \cdots, \phi_n$ \\ \hline \hline
    UZ1 & $O(\epsilon^1)$ & 0.5804\\
    UZ2 & $O(\epsilon^2)$ & 1.5386, 1.3791\\
    UZ3 & $O(\epsilon^3)$ &  0.6784, 1.3839, 0.2796 \\
    UZ4 & $O(\epsilon^4)$ & 1.9165, 1.0236, 0.7520, 1.6641\\
    UZ5 & $O(\epsilon^5)$ & 0.62858, 1.8616, 0.4527, 1.2863, 0.5936
\end{tabular}
    \caption{Phase values for UZ\textit{n}, given in units of $\pi$.}
    \label{tab:uzn}
\end{table}

\begin{table}[h]
    \centering
    \begin{tabular}{ccl}
    Pulse  &$O(\epsilon^n)$ & $\phi_1, \phi_2, \cdots, \phi_n, \phi_{n+1}$ \\ \hline \hline
    sUZ1 & $O(\epsilon^1)$ & $\frac{1}{3}, \frac{5}{3}$\\
    sUZ2 & $O(\epsilon^2)$ & 1.5672, 1.8854, 0.6364\\
    sUZ3 & $O(\epsilon^3)$ &  1.756, 1.1839, 0.0933, 0.3306 \\
    sUZ4 & $O(\epsilon^4)$ & 1.8951, 0.7211, 0.2806, 1.4335, 1.958\\
    sUZ5 & $O(\epsilon^5)$ & 0.2016, 0.6218, 1.3453, 0.4018, 1.8117, 1.6699
\end{tabular}
    \caption{Phase values for sUZ\textit{n}, given in units of $\pi$.}
    \label{tab:suzn}
\end{table}
\begin{table*}[h]
    \centering
    \begin{tabular}{c|l}
    \multirow{2}{*}{$\theta$} & $n = 2$ \\
    &$\vartheta_1, \vartheta_2, \vartheta_3$ \\\hline \hline
    $\frac{1}{4}$  & 1.18634, 0.21419, 0.523428\\
    $\frac{1}{3}$   & 1.19320, 0.23072, 0.52993 \\
    $\frac{1}{2}$  & 1.20730, 0.26434, 0.54125  \\
    $\frac{2}{3}$  & 1.22175, 0.29810, 0.55081  \\
    $\frac{3}{4}$ & 1.22900, 0.31482, 0.55512 \\
    $1$  & 1.25037, 0.36348, 0.56721 
\end{tabular}
    \caption{Angle values for RA2 (Eq. \ref{eq:ra2}), given in units of $\pi$.}
    \label{tab:ra2}
\end{table*}
\bibliography{ddrefs}